\documentstyle{ar}

\newbox\grsign \setbox\grsign=\hbox{$>$} \newdimen\grdimen \grdimen=\ht\grsign
\newbox\simlessbox \newbox\simgreatbox
\setbox\simgreatbox=\hbox{\raise.5ex\hbox{$>$}\llap
     {\lower.5ex\hbox{$\sim$}}}\ht1=\grdimen\dp1=0pt
\setbox\simlessbox=\hbox{\raise.5ex\hbox{$<$}\llap
     {\lower.5ex\hbox{$\sim$}}}\ht2=\grdimen\dp2=0pt

\def\nhi{\noindent \hangindent=0.3cm}

\input psfig.sty

\def\al{$\alpha$}
\def\alphaox{$\alpha_{\rm ox}$}
\def\bet{$\beta$}
\def\amin{$^\prime$}

\def\asec{$^{\prime\prime}$}

\def\cc{cm$^{-3}$}
\def\deg{$^{\circ}$}

\def\cc{cm$^{-3}$}
\def\e#1{$\times$10$^{#1}$}
\def\etal{{et al. }}

\def\farcs{\hbox{$.\mkern-4mu^{\prime\prime}$}}

\def\hst{{\it HST}}
\def\kms{km s$^{-1}$}
\def\lamb{$\lambda$}
\def\lax{{$\mathrel{\hbox{\rlap{\hbox{\lower4pt\hbox{$\sim$}}}\hbox{$<$}}}$}}
\def\gax{{$\mathrel{\hbox{\rlap{\hbox{\lower4pt\hbox{$\sim$}}}\hbox{$>$}}}$}}
\def\simlt{\lower.5ex\hbox{$\; \buildrel < \over \sim \;$}}
\def\simgt{\lower.5ex\hbox{$\; \buildrel > \over \sim \;$}}
\def\lum{ergs s$^{-1}$}
\def\mbh{{$M_{\rm BH}$}}
\def\micron{{$\mu$m}}

\def\percm2{cm$^{-2}$}
\def\peryr{yr$^{-1}$}

\def\solum{$L_\odot$}
\def\solmass{$M_\odot$}
\def\civ{C~{\sc IV}}

\def\feii{Fe~{\sc II}}

\def\heii{He~{\sc II}}

\def\hii{H~{\sc II}}
\def\mgii{Mg~{\sc II}}
\def\oi{[O~{\sc I}]}
\def\oii{[O~{\sc II}]}
\def\oiii{[O~{\sc III}]}
\def\oiv{[O~{\sc IV}]}

\def\nii{[N~{\sc II}]}
\def\nv{[N~{\sc V}]}

\def\neiii{[Ne~{\sc III}]}

\def\nev{[Ne~{\sc V}]}
\def\sii{[S~{\sc II}]}
\def\siii{[S~{\sc III}]}

\def\ledd{$L_{{\rm Edd}}$}

\begin{document}
\title{Nuclear Activity in Nearby Galaxies}
\markboth{Ho}{Nuclear Activity in Nearby Galaxies}
\author{Luis C. Ho
\affiliation{The Observatories of the Carnegie
Institution of Washington, 813 Santa Barbara St., Pasadena, CA 91101;
e-mail: lho@ociw.edu}}
\begin{keywords}
accretion disks, active galactic nuclei, black holes, LINERs, radio galaxies, 
Seyfert galaxies
\end{keywords}

%\noindent{Send Proofs to:}
%
%\noindent{Luis C. Ho} \\
%The Observatories of the Carnegie Institution of Washington\\
%813 Santa Barbara St. \\
%Pasadena, CA 91101 \\
%Phone: (626)-304-0248 \\
%Fax: (626)-795-8136 \\
%e-mail: lho@ociw.edu \\

\begin{abstract}
A significant fraction of nearby galaxies show evidence of weak nuclear 
activity unrelated to normal stellar processes.  Recent high-resolution, 
multiwavelength observations indicate that the bulk of this activity derives 
from black hole accretion with a wide range of accretion rates.  The low 
accretion rates that typify most low-luminosity active galactic nuclei induce 
significant modifications to their central engine.  The broad-line region and 
obscuring torus disappear in some of the faintest sources, and the optically 
thick accretion disk transforms into a three-component structure consisting of 
an inner radiatively inefficient accretion flow, a truncated outer thin disk, 
and a jet or outflow.  The local census of nuclear activity supports the 
notion that most, perhaps all, bulges host a central supermassive black hole, 
although the existence of active nuclei in at least some late-type 
galaxies suggests that a classical bulge is not a prerequisite to seed a 
nuclear black hole.  
\end{abstract}

\maketitle

\section{INTRODUCTION}

Far from being rare, exotic entities that inhabit only a tiny fraction of 
galaxies, central black holes (BHs) are now believed to be basic constituents 
of most, if not all, massive galaxies (Magorrian et al. 1998; Kormendy 2004).
Although less common in low-mass systems, central BHs also have been 
identified in some late-type, even dwarf, galaxies (Filippenko \& Ho 
2003; Barth et al. 2004; Greene \& Ho 2004, 2007b; Dong et al. 2007; Greene, 
Ho \& Barth 2008).  The realization that BH mass correlates strongly with the 
properties of the host galaxy (Kormendy 1993; Kormendy \& Richstone 1995; 
Magorrian et al.  1998; Gebhardt et al. 2000; Ferrarese \& Merritt 2000; 
Barth, Greene \& Ho 2005; Greene \& Ho 2006) has generated intense interest in 
linking BH growth 
with galaxy formation, as attested by the increasing number of conferences 
focusing on this theme (e.g., Schmitt, Kinney \& Ho 1999; Ho 2004a; 
Storchi-Bergmann, Ho \& Schmitt 2004; Merloni, Nayakshin \& Sunyaev 2005; 
Fiore 2006).  As a direct manifestation of BH accretion, and therefore BH 
growth, active galactic nuclei (AGNs) and the consequences of their energy 
feedback have figured prominently in most current ideas of structure formation 
(e.g., Granato et al.  2004; Springel, Di~Matteo \& Hernquist 2005; Hopkins et 
al. 2006).  At the same time, the community's heightened awareness of the 
importance of BHs has galvanized broad interest in the study of 
the AGN phenomenon itself.  With the BH mass known---arguably the most 
fundamental parameter of the system---what once rested on phenomenological 
analysis can now be put on a more secure physical basis.  

This review focuses on nuclear activity in nearby galaxies.  By selection, 
most of the objects occupy the faintest end of the AGN luminosity function 
and have very low accretion rates. While energetically
unimpressive, low-luminosity AGNs (LLAGNs) deserve scrutiny for several 
reasons.  By virtue of the short duty cycle of BH accretion ($\sim 10^{-2}$; 
Greene \& Ho 2007a), most AGNs spend their lives in a low state, such that the 
bulk of the population has relatively modest luminosities.  Over the past 
several decades, this attribute has led to considerable controversy regarding 
the physical origin of LLAGNs.  As absolute luminosity can no longer be used 
as a defining metric of nonstellar activity, many alternative excitation 
mechanisms have been proposed to explain LLAGNs.  Fortunately, the advent of 
new telescopes and new analysis techniques have yielded many fresh insights 
into this thorny old problem.  A major goal of this review is to summarize 
these recent developments.  Along the way, I will emphasize how the collective 
properties of LLAGNs can shed light on a poorly understood regime of the 
central engine, namely that governed by low mass accretion rate.  A key point 
I will stress is that LLAGNs are not simply scaled-down versions of their more 
familiar cousins, the classical Seyfert galaxies and quasars.  

Despite the impressive progress made in the direct detection of central BHs in 
nearby inactive galaxies, our knowledge of the 
demographics of BHs remains highly incomplete.  Direct measurements of BH 
masses based on resolved gas or stellar kinematics are still far from routine 
and are available only for about three dozen galaxies.  Certainly 
nothing approaching a ``complete'' sample exists yet.  More importantly, it is 
not obvious that the current statistics are unbiased.  As discussed by Barth 
(2004), most nearby galaxies possess chaotic gas velocity fields that defy 
simple analysis.  Stellar kinematics provide a powerful alternative, but in 
practice this technique has been limited to relatively dust-free systems and, 
for practical reasons, to galaxies of relatively high central surface 
brightness.  The latter restriction selects against the most luminous, giant 
ellipticals.  Present surveys also severely underrepresent disk-dominated 
galaxies, because the bulge component in these systems is inconspicuous and 
star formation tends to perturb the velocity field of the gas.  Finally, apart 
from galaxies within the Local Group, even the highest angular resolution 
currently achieved is inadequate to directly detect BHs with masses \lax\ 
$10^6$ \solmass.  Consequently we are nearly completely ignorant about the low 
end of the BH mass function.  Given the above limitations, it is 
desirable to consider alternative constraints on BH demography.  The commonly 
held and now well-substantiated premise that AGNs derive their energy 
output from BH accretion implies that an AGN signifies the presence of a 
central BH in a galaxy.  The AGN signature in and of itself provides little 
direct information on BH masses, but AGN statistics can inform us, effectively 
and efficiently, of some key aspects of BH demography.  For example, what 
fraction of all galaxies contain BHs?  Do BHs exist preferentially in galaxies 
of certain types?  Does environment matter?  I will discuss how studies of 
nearby AGNs have begun to answer some of these important questions.

This review is structured as follows.  I begin with an overview of the basic
methodology of the spectral classification of emission-line nuclei (\S~2) by
describing the currently adopted system, its physical motivation, the
complications of starlight subtraction, and some practical examples.  Section
3 summarizes past and current spectroscopic surveys and introduces the 
Palomar survey, covering detection rates, measurement of weak broad emission 
lines, and issues of robustness and completeless.  Host galaxy properties are
the subject of \S~4, where in addition to global and environmental effects 
I also cover results on nuclear stellar populations.  In Section 5, I devote 
considerable attention to the nuclear properties of LLAGNs in general and 
LINERs in particular, focusing on modern results obtained from 
high-resolution, multiwavelength observations from radio to hard-X-ray 
energies.  I use these data to draw inferences concerning the broad-line 
region (BLR), torus, narrow-line region (NLR), spectral energy distribution 
(SED), luminosity function, bolometric luminosities, and Eddington ratios.  
This section contains many technical details, but these will be 
essential ingredients for formulating the big picture at the end.  Section 6 
covers the controversial subject of the excitation mechanism of LINERs, the 
growing puzzle concerning the energy budget in these systems, and the nature 
of narrow-line nuclei.  The implications of LLAGNs for BH demographics are 
discussed in Section 7. Section 8 attempts to synthesize the disparate lines 
of evidence into a coherent physical framework for LLAGNs and their relation 
to other classes of objects.  Finally, Section 9 concludes with some personal 
perspectives and suggestions for future directions.

\section{SPECTRAL CLASSIFICATION OF GALACTIC NUCLEI}

\subsection{Physical Motivation}

AGNs can be identified by a variety of methods.  Most AGN surveys rely on some
aspect of the distinctive AGN spectrum, such as the presence of strong or
broad emission lines, an unusually blue continuum, or strong radio or X-ray
emission.  All of these techniques are effective, but none is free from
selection effects.  To search for AGNs in nearby galaxies, where the
nonstellar signal of the nucleus is expected to be weak relative to the host
galaxy, the most effective and least biased method is to conduct a
spectroscopic survey of a complete, optical-flux limited sample of galaxies.
To be sensitive to weak emission lines, the survey must be deep and of
sufficient spectral resolution.  To obtain reliable line intensity ratios on
which the principal nuclear classifications are based, the data must have
accurate relative flux calibration, and one must devise a robust scheme to
correct for the starlight contamination. 

The most widely used system of spectral classification of emission-line nuclei
follows the method promoted by Baldwin, Phillips \& Terlevich (1981), and
later modified by Veilleux \& Osterbrock (1987).  The basic idea is that the
relative strengths of certain prominent emission lines can be used to probe
the nebular conditions of a source.  In the context of the present discussion,
the most important diagnostic is the source of excitation, which broadly falls
into two categories: stellar photoionization or photoionization by a centrally
located, spectrally hard radiation field, such as that produced by the
accretion disk of a massive BH.  How does one distinguish between the two?  The
forbidden lines of the doublet \oi\ \lamb\lamb 6300, 6364 arise from
collisional excitation of O$^0$ by hot electrons.  Since the ionization
potential of O$^0$ (13.6 eV) is nearly identical to that of hydrogen, in an
ionization-bounded nebula \oi\ is produced predominantly in the ``partially
ionized zone,'' wherein both neutral oxygen and free electrons coexist.  In
addition to O$^0$, the conditions of the partially ionized zone are also
favorable for S$^+$ and N$^+$, whose ionization potentials are 23.3 eV and
29.6 eV, respectively.  Hence, in the absence of abundance anomalies, \nii\
\lamb\lamb 6548, 6583 and \sii\ \lamb\lamb 6716, 6731 are strong (relative to,
say, H\al) whenever \oi\ is strong, and {\it vice versa}.

In a nebula photoionized by young, massive stars, the partially ionized zone
is very thin because the ionizing spectrum of OB stars contains few photons
with energies greater than 13.6 eV.  Hence, in the optical spectra of \hii\
regions and starburst nuclei (hereinafter \hii\ nuclei\footnote{As originally
defined (Weedman et al. 1981), a star{\it burst}\ nucleus is one whose current 
star formation rate is much higher than its past average rate.  Since in 
general we do not know the star formation history of any individual object, I 
will adopt the more general designation of ``\hii\ nucleus.''}) the 
low-ionization transitions \nii, \sii, and especially \oi\ are very weak.  By 
contrast, a harder radiation field, such as that of an AGN power-law continuum 
that extends into the extreme-ultraviolet (UV) and X-rays, penetrates much 
deeper into an optically thick cloud, creating an extensive partially ionized 
zone and hence strong low-ionization forbidden lines.  A hard AGN radiation 
field also boosts the production of collisionally excited forbidden line 
emission because its high thermal energy deposition rate enhances the gas 
temperature.

\begin{figure}
\psfig{figure=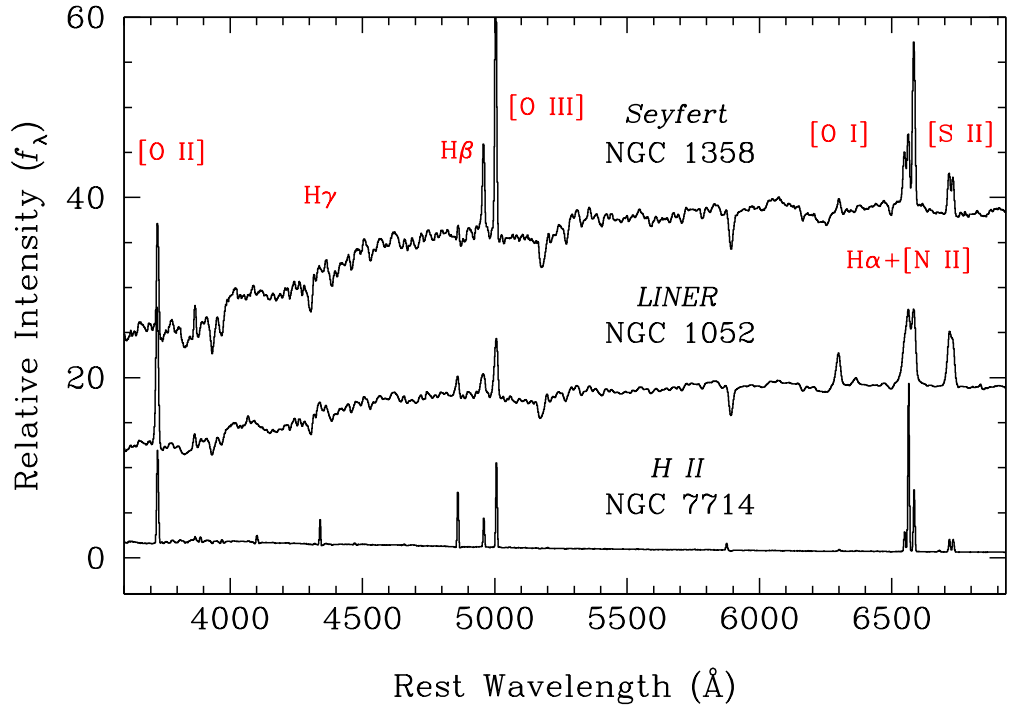,width=14cm,angle=270}
\caption{Sample optical spectra of the various classes of emission-line 
nuclei.  Prominent emission lines are identified.  (Based on Ho, Filippenko \& 
Sargent 1993 and unpublished data.)
}
\end{figure}

\subsection{Sample Spectra}
The spectra shown in Figure~1 illustrate the empirical distinction between
AGNs and \hii\ nuclei.  In NGC~7714, which has a well-known starburst nucleus
(Weedman et al. 1981), \oi, \nii, and \sii\ are weak relative to H\al.  The
\oiii\ \lamb\lamb 4959, 5007 doublet is quite strong compared to \oii\ \lamb
3727 or H\bet\ because the metal abundance of NGC~7714's nucleus is rather
low, although the ionization level of \hii\ nuclei can span a wide range,
depending on metallicity (Ho, Filippenko \& Sargent 1997c; Kewley et al. 
2001; Groves, Heckman \& Kauffmann 2006).  On the other hand, the 
low-ionization lines are markedly stronger in the other two objects shown, 
both of which qualify as AGNs.  NGC~1358 has a
``high-ionization'' AGN or ``Seyfert'' nucleus.  NGC~1052 is the prototype of 
the class known as ``low-ionization nuclear emission-line regions'' or 
LINERs (Heckman 1980b). The ionization level can be judged by the relative
strengths of the oxygen lines, but in practice is most easily gauged by the
\oiii/H\bet\ ratio.  In the commonly adopted system of Veilleux \& Osterbrock
(1987), the division between Seyferts and LINERs occurs at \oiii\ \lamb
5007/H\bet\ = 3.0.  Ho, Filippenko \& Sargent (2003) stress, however, that
this boundary has no strict physical significance.  The ionization level of the
NLR in large, homogeneous samples of AGNs spans a wide and
apparently continuous range; there is no evidence for any clear-cut transition 
between Seyferts and LINERs (Ho, Filippenko \& Sargent 2003), although with 
sufficient numbers, the two classes do delineate two distinct loci in optical 
diagnostic diagrams (Kewley et al. 2006).

The classification system discussed above makes no reference to the profiles
of the emission lines.  Luminous AGNs such as quasars and many classical
Seyfert galaxies exhibit permitted lines with a characteristically broad
component, with  full width at half-maximum (FWHM) widths of $\sim 1000$ to 
$10,000$ \kms.  This component arises from the BLR, which is thought to be 
physically distinct from the NLR responsible for the narrow lines.  Following 
Khachikian \& Weedman (1974), it is customary to refer to Seyferts with and 
without (directly) detectable broad lines as ``type~1'' and ``type~2'' 
sources, respectively.   As discussed in \S~3.4, this nomenclature can also be 
extended to include LINERs.

\begin{figure}
\psfig{figure=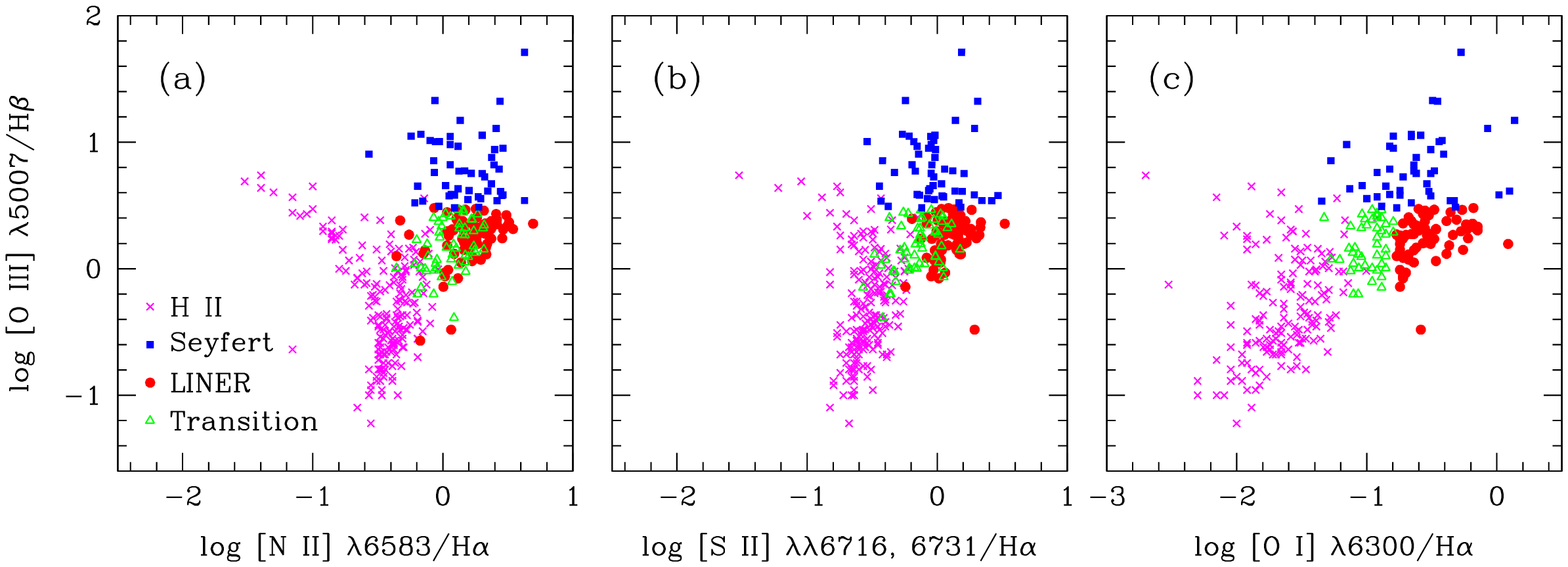,width=14cm,angle=270}
\caption{Diagnostic diagrams plotting ({\it a}) log \oiii\ \lamb 5007/H\bet\
versus log \nii\ \lamb 6583/H\al, ({\it b}) log \oiii\ \lamb 5007/H\bet\
versus log \sii\ \lamb\lamb 6716, 6731/H\al, and ({\it c}) log \oiii\ \lamb
5007/H\bet\ versus log \oi\ \lamb 6300/H\al.  
(Adapted from Ho, Filippenko \& Sargent 1997a.)
}
\end{figure}

\subsection{Diagnostic Diagrams}

The classification system of Veilleux \& Osterbrock (1987), which I
adopt throughout this paper, is based on two-dimensional line-intensity ratios
constructed from \oiii\ \lamb 5007, H\bet\ \lamb 4861, \oi\ \lamb 6300, H\al\
\lamb6563, \nii\ \lamb 6583, and \sii\ \lamb\lamb 6716, 6731 (here H\bet\ and
H\al\ refer only to the narrow component of the line).  The main virtues of
this system, shown in Figure~2, are (1) that it uses relatively strong lines,
(2) that the lines lie in an easily accessible region of the optical spectrum,
and (3) that the line ratios are relatively insensitive to reddening
corrections because of the close separation of the lines.  The definitions of 
the various classes of emission-line objects are given in Ho, Filippenko \& 
Sargent (1997a)\footnote{The classification criteria adopted here differ 
slightly from those proposed by Kewley et al. (2001), Kauffmann et al. (2003),
or Stasi\'nska et al. (2006), but this difference has little effect on the 
general conclusions.}.  In addition to the three main classes 
discussed thus far---\hii\ nuclei, Seyferts, and LINERs---Ho, Filippenko \& 
Sargent (1993) identified a group of ``transition objects'' whose \oi\ 
strengths are intermediate between those of \hii\ nuclei and LINERs.  Since 
they tend to emit weaker \oi\ emission than classical LINERs, previous authors 
have called them ``weak-\oi\ LINERs'' (Filippenko \& Terlevich 1992; 
Ho \& Filippenko 1993).  Ho, Filippenko \& Sargent (1993) postulated 
that transition objects are composite systems having both an \hii\ region and 
a LINER component; I will return to the nature of these sources in \S~6.5.

Note that my definition of LINERs differs from that originally proposed by
Heckman (1980b), who used solely the oxygen lines: \oii\ \lamb 3727 $>$
\oiii\ \lamb 5007 and \oi\ \lamb 6300 $>$ 0.33 \oiii\ \lamb 5007.  The two
definitions, however, are nearly equivalent.  Inspection of the full optical
spectra of Ho, Filippenko \& Sargent (1993), for example, reveals that 
emission-line nuclei classified as LINERs based on the Veilleux \& Osterbrock 
diagrams almost always also satisfy Heckman's criteria.  This is a consequence 
of the inverse correlation between \oiii/H\bet\ and \oii/\oiii\ in 
photoionized gas with fairly low excitation.

\subsection{Starlight Subtraction}

The scheme described above, while conceptually simple, overlooks one key
practical complication.  The integrated spectra of galactic nuclei include
starlight, which in most nearby systems overwhelms the nebular line
emission (Figure~1).  Any reliable measurement of the emission-line spectrum of 
galactic nuclei, therefore, {\it must}\ properly account for the starlight 
contamination.

An effective strategy for removing the starlight from an integrated spectrum
is that of ``template subtraction,'' whereby a template spectrum devoid of
emission lines is suitably scaled to and subtracted from the spectrum of
interest to yield a continuum-subtracted, pure emission-line spectrum.  A
number of approaches have been adopted to construct the template.  These
include using (1) the spectrum of an off-nuclear position within the same
galaxy (e.g., Storchi-Bergmann, Baldwin \& Wilson 1993); (2) the spectrum of 
a different galaxy devoid of emission lines (e.g., Costero \& Osterbrock 1977; 
Filippenko \& Halpern 1984; Ho, Filippenko \& Sargent 1993); (3) a weighted 
linear combination of the spectra of a number different galaxies, chosen to 
best match the stellar population and velocity dispersion (Ho, Filippenko \& 
Sargent 1997a); (4) a variant of (3), but employing a stellar library and 
simultaneously fitting for the emission lines and accounting for dust 
reddening (Sarzi et al. 2007); (5) a mean spectrum derived from a 
principal-component analysis of a large set of galaxies (Hao et al. 2005a); 
and (6) a model spectrum constructed from population synthesis techniques, 
using as input a library of spectra of either individual stars (e.g., Keel 
1983c), synthesis models (e.g., Tremonti et al. 2004; Sarzi et al. 2005), or 
star clusters (e.g., Bonatto, Bica \& Alloin 1989).  Some studies (e.g., Kim 
et al. 1995) implicitly assume that only the hydrogen Balmer lines are 
contaminated by starlight and that the absorption-line component can be 
removed by assuming a constant equivalent width (EW = $2-3$ \AA).  This 
procedure is inadequate for a number of reasons.  First, the stellar 
population of nearby galactic nuclei, although relatively 

\clearpage
\begin{figure}
\psfig{figure=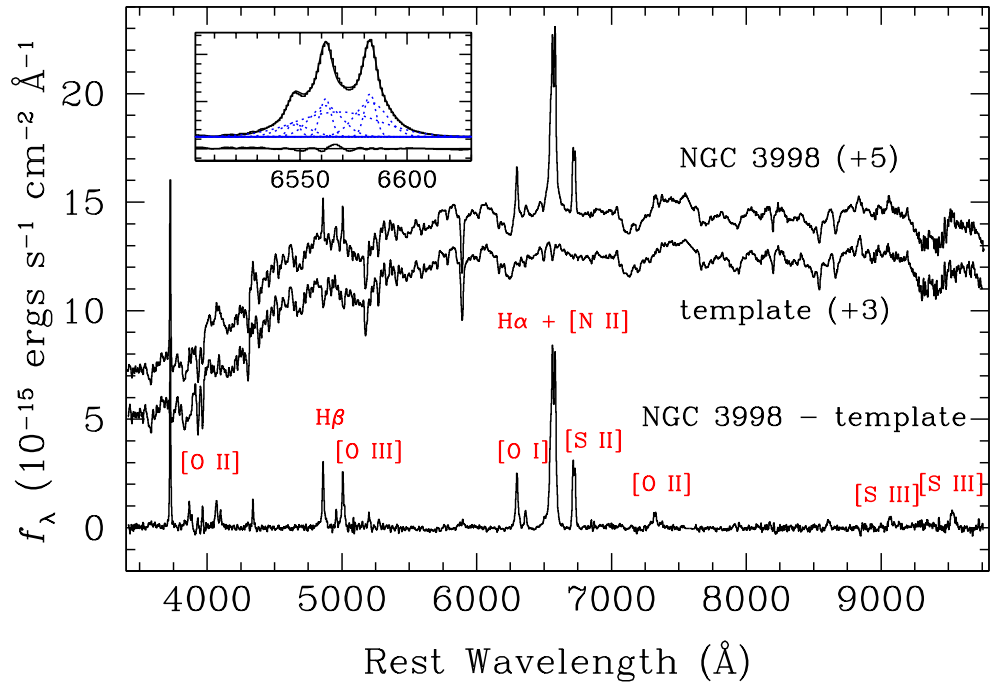,width=13.7cm,angle=270}
\caption{Illustration of starlight removal for NGC~3998 using the template
galaxy NGC~3115.  Prominent emission lines are labeled. The insert shows an 
expanded view of the H\al+\nii\ region and a multi-Gaussian decomposition 
leading to the detection of a broad H\al\ component.  (Adapted from Ho, 
Filippenko \& Sargent 1993 and Ho et al. 1997e.)
}
\end{figure}

\noindent
uniform, is by no 
means invariant (Ho, Filippenko \& Sargent 2003).  Second, the equivalent 
widths of the different Balmer absorption lines within each galaxy are 
generally not constant.  Third, the Balmer absorption lines affect not only 
the strength but also the shape of the Balmer emission lines.  And finally, 
starlight contaminates more than just the Balmer lines.

Figure~3 illustrates the starlight subtraction process for the LINER NGC~3998.  Note that in the original observed spectrum, many of the weaker emission lines 
were hardly visible, whereas after starlight subtraction, they can easily be 
measured.  The intensities of even strong lines such H\bet\ and \oiii\ 
\lamb\lamb4959, 5007 are modified. Importantly, starlight correction is 
essential for properly identifying the weak broad H\al\ component in NGC~3998.

\subsection{Other Classification Criteria}

Although the traditional optical classification system described above is 
the most widely used, there are instances when features in other spectral 
regions may be more practical or useful.  Surveys of intermediate-redshift 
galaxies, for example, cannot routinely access the H\al\ region, and under 
such circumstances it is desirable to devise a classification system based 
only on the blue part of the spectrum.  Diagnostic diagrams proposed by 
Rola, Terlevich \& Terlevich (1997) based on the strong lines \oii\ 
\lamb3727, \neiii\ \lamb\lamb 3869, 3968, H\bet, and \oiii\ \lamb\lamb 4959, 
5007 provide moderately effective discrimination between starbursts and 
AGNs.  A number of redshift surveys have searched for narrow-line AGNs based 
on the presence of \nev\ \lamb\lamb 3346, 3426 (e.g., Hall et al. 2000; Barger 
et al.  2001; Szokoly et al. 2004).  With an ionization potential of 97 eV, 
\nev\ unambiguously arises from nonstellar excitation, but the practical 
difficulty is that these lines are quite weak (strength $\sim$10\% of 
\oiii\ \lamb5007) and often can only be detected with confidence in 
stacked spectra (e.g., Zakamska et al. 2003).

The long-standing controversy over the relevance of shock excitation also 
has led to the development of line diagnostics outside of the traditional 
optical window.  D\'\i az, Pagel \& Wilson (1985; see also D\'\i az, Pagel 
\& Terlevich 1985; Kirhakos \& Phillips 1989) suggested that \siii\ 
\lamb\lamb9069, 9532, in combination with the optical lines of \oii, \oiii, 
and \sii, are effective in identifying shock-excited nebula.  Since shock 
heating on average achieves higher equilibrium electron temperatures than 
photoionization, high-ionization UV lines such as \nv\ \lamb 1240 and 
\civ\ \lamb 1549 can serve as a powerful discriminant between these two 
processes (e.g., Allen, Dopita \& Tsvetanov 1998).  The limited availability 
of UV spectra, however, has restricted the wide use of these diagnostics.

Rapid progress in infrared (IR) technology has offered an important new window 
that is not only less affected by dust but also potentially has distinctive 
diagnostic power.  Alonso-Herrero et al. (1997) show that 
\feii\ 1.644/Br$\gamma$ can serve as an effective substitute for the 
conventional \oi\ \lamb6300/H\al\ ratio.  Unfortunately, other strong near-IR 
features, notably the vibrational lines of H$_2$, are less useful because they 
can be excited by multiple mechanisms (Larkin et al. 1998).  The mid-IR regime 
is much more promising, particularly with the sensitivity and wide bandpass 
afforded by {\it Spitzer}.  For the first time, many of the diagnostic lines 
previously discussed in a theoretical context (Spinoglio \& Malkan 1992; Voit 
1992) actually can now be measured (e.g., Bendo et al. 2006; Dale et al. 2006; 
Sturm et al. 2005, 2006; Rupke et al.  2007).  In addition to high-ionization 
lines such as \neiii\ \lamb15.5 \micron, \nev\ \lamb14.3 \micron, and \oiv\ 
\lamb 25.9 \micron, the low-ionization transitions of \feii\ \lamb26.0 
\micron\ and [Si~{\sc II}] \lamb34.8 \micron\ may prove to be especially useful as 
they can constrain models of photo-dissociation and X-ray dissociation 
regions.  The hard radiation field of AGNs, even of low-luminosity objects 
such as LINERs, appears to leave an imprint on the detailed emission spectrum 
of polycyclic aromatic hydrocarbons (Sturm et al. 2006; Smith et al. 2007).

Finally, a comment on nomenclature.  It is important to stress that the 
classification scheme outlined above, physically motivated by the desire to 
separate objects by their source of excitation, is based strictly on the 
characteristics of the narrow emission lines and not on ancillary attributes 
such as luminosity, presence of broad emission lines, galaxy morphology, or 
radio properties.  Although one still customarily draws a quaint distinction 
between quasars and Seyferts based on luminosity, it is widely acknowledged 
that this division is largely historical.  In terms of their position on the 
line-ratio diagrams, quasars fall on the high-ionization branch and thus can be 
classified as Seyferts.  The same holds for many broad-line and narrow-line 
radio galaxies, including most Fanaroff \& Riley (FR; 1974) type~II radio 
sources, whose high luminosities generally translate directly into a high 
degree of ionization.  By the same token, most FR~I sources, because of their 
low luminosity, typically have fairly low-ionization spectra, and hence 
technically qualify as LINERs.  FR~I radio galaxies and LINERs are not 
separate beasts (cf. Falcke, K\"ording \& Markoff 2004; Chiaberge, Capetti 
\& Macchetto 2005).  Strong historical prejudice also compels many to regard 
Seyfert nuclei as invariably radio-quiet sources that reside exclusively in 
spiral hosts, when, in fact, neither rule strictly holds (Ho \& Peng 2001).  
Despite claims to the contrary (Krolik 1998; Sulentic, Marziani \& 
Dultzin-Hacyan 2000), broad emission lines emphatically are {\it not}\ 
solely confined to Seyfert nuclei (\S~3.4).  This misconception has led 
some people to define the Seyfert and LINER classes by their presence or 
absence of broad emission lines.

\section{SURVEYS OF NEARBY GALACTIC NUCLEI}

\subsection{The Palomar Survey}

The earliest redshift surveys already indicated that the spectra of galaxy 
centers often show strong emission lines (e.g., Humason, Mayall \& Sandage 
1956).  In many instances, the spectra revealed abnormal line-intensity 
ratios, most notably the unusually great strength of \nii\ relative to H\al\ 
(Burbidge \& Burbidge 1962, 1965; Rubin, Ford \& Thonnard 1980; Rose \& Searle 
1982).  That the optical emission-line spectra of some nuclei show patterns of 
low ionization was noticed from time to time, primarily by Osterbrock and his 
colleagues (e.g., Osterbrock \& Dufour 1973; Osterbrock \& Miller 1975; Koski 
\& Osterbrock 1976; Costero \& Osterbrock 1977; Grandi \& Osterbrock 1978; 
Phillips 1979), but also by others (e.g., Disney \& Cromwell 1971; Danziger, 
Fosbury \& Penston 1977; Fosbury \etal 1977, 1978; Penston \& Fosbury 1978; 
Stauffer \& Spinrad 1979).

The activity in this field culminated in the 1980s, beginning with the
recognition (Heckman, Balick \& Crane 1980; Heckman 1980b) of LINERs as a
major constituent of the extragalactic population, and then followed by further
systematic studies of larger samples of galaxies (Stauffer 1982a, 1982b; Keel
1983b, 1983c; Phillips \etal 1986; V\'eron \& V\'eron-Cetty 1986; 
V\'eron-Cetty \& V\'eron 1986; see Ho 1996 for more details).  These surveys 
established three important results. (1) A large fraction of local galaxies 
contain emission-line nuclei. (2) Many of these sources are LINERs.  And (3) 
LINERs may be accretion-powered systems.

Despite the successes of these seminal studies, there was room for improvement.
Although most of the surveys attempted some form of starlight subtraction, the
accuracy of the methods used was limited (see discussion in
Ho, Filippenko \& Sargent 1997a), the procedure was sometimes inconsistently 
applied, and in some of the surveys, starlight subtraction was altogether 
neglected.  The problem is exacerbated by the fact that the apertures used for 
the observations were quite large, thereby admitting an unnecessarily large 
amount of starlight.  Furthermore, most of the data were collected with rather 
low spectral resolution (FWHM $\approx$ 10 \AA).  Besides losing useful
kinematic information, blending between the emission and absorption
components further compromises the ability to separate the two.

Thus, it was clear that much would be gained from a survey having greater
sensitivity to the detection of emission lines.  The sensitivity could be
improved in at least four ways: by taking spectra with higher signal-to-noise 
ratio and spectral resolution, by using a narrower slit to better isolate the 
nucleus, and by employing more effective methods to handle the starlight 
correction.

The Palomar spectroscopic survey of nearby galaxies (Filippenko \& Sargent
1985; Ho, Filippenko \& Sargent 1995, 1997a, 1997b, 1997c, 1997d, 2003; Ho 
et al. 1997e) was designed with these goals in mind.  Using a double CCD 
spectrograph mounted on the Hale 5-m reflector, high-quality, 
moderate-resolution, long-slit spectra were obtained for a magnitude-limited 
($B_T\,\leq$ 12.5 mag) sample of 486 northern ($\delta$ $>$ 0\deg) galaxies.  
Drawn from the Revised Shapley-Ames (RSA) Catalog of Bright Galaxies (Sandage 
\& Tammann 1981), the bright magnitude limit ensured that the sample had a 
high degree of completeness.  The spectra simultaneously cover the wavelength 
ranges $6210-6860$ \AA\ with $\sim$2.5 \AA\ resolution (FWHM) and $4230-5110$ 
\AA\ with $\sim$4 \AA\ resolution.  Most of the observations were obtained 
with a narrow (1\asec$-$2\asec) slit, and relatively long exposure times 
gave high signal-to-noise ratios.  This survey still contains 
the largest database to date of homogeneous and high-quality optical spectra 
of nearby galaxies.  It is also the most sensitive; the detection limit for 
emission lines is EW $\approx$ 0.25 \AA, roughly an order-of-magnitude 
improvement compared to previous or subsequent work.  The selection criteria 
ensure that the sample gives a fair representation of the local 
($z\,\approx\,0$) galaxy population, and the proximity of the objects (median 
distance = 17 Mpc) results in relatively good spatial resolution (typically 
\lax\ 200 pc)\footnote{A distance scale based on $H_0$ = 75 \kms~Mpc$^{-1}$ is 
assumed throughout}.  These properties of the Palomar survey make it 
ideally suited to address issues on the demographics and physical properties 
of nearby, and especially low-luminosity, AGNs.  Unless otherwise noted, most 
of the results presented in this paper will be taken from the Palomar survey.  

The Palomar survey has one other virtue that is not widely appreciated.
Because the sample is large and essentially unbiased with respect to 
nuclear or global properties, it is ideally suited for comparative studies 
of various subpopulations.  Examples include efforts to discern differences 
between type~1 versus type~2 sources to test AGN unification, to ascertain the 
influence of bars or environment on nuclear activity, or to test for subtle 
differences between the different AGN classes.  The robustness of these and 
similar studies almost always hinges on the availability of proper control 
samples.  With the Palomar survey, there is no need to construct a separate 
control sample, which is always a difficult and somewhat dubious undertaking, 
because the control sample is built into the survey.

\subsection{Other Surveys} 

For completeness, I mention several other sources of nearby AGNs that have 
been widely used by the community.  The AGN sample culled from the CfA 
Redshift Survey (Huchra \& Burg 1992) has been an important resource for a 
long time.  Comprising 47 relatively bright Seyferts and a handful of LINERs, 
the CfA sample in many ways complements the Palomar sample at the bright end 
of the luminosity function.  However, as discussed in Ho \& Ulvestad (2001), 
the selection effects of the CfA sample are not easy to quantify because of 
the subjective and somewhat nonstandard manner in which AGNs were picked from 
the parent survey.  Prior to the full publication of the Palomar survey, 
Maiolino \& Rieke (1995) assembled a compilation of 91 Seyferts from a 
literature search of the galaxies in the RSA.  These ``RSA Seyferts'' have 
subsequently been used in a number of follow-up studies.  The substantial 
improvement in the data quality and analysis of the Palomar survey has 
resulted in many revised classifications of the RSA galaxies.  Lastly, a 
cautionary note. Many investigators rely on literature compilations, such as 
those assembled in V\'eron-Cetty \& V\'eron's (2006) catalog or the
NASA/IPAC Extragalactic Database, as their source for AGN classifications.  
This is dangerous.  The classifications in these compilations are 
highly heterogeneous and in some cases wrong.

The sample of nearby AGNs emerging from the Sloan Digital Sky Survey (SDSS)
(Kauffmann et al. 2003; Hao et al. 2005b; Kewley et al. 2006) far surpasses 
that of the Palomar survey in number but not in sensitivity.  Moreover, 
because SDSS samples more distant galaxies, the 3\asec-diameter fibers used in 
the survey subtend a physical scale of $\sim$5.5 kpc at the typical redshift 
$z \approx 0.1$, 30 times larger than in the Palomar survey.  The SDSS spectra 
therefore include substantial contamination from off-nuclear emission, which 
dilutes and, in some cases, inevitably confuses the signal from the nucleus.  
Contamination by host galaxy emission has two consequences.  First, only 
relatively bright nuclei have enough contrast to be detected.  But second, 
contamination can introduce a more pernicious systematic effect 
that can be hard to quantify.  Apart from normal \hii\ regions, galactic disks 
are known to contain {\it extended}\ emission-line regions that exhibit 
low-ionization, LINER-like spectra.  They can be confused with genuine {\it 
nuclear}\ LINERs.  Examples include gas shocked by supernova remnants (e.g., 
Dopita \& Sutherland 1995), ejecta from starburst-driven winds (Armus, Heckman 
\& Miley 1990), and diffuse, warm ionized plasma (e.g., Collins \& 
Rand 2001).  Massive, early-type galaxies, though generally lacking in ongoing 
star formation, also often possess X-ray emitting atmospheres that exhibit 
extended, low-ionization emission-line nebulae (e.g., Fabian et al. 1986; 
Heckman et al. 1989).  These physical processes, while interesting in their 
own right, are not directly related, and thus irrelevant, to the AGN 
phenomenon.  Thus, LINERs selected from samples of distant galaxies should be 
regarded with considerable caution. This comment applies also 
to LINERs selected from samples of IR-bright galaxies (e.g., Kim et 
al. 1995; Kewley et al. 2001; Corbett et al. 2003), which, in addition to being 
relatively distant and maximally confused with starburst processes, have the 
additional disadvantage of often being merging or interacting systems, wherein 
shocks undoubtedly generate extended LINER-like emission.  I strongly 
recommend that researchers avoid IR-selected samples if they are interested in 
investigating LINERs as an accretion phenomenon.  Many of the objects in the 
catalog of LINERs compiled by Carrillo et al. (1999), which has been the basis 
of several recent studies (Satyapal, Sambruna \& Dudik 2004; Dudik et al. 
2005; Gonz\'alez-Mart\'\i n et al. 2006), suffer precisely from this 
complication and should be used judiciously.

\begin{figure}
\hskip 0.0truein
\psfig{figure=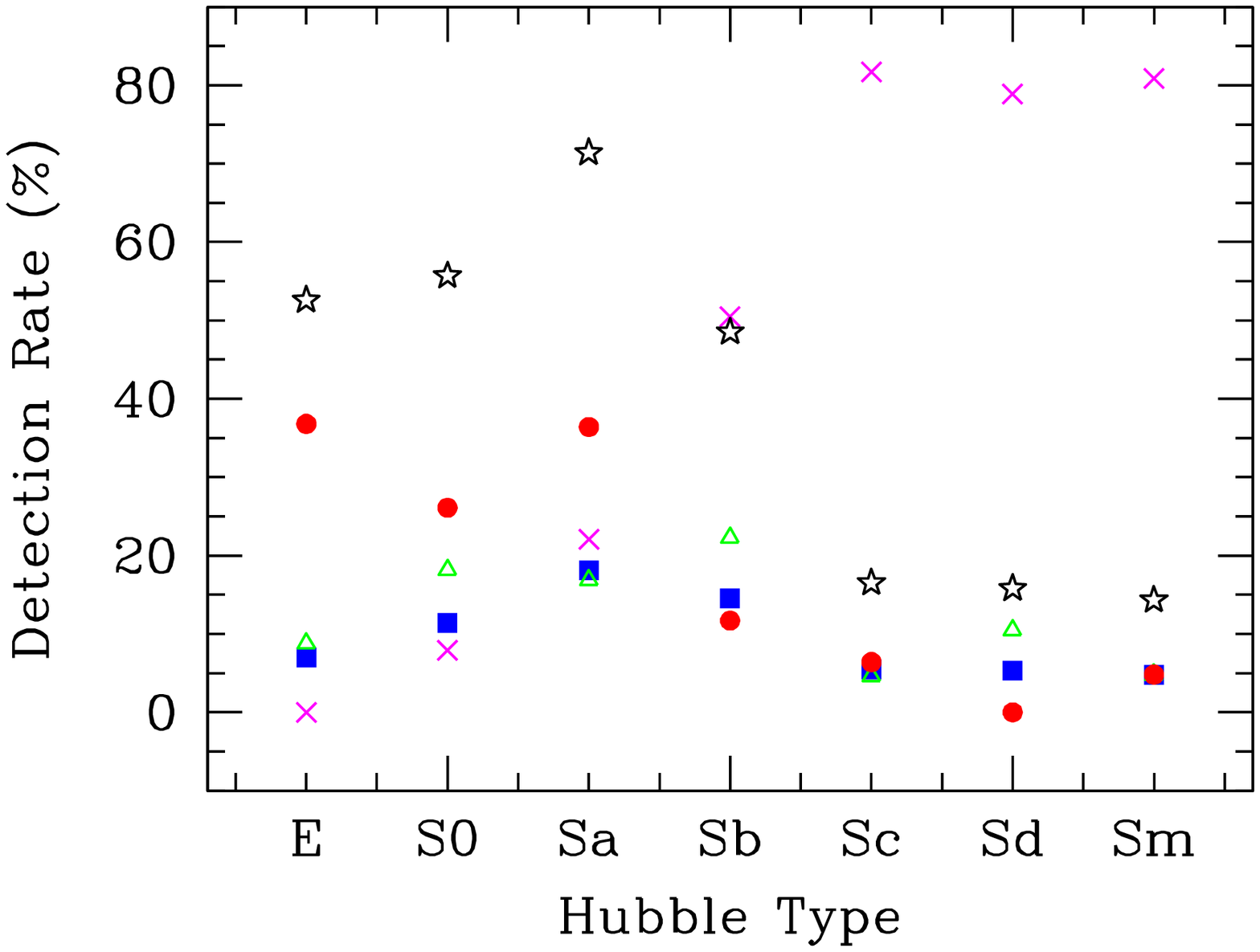,width=6.9cm,angle=0}
\hskip -0.2truein
\psfig{figure=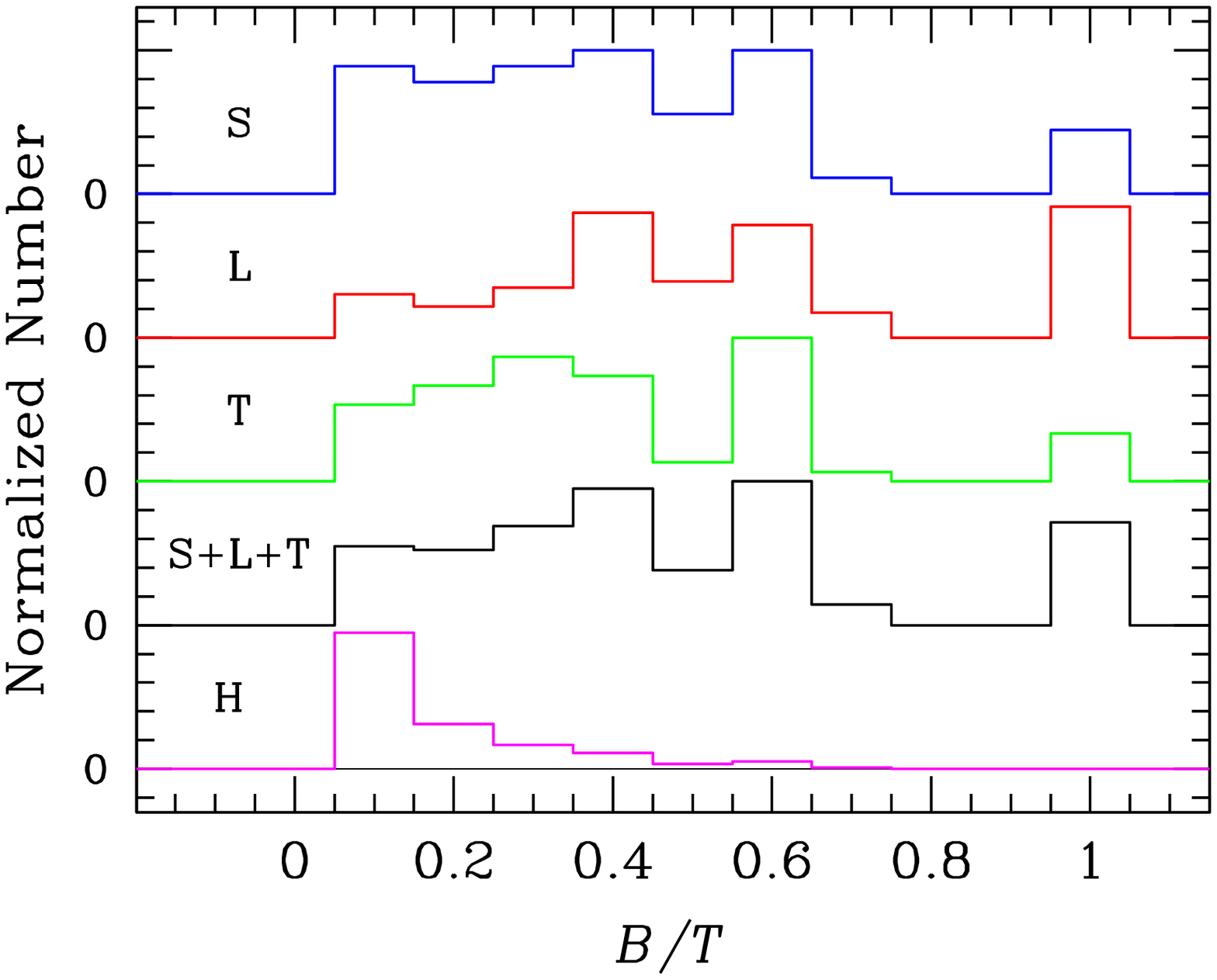,width=6.9cm,angle=0}
\caption{{\it Left}: Detection rate of emission-line nuclei as a function of 
Hubble type.  The different classes of nuclei are marked as follows: Seyferts = 
blue squares, LINERs = red circles, transition objects = green triangles,
LINERs + transition objects + Seyferts = black stars, \hii\ nuclei = magenta 
crosses.  {\it Right}: Distribution of bulge-to-total ($B/T$) light ratios, 
derived from the morphological type and its statistical dependence on $B/T$. 
The histograms have been shifted vertically for clarity.
(Adapted from Ho, Filippenko \& Sargent 1997a, 1997b.)
}
\end{figure}

\subsection{Detection Rates}

In qualitative agreement with early work, the Palomar survey shows that a 
substantial fraction (86\%) of all galaxies contain detectable emission-line 
nuclei (Ho, Filippenko \& Sargent 1997b).  The detection rate is essentially 
100\% for all disk (S0 and spiral) galaxies, and $>$50\% for elliptical 
galaxies.  One of the most surprising results is the large fraction of objects 
classified as AGNs or AGN candidates, as summarized in Figure~4.  
Summed over all Hubble types, 43\% of all galaxies that fall in the survey 
limits can be considered ``active.'' This percentage becomes even more 
remarkable for galaxies with an obvious bulge component, rising to 
$\sim$50\%$-$70\% for Hubble types E$-$Sb.  By contrast, the detection rate 
of AGNs drops dramatically toward later Hubble types (Sc and later), which 
almost invariably (80\%) host \hii\ nuclei.  This strong dependence of nuclear 
spectral class on Hubble type has been noticed in earlier studies (Heckman 
1980a; Keel 1983b), and further confirmed in SDSS (Kauffmann et al. 2003; 
Miller et al. 2003).  A qualitatively similar conclusion, cast in terms of 
host galaxy stellar mass rather than Hubble type, is reached by Gallo et al. 
(2008) and Decarli et al. (2007).  Decarli et al. also claim that the 
occurrence of AGN activity in Virgo cluster spirals does not depend on 
morphological type, but it must be noted that the sources of spectroscopy and 
nuclear classification employed in that study are very heterogeneous.

Within the parent galaxy sample, 11\% have Seyfert nuclei, at least doubling 
estimates based on older (Stauffer 1982b; Keel 1983b; Phillips, Charles \& 
Baldwin 1983) or shallower (Huchra \& Burg 1992; Maia, Machado \& Willmer 
2003; Gronwall et al. 2004; Hao et al. 2005a) surveys.  LINERs constitute the 
dominant population of AGNs.  ``Pure'' LINERs are present in $\sim$20\% of all 
galaxies, whereas transition objects, which by assumption also contain a LINER 
component, account for another $\sim$13\%. Thus, if all LINERs can be regarded 
as genuine AGNs (see \S~6), they truly are the most populous constituents, 
making up 1/3 of all galaxies and 2/3 of the AGN population (here taken to 
mean all objects classified as Seyferts, LINERs, and transition objects).

Within the magnitude range $14.5 < r < 17.7$ in SDSS, Kauffmann et al. (2003) 
report an overall AGN fraction (for narrow-line sources) of $\sim$40\%, of 
which $\sim$10\% are Seyferts.  The rest are LINERs and transition objects.  
Using a different method of starlight subtraction, Hao et al. (2005b) obtain 
very similar statistics for their sample of Seyfert galaxies.  Roughly 30\% of 
the galaxies on the red sequence in SDSS exhibit LINER-like emission (Yan et 
al.  2006).  Although these detection rates broadly resemble those of the 
Palomar survey, one should recognize important differences between the two 
surveys.  The Palomar objects extend much farther down the luminosity function 
than the SDSS.  The emission-line detection limit of the Palomar survey, EW = 
0.25 \AA, is roughly 10 times fainter than the cutoff chosen by Hao et al. 
(2005b). The faint end of the Palomar H\al\ luminosity function reaches 
$\sim$1\e{37} \lum\ (\S~5.9), again a factor of 10 lower than the SDSS 
counterpart.  Given that LINERs selected from SDSS are highly susceptible to 
extranuclear contamination, as discussed earlier, it is in fact quite 
surprising---and perhaps fortuitous---that the detection rates of these 
objects agree so well between the two surveys.  

\subsection{Broad Emission Lines}

Broad emission lines, a defining attribute of classical Seyferts and quasars,
are also found in nuclei of much lower luminosities.  The well-known case of
the nucleus of M81 (Peimbert \& Torres-Peimbert 1981; Filippenko \& Sargent 
1988), for example, has a broad (FWHM $\approx$ 3000 \kms) H\al\ line with a 
luminosity of $2 \times\,10^{39}$ \lum\ (Ho, Filippenko \& Sargent 
1996), and many other less conspicuous cases have been discovered in the 
Palomar survey (Ho et al. 1997e).  Searching for broad H\al\ emission in 
nearby nuclei is nontrivial, because it entails measurement of a (generally) 
weak, low-contrast, broad emission feature superposed on a complicated stellar 
background.  Thus, the importance of careful starlight subtraction cannot be 
overemphasized.  Moreover, even if this could be accomplished perfectly, 
one still has to contend with deblending the H\al +\nii\ complex.  The narrow 
lines in this complex are often heavily blended together, and rarely do the 
lines have simple profiles.  The strategy adopted by Ho \etal (1997e) is to 
use the empirical line profile of the \sii\ lines to model \nii\ and the 
narrow component of H\al.  The case of NGC~3998 is shown in Figure~3.

Of the 221 emission-line nuclei in the Palomar survey classified as LINERs,
transition objects, and Seyferts, 33 (15\%) definitely have broad H\al\ and an
additional 16 (7\%) probably do.  Questionable detections were found in
another 8 objects (4\%).  Thus, approximately 20\%$-$25\% of all nearby AGNs
are type~1 sources.  These numbers, of course, should be regarded as lower
limits, since undoubtedly there must exist AGNs with even weaker broad-line
emission that fall below the detection threshold.  Although the numbers are 
meager, direct comparison with small-aperture {\it Hubble Space Telescope
(HST)}\ spectra (e.g., Nicholson et al. 1998; Barth et al. 2001b; Shields et 
al. 2007) reveals that the Palomar statistics on broad H\al\ detections seem 
to be quite robust.  The type~1.9 classification of almost every 
object with overlapping \hst\ data turns out to survive.  More surprising 
still, no new cases of broad H\al\ emission have turned up from \hst\ 
observations.  Given the difficulty of measuring the weak broad H\al\ feature 
on top of the dominant stellar continuum, it is likely that in general the 
line widths may have been systematically underestimated from the ground-based 
spectra.   Circumstantial support for this conjecture comes from Zhang, 
Dultzin-Hacyan \& Wang (2007), who find that Palomar LLAGNs tend to have 
smaller BH virial masses (estimated from the H\al\ linewidth and a BLR 
size-luminosity relation) than predicted from their bulge stellar velocity 
dispersion.  They conclude that the BLR size in LLAGNs may be larger than 
normal given their luminosity, but an equally plausible explanation is that 
the H\al\ linewidths have been systematically underestimated.

It is illuminating to consider the incidence of broad H\al\ emission as
a function of spectral class.  Among objects formally classified as Seyferts,
$\sim$40\% are Seyfert~1s.  The implied ratio of Seyfert~1s to Seyfert~2s 
(1:1.6) has important consequences for some models concerning the evolution 
and small-scale geometry of AGNs (e.g., Osterbrock \& Shaw 1988).  Within the 
Palomar sample, nearly 25\% of the ``pure'' LINERs have detectable broad H\al\ 
emission.  By direct analogy with the historical nomenclature established for 
Seyferts, LINERs can be divided into ``type~1'' and ``type~2'' sources 
according to the presence or absence of broad-line emission, respectively (Ho,
Filippenko \& Sargent 1997a; Ho et al. 1997e).  The detection rate of broad 
H\al, however, drops drastically for transition objects.  The cause for this 
dramatic change is unclear.  In these objects the broad-line component may 
simply be too weak to be detected in the presence of substantial contamination 
from the \hii\ region component, or it may be intrinsically absent (\S\S~5.5, 
6.5).

\subsection{Robustness and Completeness}

To gain confidence in the AGN statistics based on optical spectroscopy, one 
must have some handle on whether the existing AGN detections are trustworthy 
and whether there are many AGNs that have been missed.  The robustness issue 
hinges on the question of whether the weak, nearby sources classified as AGNs 
are truly accretion-powered.  As I argue in \S~6, this appears largely to be the
case.  The completeness issue can be examined in two regimes.  Among galaxies 
with prominent bulges (Sbc and earlier), for which the spectroscopic AGN 
fractions are already very high ($\sim$50\%$-$75\%), there is not much room 
for a large fraction of missing AGNs, although it is almost certain that some 
have indeed eluded detection in the optical (e.g., Tzanavaris \& 
Georgantopoulos 2007).  The same does not necessarily hold for galaxies of 
Hubble types Sc and later.  While the majority of these systems are 
spectroscopically classified as \hii\ nuclei, one must be wary that weak AGNs, 
if present, may be masked by brighter off-nuclear \hii\ regions or \hii\ 
regions projected along the line of sight.  After all, some very late-type 
galaxies {\it do}\ host {\it bona fide}\ AGNs (see \S~7).

The AGN content of late-type galaxies can independently be assessed by using
a diagnostic less prone to confusion by star-forming regions, namely by 
looking for compact, nuclear radio or X-ray cores.  Ulvestad \& Ho (2002) 
performed a Very Large Array (VLA) survey for radio cores in a 
distance-limited sample of 40 Palomar Sc galaxies classified as hosting \hii\ 
nuclei.  To a sensitivity limit of $P_{\rm rad} \approx 10^{18}-10^{20}$ 
W~Hz$^{-1}$ at 5~GHz, and a resolution of $\Delta \theta$ = 1\asec, they found 
that {\it none} of the galaxies contains a compact radio core.  The VLA study 
of Filho, Barthel \& Ho (2000) also failed to detect radio cores in a more 
heterogeneous sample of 12 \hii\ nuclei.

Information on nuclear X-ray cores in late-type galaxies is much more limited
because to date there has been no systematic investigation of these systems
with {\it Chandra}.  A few studies, however, have exploited the High
Resolution Imager (HRI) on {\it ROSAT} to resolve the soft X-ray (0.5$-$2 keV)
emission in late-type galaxies (Colbert \& Mushotzky 1999; Lira, Lawrence \&
Johnson 2000; Roberts \& Warwick 2000).  Although the resolution of the HRI
($\sim$5\asec) is not ideal, it is nonetheless quite effective for identifying
point sources given the relatively diffuse morphologies of late-type galaxies.
Compact X-ray sources, often quite luminous (\gax\ 10$^{38}$ \lum), are
frequently found, but generally they do {\it not} coincide with the galaxy 
nucleus; the nature of these off-nuclear ``ultraluminous X-ray sources'' is 
discussed by Fabbiano (2006).

To summarize: unless \hii\ nuclei in late-type galaxies contain radio and
X-ray cores far weaker than the current survey limits---a possibility worth
exploring---they do not appear to conceal a significant population of
undetected AGNs.

\section{HOST GALAXY PROPERTIES}

\subsection{Global Parameters}

The near dichotomy in the distribution of Hubble types for galaxies hosting
active versus inactive nuclei (Figure~4) leads to the expectation that the two
populations ought to have fairly distinctive global, and perhaps even nuclear,
properties.  Moreover, a detailed examination of the host galaxies of AGNs
may shed light on the origin of their spectral diversity.  These issues were
examined by Ho, Filippenko \& Sargent (2003) using the database from the 
Palomar survey.  The host galaxies of Seyferts, LINERs, and transition objects 
display a remarkable degree of homogeneity in their large-scale properties, 
{\it after} factoring out spurious differences arising from small mismatches 
in Hubble type distribution.  The various nuclear types have slightly different
Hubble distributions, which largely control many of the statistical properties 
of the host galaxies.  Unless this effect is taken into account, one can 
arrive at erroneous conclusions about the intrinsic differences of the AGN 
populations.  This is a crucial step, one that is often not appreciated.
All three classes have essentially identical total luminosities ($\sim L^*$), 
bulge luminosities, sizes, and neutral hydrogen content.  Moreover, no obvious 
differences are found in terms of integrated optical colors or far-IR
luminosities and colors, which implies very similar global stellar content and 
current star formation rates.  No clear differences in environment can be 
seen either.  The only exception is that, relative to LINERs, transition 
objects may show a mild enhancement in the level of global star formation, and 
they may be preferentially more inclined.  This is consistent with the 
hypothesis that the transition class arises from spatial blending of emission 
from a LINER and \hii\ regions.  The velocity field of the ionized gas within 
the nuclear region, as measured by the width and asymmetry of the narrow 
emission lines, is crudely similar among the three AGN classes, an observation 
that argues against the proposition that fast shocks primarily drive the 
spectral variations (\S~6.2).

\subsection{Nuclear Stellar Populations}

The uniformity in the global stellar populations among the three AGN classes 
extends to circumnuclear and even nuclear scales.  The Palomar spectra cover a 
suite of stellar absorption-line indices and nuclear continuum colors, which 
collectively can be used to obtain crude constraints on the age and 
metallicity of the stars within the central 2\asec\ ($\sim 200$ pc).  After 
isolating a subsample that mitigates the Hubble type dependence, Ho, Filippenko,
\& Sargent (2003) find that Seyferts, LINERs, and transition objects have very 
similar stellar content.  The same holds true when comparing type~1 and type~2 
objects, both for LINERs and Seyferts.  With a few notable exceptions such as 
NGC~404 and NGC~4569 (Maoz et al. 1998; Barth \& Shields 2000; Gabel \& 
Bruhweiler 2002) or NGC~4303 (Colina et al. 2002), the stellar population 
always appears evolved.  Similar findings have been reported for smaller 
samples of LINERs, based on both optical (Boisson et al. 2000; Serote-Roos \& 
Gon\c{c}alves 2004; Zhang, Gu \& Ho 2008) and near-IR spectroscopy (Larkin et 
al. 1998; Bendo \& Joseph 2004).  The optical regime is not well suited to 
detect very young, ionizing stars.  However, the Palomar spectra do cover the 
broad \heii\ \lamb4686 emission bump, a feature indicative of Wolf-Rayet stars 
commonly seen in very young ($3-6$ Myr) starbursts.  Notwithstanding the 
difficulty of detecting this feature on top of a dominant old population, it 
is noteworthy that not a {\it single}\ case has been seen among the sample of 
over 200 Palomar LLAGNs.  By contrast, the Wolf-Rayet bump has been found in a 
number of the \hii\ nuclei (Sargent \& Filippenko 1991; Ho, Filippenko \& 
Sargent 1995), which, as a class compared to the LLAGNs, exhibit markedly 
younger stars, as evidenced by their blue continuum, strong H\bet\ and 
H$\delta$ absorption, and weak metal lines (Ho, Filippenko \& Sargent 2003).  
The general dearth of young, massive stars in LLAGNs presents a serious 
challenge to proposals that seek to account for the excitation of their line 
emission in terms of starburst models.

Closer in, on scales \lax\ 10 pc accessible by {\it HST},
Sarzi et al.  (2005) studied the nuclear stellar population for a 
distance-limited subsample of 18 Palomar LLAGNs.  Their population synthesis 
analysis shows that the majority (80\%) of the objects have predominantly old 
(\gax\ 5 Gyr), mildly reddened stars of near-solar metallicity, the only 
exceptions being 3 out of 6 transition objects and 1 out of 4 LINER~2s that 
require a younger (\lax\ 1 Gyr) component.  In no case, however, is the younger 
component ever energetically dominant, falling far short of being able to 
account for the ionization budget for the central region.  

The results of Ho, Filippenko \& Sargent (2003) have been disputed by 
Cid~Fernandes et al.  (2004), who obtained new ground-based spectra for a 
subset of the Palomar LINERs and transition objects.  Gonz\'alez~Delgado et 
al. (2004), in a study similar to that of Sarzi et al. (2005), further 
analyzed STIS spectra of some of these.  An important improvement of their 
ground-based data is that they extend down to $\sim 3500$ \AA, covering the 
4000~\AA\ break and the higher-order Balmer lines, which are sensitive probes 
of intermediate-age ($\sim 10^7-10^9$ yr) stars.  While LINERs are 
predominantly old, roughly half of the transition nuclei show significant 
higher-order Balmer lines.  Again, there are virtually no traces of Wolf-Rayet 
features.  These authors propose that the ionization mechanism of transition 
sources must be somehow linked to the intermediate-age stellar population.  

I disagree with their assessment.  Figure~1 of Cid~Fernandes et al. (2004) 
clearly shows that, as in the parent Palomar sample, the Hubble type 
distribution of the LINERs is skewed toward much earlier types than that of the 
transition objects.  Any meaningful comparison of the stellar population, which 
strongly depends on Hubble type, must take this into account.  The strategy 
employed by Ho, Filippenko \& Sargent (2003) is to restrict the two-sample 
comparisons to a relatively narrow range of Hubble types (Sab$-$Sbc). Within 
this domain, LINERs and transition objects (as well as Seyferts) have 
statistically indistinguishable stellar indices and continuum colors.  Given 
the limited size of Cid~Fernandes et al.'s sample, it is not possible to adopt 
the same strategy.  To minimize the Hubble type bias, I examined the subsample 
of 30 spiral galaxies in their study.  Out of 17 transition objects, 15 (88\%) 
contain intermediate-age stars according to their Table~3; but so do 10 out of 
the 13 (77\%) LINERs in this subgroup.  This simple exercise underscores the 
importance of sample selection effects, and leaves me unconvinced 
that transition objects have a younger stellar population than LINERs.  True, 
both classes evidently do contain detectable amounts of intermediate-age 
stars---a qualitatively different conclusion than was reached in the Palomar 
survey, whose spectral coverage was not well-suited to detect this 
population---but the fact remains that in a {\it relative}\ sense all three 
LLAGN classes in the Palomar survey have statistically similar populations.
If the poststarburst component is responsible for the nebular emission in 
LLAGNs, we might expect the intensity of the two to be correlated, by analogy
with what has been found for higher luminosity AGNs (Kauffmann et al. 2003).
I searched for this effect in the final sample presented in Gonz\'alez~Delgado 
et al. (2004), but did not find any correlation.  Clearly we wish to know what 
factors drive the spectral variations in LLAGNs; whatever they are (\S~6), 
they are unlikely to be related to stellar population.

\subsection{Influence of Bars and Environment}

Numerical simulations (e.g., Heller \& Shlosman 1994; see review in Kormendy 
\& Kennicutt 2004) suggest that large-scale 
stellar bars can be highly effective in delivering gas to the central few 
hundred parsecs of a spiral galaxy, thereby potentially leading to rapid star 
formation.  Further instabilities result in additional inflow to smaller 
scales, which may lead to increased BH fueling and hence elevated nonstellar 
activity in barred galaxies compared to unbarred galaxies.  As discussed in 
\S~3.1, the Palomar sample is ideally suited for statistical comparisons of 
this nature, which depend delicately on issues of sample 
completeness and the choice of control sample.  Ho, Filippenko \& Sargent 
(1997d) find that while the presence of a bar indeed does enhance both the 
probability and rate of star formation in galaxy nuclei, it appears to have no 
impact on either the frequency or strength of AGN activity.  Bearing in mind 
the substantial uncertainties associated with sample selection, as well as the 
method and wavelength used to identify bars (Laurikainen, Salo \& Buta 2004), 
other studies broadly come to a similar conclusion (see review by Combes 2003), 
although Maia, Machado \& Willmer (2003) claim, on the basis of a 
significantly larger and somewhat more luminous sample drawn from the Southern 
Sky Redshift Survey, that Seyfert galaxies are preferentially more barred.

In the same vein, dynamical interactions with neighboring companions should
lead to gas dissipation, enhanced nuclear star formation, and perhaps central
fueling (e.g., Hernquist 1989).  Schmitt (2001) and Ho, Filippenko \& Sargent 
(2003) studied this issue using the Palomar data, parameterizing the nearby 
environment of each object by its local galaxy density and the distance to its 
nearest sizable neighbor.  After accounting for the well-known 
morphology-density relation, it was found that the local environment, like 
bars, has little impact on AGNs, at least in the low-luminosity regime sampled 
locally.  These findings broadly agree with the results from SDSS (Miller et 
al. 2003; Li et al. 2008).  Kauffmann et al. (2004), Wake et al. (2004), and 
Constantin \& Vogeley (2006), in fact, report a drop in the fraction of 
high-luminosity AGNs for dense environments.

\section{NUCLEAR PROPERTIES}

\subsection{Ionizing Continuum Radiation}

AGNs, at least when unobscured, reveal themselves as pointlike nuclear sources 
with power-law spectra at optical and UV wavelengths, typically described
by a continuum flux density $f_\nu \propto \nu^{\alpha}$, with $\alpha \approx 
-0.5$ (e.g., Vanden~Berk et al. 2001).  In unbeamed sources, this featureless 
continuum traces the low-frequency tail of the ``big blue bump'' (Shields 
1978; Malkan \& Sargent 1982), which supplies the bulk of the ionizing 
photons.  This feature is extremely difficult to detect in LLAGNs, both 
because the big blue bump is weak or absent (\S~5.8) and because the sources 
are exceedingly faint.  The optical nuclei of LINERs can have $M_B$ \gax\ 
$-10$ mag (Ho 2004b), at least $10^{4}$ times fainter than their (usually 
bulge-dominated) hosts ($M_B \simeq M^* \approx -20$ mag).  To overcome this 
contrast problem, searches for nuclear point sources in the optical and 
near-IR have relied on \hst\ images (e.g., Chiaberge, Capetti \& Celotti 1999; 
Quillen et al. 2001; Verdoes~Kleijn et al. 2002; Chiaberge, Capetti \& 
Macchetto 2005; Balmaverde \& Capetti 2006; Gonz\'alez-Mart\'\i n et al. 
2006).  But resolution alone is not enough.  Given the extreme faintness of 
the nucleus, the intrinsic cuspiness of the underlying bulge profile, 
complexities of the point-spread function, and the often irregular background 
marred by circumnuclear dust features, one must pay very close attention to 
{\it how}\ the measurements are made.  Simple aperture photometry or searching 
for central excess emission can yield very misleading results.  The most robust 
technique to extract faint nuclei in the presence of these complications 
employs two-dimensional, multi-component fitting (Ho \& Peng 2001; 
Ravindranath et al.  2001; Peng et al. 2002).  Using this method, nuclear 
sources with optical magnitudes as faint as $\sim 20$ have been measured, with 
limits down to $\sim 22-23$ mag possible for nearby galaxies.  Due to the 
computational requirements of two-dimensional fitting, however, not many 
LLAGNs have yet been analyzed in this manner, and fewer still have enough 
photometric points to define even a crude spectral slope.

In a few cases, the optical featureless continuum has been detected 
spectroscopically.  From the ground, this was only possible for a couple of 
the brightest sources.  The stellar features of NGC~7213 (Halpern \& 
Filippenko 1984) and Pictor~A (Carswell et al. 1984; Filippenko 1985) show 
dilution by a featureless continuum, which can be described approximately by a 
power law with a spectral index of $\alpha \approx -1.5$.  The 
nuclear continuum is much more readily seen in small-aperture spectra that 
help to reject the bulge starlight.  \hst\ spectra have isolated the optical 
continuum in several LINERs (Ho, Filippenko \& Sargent 1996; Nicholson et al.  
1998; Ho et al. 2000; Shields et al. 2000; Barth et al.  2001a; Sabra et al. 
2003), although in most objects it remains too faint to be detected 
spectroscopically (Sarzi et al. 2005).  In all well-studied cases, the optical 
continuum is quite steep, with $\alpha \approx -1$ to $-2$.  This range 
in spectral slopes is consistent with the broad-band optical (Verdoes~Kleijn 
et al. 2002) and optical-UV colors (Chiaberge et al. 2002) of the cores 
frequently detected in the LINER nuclei of FR~I radio galaxies.

The predominantly old population of present-day bulges ensures that the 
stellar contamination largely disappears in the UV, especially at high 
resolution.  A number of attempts have been made to detect UV emission in 
LINERs using {\it IUE}, but most of these efforts yielded ambiguous results 
(see review in Filippenko 1996), and real progress had to await the {\it HST}.
Two dedicated \hst\ UV ($\sim 2300$ \AA) imaging studies have been completed.  
Using the pre-COSTAR FOC, Maoz et al. (1996) surveyed a complete sample of 110 
large, nearby galaxies, and among the subset with spectral classifications 
from Palomar, Maoz et al. (1995) discovered that $\sim$25\% of the LINERs show 
an unresolved UV core.  Barth et al. (1998) found similar statistics in a more 
targeted WFPC2 study.  They also made the suggestion, later 
confirmed by Pogge et al. (2000), that dust obscuration is probably the main 
culprit for the nondetection of UV emission in the majority of LINERs.  The 
implication is that UV emission is significantly more common in LINERs than 
indicated by the detection rates.  In some type~2 objects (e.g., NGC~4569 
and NGC~6500), the UV emission is spatially extended and presumably not 
related to the nuclear source.  Second-epoch UV observations with the 
ACS/HRC revealed that nearly all of the UV-bright sources exhibit long-term 
variability (Maoz et al. 2005), an important result that helps assuage fears 
that the UV emission might arise mainly from young stars (Maoz et al. 1998).  
Importantly, both type~1 {\it and}\ type~2 LINERs vary.  UV variability has 
also been discovered serendipitously in a few other sources (Renzini et al. 
1995; O'Connell et al. 2005).

%%%%%%%%%%%%%%%%%%%%%%%%%%%%%%%%%%%%%%%%%%%%%%%%%%%%%%%%%%%%%%%%%%%%%%%%%%
%%BoundingBox: 80 150 350 770
\begin{figure*}[t]
\centerline{\psfig{file=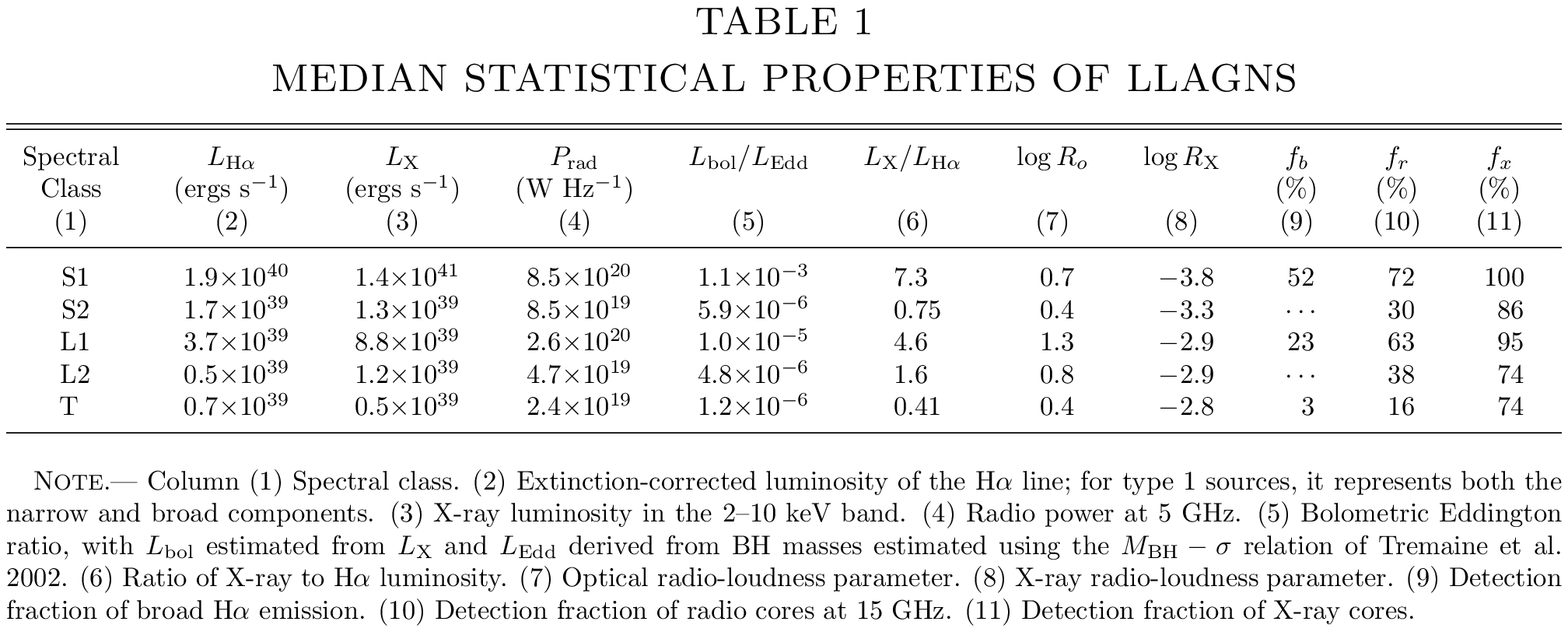,width=14.0cm,angle=90}}
\end{figure*}
%%%%%%%%%%%%%%%%%%%%%%%%%%%%%%%%%%%%%%%%%%%%%%%%%%%%%%%%%%%%%%%%%%%%%%%%%%

\subsection{Radio Cores}

AGNs, no matter how weak, are almost never silent in the radio.  Barring 
chance superposition with a supernova remnant, the presence of a compact radio 
core is therefore a good AGN indicator.  Because of the 
expected faintness of the nuclei, however, any search for core emission must 
be conducted at high sensitivity, and arcsecond-scale angular resolution or 
better is generally needed to isolate the nucleus from the surrounding host, 
which emits copious diffuse synchrotron radiation.  In practice, this 
requires an interferometer such as the VLA.

The prevalence of weak AGNs in nearby early-type galaxies has been established
from the VLA radio continuum studies of Sadler, Jenkins \& Kotanyi (1989) and 
Wrobel \& Heeschen (1991), whose 5~GHz surveys with $\Delta \theta \approx$
5\asec\ report a high incidence ($\sim30$\%--40\%) of radio cores in complete, 
optical flux-limited samples of elliptical and S0 galaxies.  Interestingly, 
the radio detection rate is similar to the detection rate of optical 
emission lines (Figure~4), and the optical counterparts of the radio cores are 
mostly classified as LINERs (Phillips et al. 1986; Ho 1999a).  Conversely, 
Heckman (1980b) showed that LINERs tend to be associated with compact radio 
sources.  The radio powers are quite modest, generally in the range of 
$10^{19}-10^{21}$ W Hz$^{-1}$ at 5 GHz.  When available, the spectral indices 
tend to be relatively flat (e.g., Wrobel 1991; Slee et al. 1994).  With the 
exception of a handful of well-known radio galaxies with extended jets (Wrobel 
1991), most of the radio emission is centrally concentrated.   

No comparable radio survey has been done for spiral galaxies.  Over the last 
few years, however, a number of studies, mostly using the VLA, have 
systematically targeted sizable subsets of the Palomar galaxies, to the point 
that by now effectively the entire Palomar AGN sample has been surveyed at 
arcsecond ($\Delta \theta \approx$ 0\farcs15$-$2\farcs5) resolution (Filho, 
Barthel \& Ho 2000, 2002a, 2006; Nagar et al. 2000, 2002; Ho \& Ulvestad 
2001; Filho et al. 2004; Nagar, Falcke \& Wilson 2005; Krips et al. 2007).  
Because the sensitivity, resolution, and observing frequency varied from study 
to study, each concentrating on different subclasses of objects, it is 
nontrivial to combine the literature results.  The only survey that samples a 
significant fraction of the three LLAGN classes at a uniform sensitivity and 
resolution is that by Nagar et al. (2000, 2002; Nagar, Falcke \& Wilson 
2005), which was conducted at 15~GHz and $\Delta \theta$ = 0\farcs15.  The 
main drawback is that the sensitivity of this survey (rms $\approx$ 0.2 mJy) 
is rather modest, and mJy-level sources can be missed if they possess 
relatively steep spectra.  Despite these limitations, Nagar et al. detected a 
compact core, to a high level of completeness, in 44\% of the LINERs. 
Importantly, to the same level of completeness, the Seyferts exhibit a very 
similar detection rate (47\%).  LINER~2s have a lower detection rate than 
LINER~1s (38\% versus 63\%; see Table~1), but the same pattern is reflected 
almost exactly within the Seyfert population (detection rate 30\% for type~2s 
versus 72\% for type~1s).  Transition objects, on the other hand, clearly 
differ, showing a markedly lower detection rate of only 16\%, consistent with 
the 8.4~GHz survey of Filho, Barthel \& Ho (2000, 2002a, 2006), where the 
detection rate is $\sim 25$\%.  The statistical differences in the Hubble type 
distributions of the three AGN classes (Ho, Filippenko \& Sargent 2003) 
slightly complicate the interpretation of these results.  To the extent that 
radio power shows a mild dependence on bulge strength or BH mass (Nagar, 
Falcke \& Wilson 2005; see Ho 2002a), the detection rates, strictly speaking, 
should be renormalized to account for the differences in morphological types 
among the three classes.  This effect, however, will not qualitatively change 
the central conclusion: if a compact radio core guarantees AGN pedigree, then 
LINERs, of either type~1 or type~2, are just as AGN-like as Seyferts, whereas 
a significant fraction of transition objects (roughly half) may be unrelated 
to AGNs.

The detection rates from the Nagar et al. survey can be viewed as firm lower 
limits.  At $\Delta \theta$ = 1\asec\ and rms = 0.04 mJy at 1.4 and 5 GHz, for 
example, the detection rate for the Palomar Seyferts rises to 75\% (Ho \& 
Ulvestad 2001).  Although no lower frequency survey of LINERs has been 
completed so far (apart from the lower resolution studies of Sadler, Jenkins, 
\& Kotanyi 1989 and Wrobel \& Heeschen 1991 confined to early-type galaxies), 
the preliminary study by Van~Dyk \& Ho (1998) of 29 LINERs at 5 and 3.6 GHz 
($\Delta \theta$ = 0\farcs5; rms = 0.05$-$0.1 mJy) yielded a detection rate of 
over 80\%, again suggesting that LINERs and Seyferts have a comparably high 
incidence of radio cores.

Importantly, a sizable, flux-limit subset of the 15 GHz detections has been 
reobserved with the Very Long Baseline Array at 5 GHz, and essentially 
{\it all}\ of them have been detected at milliarcsecond resolution (Nagar, 
Falcke \& Wilson 2005).  The high brightness temperatures (\gax\ 
$10^6-10^{11}$~K) leaves no doubt that the radio cores are nonthermal and 
genuinely associated with AGN activity.

Where multifrequency data exist, their spectra tend to be flat or even mildly 
inverted ($\alpha \approx -0.2$ to $+0.2$; Ho et al. 1999b; Falcke et al. 2000; 
Nagar et al. 2000; Nagar, Wilson \& Falcke 2001; Ulvestad \& Ho 2001b; 
Anderson, Ulvestad \& Ho 2004; Doi et al. 2005; Krips et al. 2007), seemingly 
more optically thick than Seyferts (median $\alpha = -0.4$; Ulvestad \& Ho 
2001a), and variability on timescales of months is common (Nagar et al. 2002; 
Anderson \& Ulvestad 2005).   Both of these characteristics suggest that the 
radio emission in LINERs is mainly confined to a compact core or base of a 
jet.  Seyfert galaxies contain radio cores as well, but they are often 
accompanied by linear, jetlike features resolved on arcsecond scales (e.g., 
Ulvestad \& Wilson 1989; Kukula et al. 1995; Ho \& Ulvestad 2001; Gallimore et 
al. 2006).  This extended component appears to be less prevalent in LINERs, 
although a definitive comparison must await a survey matched in resolution, 
sensitivity, and wavelength with that performed for the Seyferts (Ho \& 
Ulvestad 2001).  Higher resolution images on milliarcsecond scales
do resolve elongated structures akin to subparsec-scale jets, but most of the 
power is concentrated in a compact, high-brightness temperature core 
(Bietenholz, Bartel \& Rupen 2000; Falcke et al. 2000; Ulvestad \& Ho 2001b; 
Filho, Barthel \& Ho 2002b; Anderson, Ulvestad \& Ho 2004; Filho et al. 2004; 
Krips et al.  2007).  The comprehensive summary presented in Nagar, Falcke \& 
Wilson (2005) indicates that the incidence of milliarcsecond-scale radio cores 
is similar for LINERs and Seyferts, but that subparsec-scale jets occur more 
frequently in LINERs.

\subsection{X-ray Cores}

X-ray observations provide another very effective tool to isolate LLAGNs and 
to diagnose their physical properties.  Ultra-faint LLAGNs can be identified 
where none was previously known in the optical (e.g., Loewenstein et al. 
2001; Ho, Terashima \& Ulvestad 2003; Fabbiano et al. 2004; Pellegrini et al. 
2007; Wrobel, Terashima \& Ho 2008).  Here, too, sensitivity and resolution 
are critical, as the central regions of galaxies contain a plethora of 
discrete nonnuclear sources, often suffused with a diffuse thermal plasma.  
{\it Chandra}, whose ACIS camera delivers $\sim$0\farcs5 images, is the 
instrument of choice, although in some instances even data at $\sim$5\asec\ 
resolution (e.g., {\it ROSAT} HRI) can still provide meaningful constraints, 
especially if accompanied by spectral information (e.g., {\it XMM-Newton}).

As in the radio, no truly unbiased high-resolution X-ray survey has yet been 
performed of an optical flux-limited sample of nearby galaxies.  The closest 
attempt was made by Roberts \& Warwick (2000), who searched for X-ray nuclear 
sources in 83 Palomar galaxies ($\sim 20$\% of the total sample) having 
archival {\it ROSAT}\ HRI data.  This subset is probably not unbiased, but it 
does encapsulate all the nuclear spectral classes in the Palomar survey.  
In total, X-ray cores were detected in 54\% of the sample, with Seyferts and 
LINERs (including transition objects) both showing a higher detection rate 
($\sim$70\%) than absorption (30\%) or \hii\ nuclei (40\%).  The high detection
rate among the optically classified LLAGNs agrees well with other {\it ROSAT}\ 
studies of Palomar sources (Koratkar et al. 1995; Komossa, B\"ohringer \&
Huchra 1999; Halderson et al. 2001; Roberts, Schurch \& Warwick 2001), but 
the nonnegligible detection rate among the inactive members suggests that 
a significant fraction of the ``core'' flux may be nonnuclear emission [X-ray 
binaries (XRBs) and diffuse gas] insufficiently resolved by {\it ROSAT}.  

Observations with {\it Chandra}\ (e.g., Ho et al. 2001;  Eracleous et al. 2002)
confirm the suspicion that earlier X-ray studies may have suffered from 
confusion with extranuclear sources (Figure~5).  Importantly, the sharp 
resolution and low background noise of ACIS enable faint point sources to be 
detected with brief (few ks) exposures.  This makes feasible, for the first 
time, X-ray surveys of large samples of galaxies selected at non-X-ray 
wavelengths.  In a snapshot survey of a distance-limited sample of Palomar 
LLAGNs, Ho et al. (2001) find that $\sim$75\% of LINERs, both type~1 and 
type~2, contain X-ray cores, some as faint as $\sim 10^{38}$ \lum\ in the 
$2-10$ keV band.  Terashima \& Wilson (2003b) report an even 

\begin{figure}
\psfig{figure=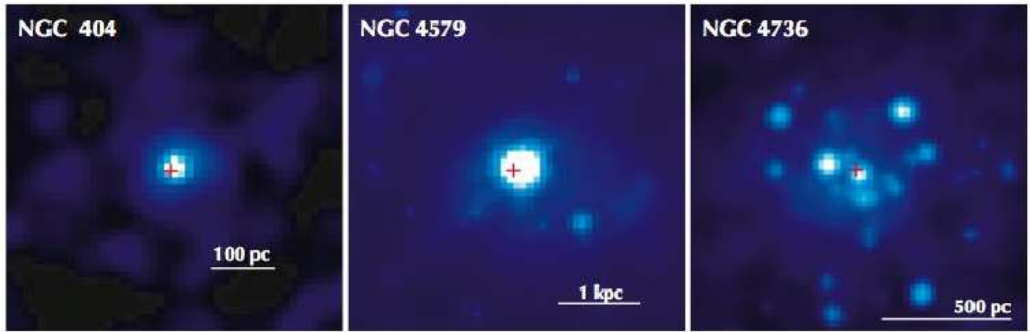,width=13.7cm,angle=0}
\caption{{\it Chandra}/ACIS images of three LLAGNs, illustrating the diversity
and complexity of the X-ray morphologies of their circumnuclear regions.
The cross marks the near-IR position of the nucleus.  (Courtesy of H.M.L.G.
Flohic and M. Eracleous.)
}
\end{figure}

\noindent
higher detection 
rate (100\%) for a sample of LINERs chosen for having a flat-spectrum radio 
core.  To date, roughly 50\% of the entire Palomar sample, among them 40\% of 
the AGNs, have been observed by {\it Chandra}.  This rich archival resource 
has been the basis of a number of recent investigations focused on quantifying 
the AGN content of LINERs, chief among them Satyapal, Sambruna \& Dudik 
(2004), Dudik et al.  (2005), Pellegrini (2005), Satyapal et al. (2005),
Flohic et al. (2006), and Gonz\'alez-Mart\'\i n et al. (2006).  A common 
conclusion that can be distilled from these studies is that the incidence of 
X-ray cores among LINERs is quite high, ranging from $\sim$50\% to 70\%, down 
to luminosity limits of $\sim 10^{38}$ \lum.  The incidence of X-ray cores in 
LINERs is somewhat lower than, but still compares favorably to, that found in 
Palomar Seyferts ($\sim 90$\%), the vast majority of which now have suitable 
X-ray observations, as summarized in Cappi et al. (2006) and Panessa et al. 
(2006).  While the impact of selection biases cannot be assessed easily,
they are probably not very severe because most of the observations 
were not originally intended to study LINERs, nor were they targeting famous 
X-ray sources.

It is of interest to ask whether the incidence of X-ray cores in LINERs 
depends on the presence of broad H\al\ emission.  The moderate-resolution 
{\it ROSAT}/HRI studies of Roberts \& Warwick (2000) and Halderson et al. 
(2001) showed roughly comparable detection rates for type~1 and type~2 
LINERs, suggesting that the two classes are intrinsically similar and that 
obscuration plays a minor role in differentiating them.  On the other hand, 
detailed X-ray spectral analysis has raised the suspicion that LINER~2s may be 
a highly heterogeneous class, with the bulk of the X-ray emission possibly 
arising from stellar processes.  An important caveat is that these studies 
were based on large-beam observations, mostly using {\it ASCA}\ (Terashima, 
Ho \& Ptak 2000; Terashima et al. 2000a, 2002; Roberts, Schurch \& Warwick 
2001) and the rest using {\it BeppoSAX}\ (Georgantopoulos et al. 2002; 
Pellegrini et al. 2002).  A clearer, more consistent picture emerges from the 
recent {\it Chandra}\ work cited above.  Although the individual samples 
remain small, most {\it Chandra}\ surveys detect LINER~2s with roughly similar 
frequency as LINER~1s, $\sim$50\%$-$60\%.  To gain a more comprehensive 
census, I have assembled {\it Chandra}\ measurements for all Palomar LINERs 
from the literature, along with unpublished material for a significant number 
of additional objects in public archives, which were analyzed following Ho 
et al. (2001).  Although clearly heterogeneous and incomplete, the final 
collection of 64 LINERs (20 type~1, 44 type~2) does constitute 70\% of the 
entire Palomar sample.  The detection rate among all LINERs is 86\%, broken 
down into 95\% for LINER~1s and 74\% for LINER~2s.  For completeness, note 
that a similar exercise for 36 transition objects (55\% of the parent sample) 
yields a detection rate of 74\%, identical to that of LINER~2s and only 
marginally lower than that of Seyfert~2s (86\%; Table~1).

The X-ray spectral properties of LLAGNs, particularly LINERs, have most 
thoroughly been investigated using {\it ASCA}\ (Yaqoob et al. 1995; Ishisaki et 
al. 1996; Iyomoto et al. 1996, 1997, 1998a, 1998b; Ptak et al. 1996, 1999; 
Terashima et al. 1998a, 1998b, 2000a, 2000b, 2002; Ho et al. 1999a; Terashima, 
Ho \& Ptak 2000; Roberts, Schurch \& Warwick 2001), with important 
contributions from {\it BeppoSAX} (Pellegrini et al.  2000a, 2000b, 2002; 
Iyomoto et al.  2001; Georgantopoulos et al. 2002; Ptak et al. 2004).  A
seminal study on M81 was done using {\it BBXRT}\ (Petre et al. 1993). Although 
the nuclear 
component was not spatially isolated because of the poor angular resolution of 
these telescopes, they had sufficient effective area to yield good photon 
statistics over the energy range $\sim 0.5-10$ keV to {\it spectrally}\ 
isolate the hard, power-law AGN signal.  The most salient properties are the 
following.  (1) Over the region $\sim 0.5-10$ keV, the continuum can be fit 
with a power law with an energy index of $\alpha \approx -0.4$ to $-1.2$.  
Although this range overlaps with that seen in more luminous sources, the 
typical value of $\sim -0.8$ in LLAGNs may be marginally flatter than in 
Seyfert~1s ($\langle\alpha\rangle = -0.87\pm0.22$; Nandra et al. 1997b) or 
radio-quiet quasars ($\langle\alpha\rangle = -0.93\pm0.22$; Reeves \& Turner 
2000), perhaps being more in line with radio-loud quasars ($\langle \alpha 
\rangle = -0.6\pm0.16$; Reeves \& Turner 2000). (2) With a few notable 
exceptions (e.g., M51: Fukazawa et al. 2001, Terashima \& Wilson 2003a; 
NGC~1052: Weaver et al. 1999, Guainazzi et al. 2000; NGC~4258: Makishima et 
al. 1994, Fiore et al. 2001; NGC~4261: Matsumoto et al.  2001), the power-law 
component shows very little intrinsic absorption.  This trend conflicts with 
the tendency for the degree of obscuration to increase with decreasing 
luminosity (e.g., Lawrence \& Elvis 1982).  (3) Signatures of X-ray 
reprocessing by material from an optically thick accretion disk, in the form 
of Fe~K$\alpha$ emission or Compton reflection (Lightman \& White 1988; George
\& Fabian 1991), are weak or absent; the weakness of the Fe~K$\alpha$ line in 
LLAGNs runs counter to the inverse correlation between iron line strength and 
luminosity observed in higher luminosity AGNs (Nandra et al. 1997b). (4) In 
the few cases where Fe~K$\alpha$ emission has been detected, it is always 
narrow.  (5) Apart from the hard power law, most objects require an extra soft 
component at energies \lax\ 2 keV that can be fit by a thermal plasma model 
with a temperature of $kT \approx 0.4-0.8$ keV and near-solar abundances.  (6) 
Contrary to the trend established for luminous sources (Nandra et al. 1997a), 
short-term, large-amplitude X-ray variability is rare in LLAGNs (Ptak et al. 
1998).

More recent observations with {\it Chandra}\ and {\it XMM-Newton}\ have 
refined, but not qualitatively altered, the above results.  Where detailed 
spectral analysis is possible (e.g., B\"ohringer et al. 2001; Kim \& Fabbiano 
2003; Pellegrini et al. 2003a; Terashima \& Wilson 2003b; Filho et al. 2004; 
Page et al. 2004; Starling et al. 2005; Flohic et al. 2006; 
Gonz\'alez-Mart\'\i n et al. 2006; Soria et al. 2006), the hard power-law 
component (except in objects previously known to be heavily absorbed) 
continues to be relatively unabsorbed, even among many type~2 sources, and to 
show little signs of reflection.  No convincing case of a relativistic 
Fe~K$\alpha$ line has yet surfaced in an LLAGN.  The marginally broad iron 
lines discovered with {\it ASCA}\ in M81 (Ishisaki et al. 1996) and NGC~4579 
(Terashima et al. 1998a) has now been resolved into multiple components 
(Dewangan et al. 2004; Page et al. 2004; Young et al. 2007), none of which can 
be associated with a canonical disk.  At the same time, the equivalent width 
limits for even the narrow component have become impressively low (e.g., Ptak 
et al. 2004).  Interestingly, a soft thermal component is still required in 
many objects (\S~5.4), but there is no evidence for blackbody-like soft excess 
emission commonly seen in Seyferts and quasars (e.g., Turner \& Pounds 1989; 
Inoue, Terashima \& Ho 2007).

\subsection{Circumnuclear Thermal Plasma}

Early X-ray observations of LLAGNs using {\it ASCA}\ have consistently revealed
the presence of a diffuse, thermal component, typically with a temperature of 
$kT \approx 0.5$ keV (Ptak et al. 1999; Terashima et al. 2002).  The uniform 
analysis of {\it ROSAT}\ data by Halderson et al. (2001) concluded that $\sim 
80$\% of the Palomar sources contain an extended component.  However, without 
better resolution, it was impossible to know the extent of confusion with 
point sources, how much of the gas is truly associated with the nuclear region 
of the galaxy, or the density and temperature profile of the gas.

Our view of the diffuse component in the nuclear region has been dramatically 
sharpened with {\it Chandra}\ and {\it XMM-Newton}.  Not only has the near 
ubiquity of diffuse gas been confirmed in many nearby galaxies (Ho et al. 
2001; Eracleous et al. 2002; Terashima \& Wilson 2003b; Pellegrini 2005;  
Rinn, Sambruna \& Gliozzi 2005; Cappi et al. 2006; Gonz\'alez-Mart\'\i n et 
al. 2006; Soria et al. 2006), including
our own (Muno et al. 2004) and our close neighbor M31 (Garcia et al. 
2005), but quantitative, statistical properties of the gas are now becoming 
available.  In the comprehensive investigation of 19 LINERs by Flohic et al. 
(2006), the diffuse emission, detected in 70\% of the sample, is concentrated 
within the central few hundred pc.  With an average 0.5$-$2 keV luminosity of 
$\sim 10^{38}$ \lum, it accounts for more than half of the total central 
luminosity in most cases.  The average spectrum is similar to that seen in 
normal galaxies: it can be described by a thermal plasma with $kT = 0.5$ keV 
plus a power-law component with $\alpha = -0.3$ to $-0.5$.  I will return to 
the nature of the hard component in \S~6.5.  What is the origin of the thermal 
plasma?  Given what we know about the stellar populations (\S~4.2), a starburst
origin, as suggested by Gonz\'alez-Mart\'\i n et al. (2006), seems improbable.  
In normal elliptical galaxies, the X-ray--emitting gas represents the 
repository of thermalized stellar ejecta generated from mass loss from evolved 
stars and Type~Ia supernovae (e.g., Awaki et al. 1994).  There is no reason 
not to adopt the same picture to explain the hot plasma in LINERs and other 
LLAGNs.  High-resolution X-ray spectroscopy of the highly ionized gas around 
the nucleus of M81 (Page et al. 2003; Young et al. 2007) and NGC~7213 
(Starling et al. 2005) reveals that the plasma is collisionally ionized.  
Starling et al. note that this may be a property unique to LINERs, as thermal 
gas in luminous Seyferts is usually photoionized rather than collisionally 
ionized (e.g., Kinkhabwala et al. 2002).

\subsection{Broad-line Region}

Luminous, unobscured AGNs distinguish themselves unambiguously by their
characteristic broad permitted lines.  The detection of broad H\al\ emission
in 

\clearpage
\begin{figure}
\psfig{figure=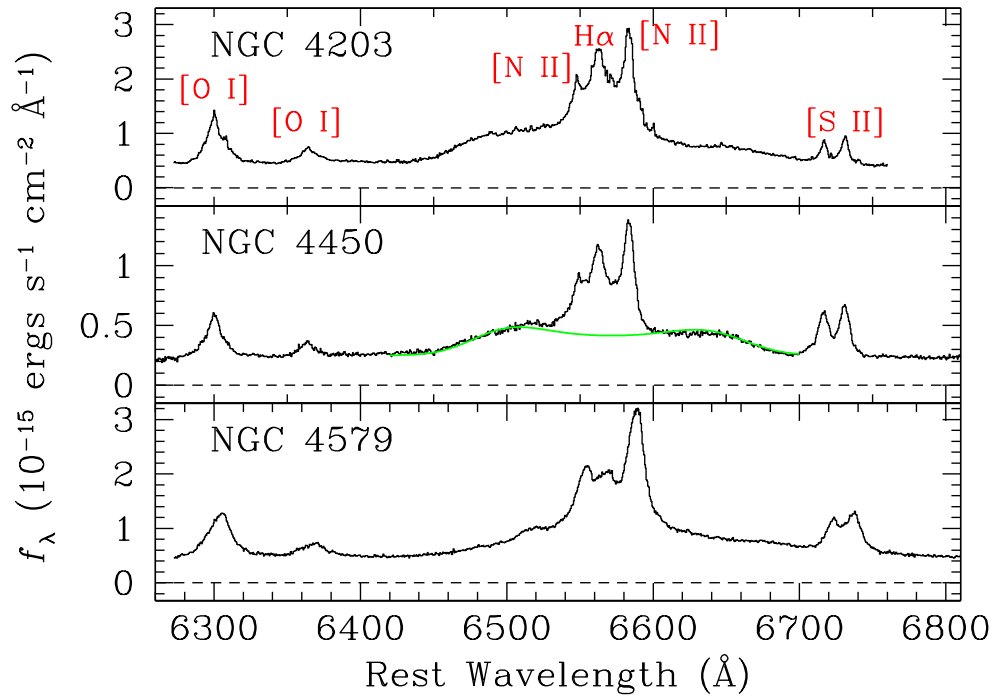,width=13.7cm,angle=270}
\caption{LINERs with broad, double-peaked H\al\ emission discovered with 
{\it HST}.  A model fit for the disk profile in NGC 4450 is shown for 
illustration (green curve).  (Adapted from Ho et al. 2000, Shields et al. 
2000, and Barth et al. 2001a.)
}
\end{figure}

\noindent
$\sim$25\% of LINERs (Ho et al. 1997e) thus constitutes strong evidence in
favor of the AGN interpretation of these sources.  LINERs, like Seyferts,
come in two flavors---some have a visible BLR (type~1), and others do not
(type~2).  The broad component becomes progressively more difficult to detect
in ground-based spectra for permitted lines weaker than H\al.  However, \hst\
spectra of LINERs, when available, show broad higher-order Balmer lines as
well as UV lines such as Ly$\alpha$, \civ\ \lamb1549, \mgii\ \lamb2800, and
\feii\ multiplets (Barth \etal 1996; Ho, Filippenko \& Sargent 1996).
A subset of LINERs contain broad lines with {\it double-peaked}\ profiles
(Figure~6), analogous to those seen in a minority of radio galaxies (Eracleous
\& Halpern 1994), where they are often interpreted as a kinematic signature of
a relativistic accretion disk (Chen \& Halpern 1989).  Most
of the nearby cases have been discovered serendipitously, either as a result
of the broad component being variable (NGC~1097: Storchi-Bergmann, Baldwin \&
Wilson 1993; M81: Bower et al. 1996; NGC~3065: Eracleous \& Halpern 2001) or
because of the increased sensitivity to weak, broad features afforded by
small-aperture measurements made with {\it HST}\ (NGC~4450: Ho et al. 2000;
NGC~4203: Shields et al. 2000; NGC~4579: Barth et al. 2001a).  Double-peaked
broad-line AGNs may be more common than previously thought, especially among
LLAGNs, perhaps as a consequence of their accretion disk structure (\S~8).

A pressing question, however, is: What fraction of the more numerous LINER~2s
are AGNs?  By analogy with the Seyfert~2 class, do LINER~2s contain a hidden
LINER~1 nucleus?  At first sight, it might seem that there is no
{\it a priori}\ reason why the orientation-dependent unification model, which
has enjoyed much success in the context of Seyfert galaxies, should not apply
equally to LINERs.  If we suppose that the ratio of LINER~2s to LINER~1s is
similar to the ratio of Seyfert~2s to Seyfert~1s---1.6:1 in the Palomar
survey---we can reasonably surmise that the AGN fraction in LINERs may be as
high as $\sim$60\%.  That at least some LINERs {\it do}\ indeed contain a
hidden BLR was demonstrated by the deep Keck spectropolarimetric observations
of Barth, Filippenko \& Moran (1999a, 1999b).  In a survey of 14 LLAGNs,
mostly LINERs, these authors detected broad H\al\ emission in three objects
($\sim 20$\%) polarized at a level of 1\%$-$3\%.  Interestingly, all three
objects are elliptical galaxies with double-sided radio jets.  NGC~315 and
NGC~1052 technically qualify as type~1.9 LINERs (Ho et al. 1997e), whereas
NGC~4261 is a LINER~2.  Although the sample is small, these observations 
prove two important points: (1) the weak broad H\al\ features detected in
direct light is not always scattered emission (Antonucci 1993), since
polarized emission was not detected in several other LINER~1.9s included in 
Barth, Filippenko \& Moran's survey; (2) an obscured nucleus does lurk in 
some LINER~2s.

At the same time, other bright LINER~2s have resisted detection by 
spectropolarimetry.  As in the case of Seyferts (Tran 2001), however, the 
nondetection of polarized broad lines does not necessarily imply that there is 
no hidden BLR.  Nevertheless, the BLR in some type~2 AGNs, especially LINERs 
but also Seyferts, may be intrinsically absent, not obscured.  In the case of 
some Seyferts, mostly weak sources, the evidence comes from low absorbing 
X-ray column densities (Bassani et al. 1999; Pappa et al. 2001; Panessa \& 
Bassani 2002; Gliozzi et al. 2004; Cappi et al. 2006; Gliozzi, Sambruna \& 
Foschini 2007; Bianchi et al. 2008; but see Ghosh et al. 2007) as well as 
optical variability (Hawkins 2004).  LINERs, as a class, very much conform to 
this picture.  As discussed further below, LINERs of either type generally 
show very little sign of absorbing or reprocessing material, and UV 
variability is common.  A few exceptions exist (e.g., NGC~1052: Guainazzi et 
al. 2000; NGC~4261: Sambruna et al. 2003, Zezas et al. 2005), but, 
interestingly, these are precisely the very ones for which Barth, Filippenko, 
\& Moran discovered hidden BLRs.  NGC~4258, also highly absorbed in the X-rays 
(Fiore et al. 2001), shows polarized narrow lines rather than broad lines 
(Barth et al. 1999).

An excellent of a LINER with a naked type~2 nucleus is the Sombrero galaxy.  
Although clearly an AGN, it shows no trace of a broad-line component, neither 
in direct light (Ho et al.  1997e), not even when very well isolated with a 
small \hst\ aperture (Nicholson et al. 1998), nor in polarized light (Barth, 
Filippenko \& Moran 1999b).  Its Balmer decrement indicates little reddening 
to the NLR.  For all practical purposes, the continuum emission from the 
nucleus looks unobscured. It is detected as a variable UV source (Maoz et al. 
1995, 2005) and in the soft and hard X-rays (Nicholson et al. 1998; Ho et al. 
2001).  The X-ray spectrum is only very mildly absorbed (Nicholson et al. 
1998; Pellegrini et al. 2002, 2003a; Terashima et al. 2002), with no signs of 
Fe~K$\alpha$ emission expected from reprocessed material, consistent with the 
modest mid-IR emission reported by Bendo et al. (2006).  In short, there is no 
sign of anything being hidden or much doing the hiding.  So where is the BLR?  
It is just not there.  

The lack of a BLR in very low-luminosity sources may be related to a physical 
upper limit in the broad-line width (Laor 2003).  If LLAGNs obey the same 
BLR-luminosity relation as in higher luminosity systems, their BLR velocity 
depends on the BH mass and luminosity.  At a limiting bolometric luminosity of 
$L_{\rm bol} \approx 10^{41.8} (M_{\rm BH}/10^8\,M_\odot)^2$ \lum, $\Delta 
v\approx 25,000$ \kms, above which clouds may not survive due to excessive 
shear or tidal forces.  Alternatively, if BLR clouds arise from condensations 
in a radiation-driven, outflowing wind (Murray \& Chiang 1997), a viewpoint 
now much espoused, then it is reasonable to expect that very low-luminosity 
sources would be incapable of generating a wind, and hence of sustaining a 
BLR.  For example, the clumpy torus model of Elitzur \& Shlosman (2006) 
predicts that the BLR can no longer be sustained for $L_{\rm bol}$ \lax\ 
$10^{42}$ \lum.  In the scenario of Nicastro (2000), the BLR originates from a 
disk outflow formed at the transition radius between regions dominated by gas 
and radiation pressure.  As this radius shrinks with decreasing 
$L_{\rm bol}/L_{\rm Edd}$, where $L_{\rm Edd} = 1.3\times 10^{38}\,(M_{\rm BH}/
M_{\odot})$ \lum, the BLR is expected to disappear for 
$L_{\rm bol}/L_{\rm Edd}$ \lax\ $10^{-3}$.  The apparent correlation between 
BLR line width and $L_{\rm bol}/L_{\rm Edd}$ qualitatively supports this 
picture (Xu \& Cao 2007).  Although the existing data are sparse, they 
indicate that LINERs generally lack UV resonance absorption features 
indicative of nuclear outflows (Shields et al. 2002).  The models by Elitzur 
\& Shlosman and Nicastro are probably correct in spirit but not in detail, 
because many of the Palomar LLAGNs plainly violate their proposed thresholds 
(\S~5.10). 

Nonetheless, the statistics within the Palomar survey already provide tentative
support to the thesis that the BLR vanishes at the lowest luminosities or 
Eddington ratios. Which of the two is the controlling variable is 
still difficult to say.  For both Seyferts and LINERs, type~1 sources are 
almost a factor of 10 more luminous than type~2 sources in terms of their 
median total H\al\ luminosity (Table~1).  (The statistical differences between 
type~1 and type~2 sources cannot be ascribed to sensitivity differences in the 
detectability of broad H\al\ emission.  Type~1 objects do have stronger line 
emission compared to the type~2s, but on average their narrow H\al\ flux and 
equivalent width are only $\sim 50$\% higher, and the two types overlap 
significantly.  Moreover, as noted in \S 3.4, the broad H\al\ detection rates 
turn out to be quite robust even in light of the much higher sensitivity 
afforded by \hst.) The differences persist after normalizing by the Eddington 
luminosities: adopting a bolometric correction of $L_{\rm bol} \approx 16 
L_{\rm X}$, $L_{\rm bol}/L_{\rm Edd}=1.1\times10^{-3}$ and $5.9\times10^{-6}$ 
for Seyfert~1s and Seyfert~2s, respectively, whereas the corresponding values 
for LINER~1s and LINER~2s are $1.0\times10^{-5}$ and $4.8\times10^{-6}$.  Two 
caveats are in order.  First, while most of the type~1 sources have X-ray 
data, only 60\% of the LINER~2s and 70\% of the Seyfert~2s do.  Second, the 
X-ray luminosities, which pertain to the $2-10$ keV band, have been corrected 
for intrinsic absorption whenever possible, but many sources are too faint for 
spectral analysis.  The lower X-ray luminosities for the type~2 sources must 
be partly due to absorption, but considering the generally low absorbing 
columns, particularly among the LINERs (Georgantopoulous et al. 2002; 
Terashima et al. 2002), it is unclear if absorption alone can erase the 
statistical difference between the two types.  The tendency for Seyfert~2s to 
have lower Eddington ratios than Seyfert~1s has previously been noted, for the 
Palomar sample (Panessa et al. 2006) and others (Middleton, Done \& Schurch 
2008).

Several authors have raised the suspicion that LINER~2s may not be
accretion-powered.  Large-aperture X-ray spectra of LINER~2s, like those of 
LINER~1s, can be fit with a soft thermal component plus a power law with 
$\alpha \approx -0.7$ to $-1.5$ (Georgantopoulos et al. 2002; Terashima et al. 
2002). But this alone does not provide enough leverage to distinguish AGNs from 
starburst galaxies, many of which look qualitatively similar over the limited 
energy range covered by these observations.  We cannot turn to the iron K\al\ 
line or variability for guidance, because LLAGNs generally exhibit neither 
(\S~5.3).  The hard X-ray emission in LINER~2s is partly extended (Terashima 
et al. 2000a; Georgantopoulos et al.  2002), but the implications of this 
finding are unclear.  Just because the X-ray emission surrounding the LLAGN is 
morphologically complex and there is evidence for circumnuclear star formation 
(e.g., NGC~4736; Pellegrini et al. 2002) does not necessarily imply that there 
is a causal connection between the starburst and the LLAGN.  Roberts, Schurch 
\& Warwick (2001) advocate a starburst connection from the observation that 
LINER~2s have a mean flux ratio in the soft and hard X-ray band ($\sim 0.7$) 
similar to that found in NGC~253.  This interpretation, however, conflicts 
with the stellar population constraints discussed in \S~4.2.  It is also not 
unique.  Luminous, AGN-dominated type~1 sources themselves exhibit a tight 
correlation between soft and hard X-ray luminosity, with a ratio not 
dissimilar from the quoted value (Miniutti et al. 2008).

An important clue comes from the fact that many LINER~2s have a lower
$L_{\rm X}/L_{{\rm H}\alpha}$ ratio than LINER~1s (Ho et al. 2001). In 
particular, the observed X-ray luminosity from the nucleus, when extrapolated 
to the UV, does not have enough ionizing photons to power the H\al\ emission 
(Terashima et al.  2000a).  This implies that (1) the X-rays are heavily 
absorbed, (2) nonnuclear processes power much of the optical line emission, 
or (3) the ionizing SED is different than assumed.  As discussed in \S~6.4, this
energy budget discrepancy appears to be symptomatic of all LLAGNs in general,
not just LINER~2s, and most likely results from a combination of the second and
third effect.  There are some indications that the SEDs of LINER~2s indeed 
differ systematically from those of LINER~1s (e.g., Maoz et al. 2005; Sturm et 
al. 2006).  In light of the evidence given in \S\S~5.3, 5.6, I consider the
first solution to be no longer tenable.  One can point to objects such as
NGC~4261 (Zezas et al. 2005) as examples of LINER~2s with strong obscuration, 
but such cases are rare.

From the point of view of BH demographics, the most pressing issue is what 
fraction of the LINER~2s should be included in the AGN tally.  Some cases are 
beyond dispute (M84, M87, Sombrero).  What about the rest?  The strongest 
argument that the majority of LINER~2s are AGN-related comes from the 
detection frequency of radio (\S~5.2) and X-ray (\S~5.3) cores, which is 
roughly 60\% of that of LINER~1s.  On the other hand, the detection rate of 
Seyfert~2s are similarly lower compared to Seyfert~1s, most likely reflecting 
the overall reduction of nuclear emission across all bands in type~2 LLAGNs 
as a consequence of their lower accretion rates.  In summary, the AGN fraction 
among LINER~2s is at least 60\%, and possibly as high as 100\%.

\subsection{Torus}

In line with the absence of a BLR discussed above and using very much the 
same set of evidence, a convincing case can be made that the torus also 
disappears at very low luminosities.  In a large fraction of nearby LINERs, 
the low absorbing column densities and weak or undetected Fe~K$\alpha$ emission 
(\S~5.3) strongly indicate that we have a direct, unobstructed view of the 
nucleus.  Ghosh et al. (2007) warn that absorbing columns can be underestimated 
in the presence of extended soft emission, especially when working with 
spectra of low signal-to-noise ratio.  While this bias no doubt enters at some 
level, cases like the Sombrero (\S~5.3) cannot be so readily dismissed.  By 
analogy with situation in luminous AGNs (e.g., Inoue, Terashima \& Ho 2007; 
Nandra et al. 2007), type~1 LLAGNs, if they possess tori, should also show 
strong, narrow fluorescent Fe~K$\alpha$ emission.  This expectation is not 
borne out by observations.  NGC~3998, which has excellent X-ray data, offers 
perhaps the most dramatic example.  Apart from showing no signs whatsoever for 
intrinsic photoelectric absorption, it also possesses one of the tightest 
upper limits to date on Fe~K$\alpha$ emission: EW $<$ 25 eV (Ptak et al. 
2004).  Our sight line to the nucleus is as clean as a whistle.  Satyapal, 
Sambruna \& Dudik (2004) claim that many LINERs have obscured nuclei, but 
this conclusion is based on IR-bright, dusty objects chosen from Carrillo et 
al. (1999); as I have discussed in \S~3.2, I regard these objects not only as 
biased, but also confusing with respect to their nuclear properties.

Palomar Seyferts, whose luminosities and Eddington ratios are about an order 
of magnitude higher than those of LINERs (\S~5.10), show markedly larger 
absorbing column densities and stronger Fe~K$\alpha$ lines.  In an 
{\it XMM-Newton}\ study of a 
distance-limited sample of 27 Palomar Seyferts, Cappi et al. (2006) detect 
strong Fe~K$\alpha$ emission in over half of objects.  The distribution of 
absorbing columns is nearly continuous, from $N_{\rm H} \approx 10^{20}$ to 
$10^{25}~{\rm cm}^{-2}$, with 30\%$-$50\% of the type~2 sources being 
Compton-thick (Panessa et al. 2006).  This seems consistent with the tendency 
for Seyferts to be more gas-rich than LINERs, to the extent that this is 
reflected in their higher NLR densities (Ho, Filippenko \& Sargent 2003).

The trend of increasing absorption with increasing luminosity or Eddington 
ratio observed in Palomar LLAGNs has an interesting parallel among radio 
galaxies.  A substantial body of recent work indicates that the nuclei of FR~I 
sources, most of which are, in fact, LINERs, are largely unobscured (e.g., 
Chiaberge, Capetti \& Celotti 1999; Donato, Sambruna \& Gliozzi 2004; 
Balmaverde \& Capetti 2006). In contrast, FR~II systems, especially those with 
broad or high-excitation lines (analogs of Seyferts), show clear signs of 
absorption and Fe~K$\alpha$ emission (Evans et al. 2006).

Even if we are fooled by the X-ray observations, substantial absorption 
must result in strong thermal reemission of ``waste heat'' in the IR.  While 
sources such as Cen~A provide a clear reminder that every rule has its 
exception (Whysong \& Antonucci 2004), the existing data do suggest that, 
as a class, FR~I radio galaxies tend to be weak mid-IR or far-IR sources 
(Haas et al. 2004; M\"uller et al. 2004).  The same holds for more nearby 
LINERs.  Their SEDs do show a pronounced mid-IR peak (\S~5.8), but as I will 
argue later, it is due to emission from the accretion flow rather than from 
dust reemission.

\subsection{Narrow-line Region Kinematics}

The kinematics of the NLR are complex.  At the smallest scales probed by \hst, 
Verdoes~Kleijn, van~der~Marel \& Noel-Storr (2006) find that the velocity 
widths of the ionized gas in the LINER nuclei of early-type galaxies can be 
modeled as unresolved rotation of a thin disk in the gravitational potential 
of the central BH.  The subset of objects with FR~I radio morphologies, on the 
other hand, exhibit line broadening in excess of that expected from purely 
gravitational motions; these authors surmise that the super-virial motions may 
be related to an extra source of energy injection by the radio jet.  
Walsh et al. (2008) use multiple-slit STIS observations to map the 
kinematics of the inner $\sim 100$ pc of the NLR in a sample of 14 LLAGNs, 
mostly LINERs.  Consistent with earlier findings (Ho et al. 2002; Atkinson et 
al. 2005),  the velocity fields are generally quite disorganized, rarely 
showing clean signatures of dynamically cold disks undergoing circular rotation.
Nevertheless, two interesting trends can be discerned.  The emission line 
widths tend to be largest within the sphere of influence of the BH, 
progressively decreasing toward large radii to values that roughly match the 
stellar velocity dispersion of the bulge.  The luminous members of the sample, 
on the other hand, show more chaotic kinematics, as evidenced by large velocity 
splittings and asymmetric line profiles, reminiscent of the pattern observed 
by Rice et al. (2006) in their sample of Seyfert galaxies.  Walsh et al. 
suggest that above a certain luminosity threshold---one that perhaps coincides 
with the LINER/Seyfert division---AGN outflows and radio jets strongly perturb 
the kinematics of the NLR.  

A large fraction ($\sim 90$\%) of the Palomar LLAGNs have robust measurements 
of integrated \nii\ \lamb6583 line widths, which enable a crude assessment of 
the dynamical state of the NLR and its relation to the bulge.  Consistent with 
what has been established for more powerful systems (Nelson \& Whittle 1996; 
Greene \& Ho 2005a), the kinematics of the ionized gas are dominated by random 
motions that, to first order, trace the gravitational potential of the stars 
in the bulge.  Among the objects with available central stellar velocity 
dispersions, $\sigma_{\rm NLR}/\sigma_* \approx 0.7-0.8$ for the weakest 
sources ($L_{{\rm H}\alpha} \approx 10^{38}$ \lum), systematically rising to 
$\sigma_{\rm NLR}/\sigma_* \approx 1.2$ in the more luminous members 
($L_{{\rm H}\alpha} \approx 10^{41.5}$ \lum).  L.C. Ho (in preparation)
speculates that the central AGN injects a source of dynamical heating of 
nongravitational origin to the NLR, either in the form of radiation pressure 
from the central continuum or mechanical interaction from radio jets.   Given 
the empirical correlation between optical line luminosity and radio power 
(e.g., Ho \& Peng 2001; Ulvestad \& Ho 2001a; Nagar, Falcke \& Wilson 2005), 
and the near ubiquity of compact radio sources, it is {\it a priori}\ 
difficult to determine which of these two sources acts as the primary driver. 
The tendency for extended radio emission to be more prevalent in Seyferts 
(\S~5.2) suggests that jets may be more important.

\subsection{Spectral Energy Distribution}

The broad-band SED provides one of the most fundamental probes of the physical 
processes in AGNs.  Both thermal and nonthermal emission contribute to the 
broad-band spectrum of luminous AGNs such as quasars and classical Seyfert 
galaxies.  In objects whose intrinsic spectrum has not been modified severely 
by relativistic beaming or absorption, the SED can be separated into several
distinctive components (e.g., Elvis et al. 1994): radio synchrotron emission 
from a jet, which may be strong (``radio-loud'') or weak (``radio-quiet''); an 
IR excess, now generally considered to be predominantly thermal 
reradiation by dust grains; a prominent optical to UV ``big blue bump,'' 
usually interpreted to be pseudo-blackbody emission from an optically thick, 
geometrically thin accretion disk (Shields 1978; Malkan \& Sargent 1982); a 
soft X-ray excess, whose origin is still highly controversial (Done et al. 
2007; Miniutti et al. 2008); and an underlying power law, which is most 
conspicuous at hard X-ray energies but is thought to extend down to IR 
wavelengths, that can be attributed to Comptonization of softer seed photons.

Within this backdrop, there were already early indications that the SEDs of 
LINERs may deviate from the canonical form.  Halpern \& Filippenko (1984) 
succeeded in detecting the featureless optical continuum in NGC~7213, and 
while these authors suggested that a big blue bump may be present in this 
object, they also noted that it possesses an exceptionally high 
X-ray--to--optical flux ratio, although perhaps one not inconsistent with the 
extrapolation of the trend of increasing X-ray--to--optical flux ratio with 
decreasing luminosity seen in luminous sources (Zamorani et al. 1981; Avni \& 
Tananbaum 1982).  A more explicit suggestion that LINERs may possess a weak UV 
continuum was made in the context of double-peaked broad-line AGNs such as 
Arp~102B and Pictor~A, whose narrow-line spectra share many characteristics 
with LINERs (Chen \& Halpern 1989; Halpern \& Eracleous 1994).  The {\it HST}\
spectrum of Arp~102B, in fact, shows an exceptionally steep optical-UV 
nonstellar continuum ($\alpha \approx -2.1$ to $-2.4$; Halpern et al. 1996).
Halpern \& Eracleous (1994) further suggested that the SEDs are flat in the 
far-IR.  In an important study of M81, Petre et al. (1993) proposed that the 
relative weakness of the UV continuum compared to the X-rays is a consequence 
of a change in the structure of the central accretion flow, from a standard 
thin disk to an ion-supported torus (see \S~8.3).  Parameterizing the 
two-point spectral index between 2500 \AA\ and 2 keV by \alphaox\ $\equiv\, 
[\log L_\nu({\rm 2500~\AA}) - \log L_\nu({\rm 2~keV})]/
[\log \nu({\rm 2500~\AA}) - \log \nu({\rm 2~keV})]$, M81 and possibly other 
LINERs (Mushotzky 1993) have \alphaox\ \gax\ $-1$, to be compared with 
\alphaox\ $\approx -1.4$ for quasars and \alphaox\ $\approx -1.2$ for Seyferts 
(Mushotzky \& Wandel 1989).

The full scope of the spectral uniqueness of LLAGNs only became evident once 
the modern, albeit still fragmentary, multiwavelength data could be assembled. 
The initial studies concentrated on individual objects, emphasizing the 
weakness of the UV bump (M81: Ho, Filippenko \& Sargent 1996; Sombrero: 
Nicholson et al. 1998) and the overall consistency of the SED with spectral 
models generated from advection-dominated accretion flows (ADAFs; see Narayan 
2002 and Yuan 2007 for reviews) as unique attributes of systems with low 
Eddington ratios (NGC~4258: Lasota et al. 1996, Chary et al. 2000; M87: 
Reynolds et al. 1996; M60: Di~Matteo \& Fabian 1997).  Ho (1999b) 
systematically investigated the SEDs of a small sample of seven LLAGNs with 
available BH mass estimates and reliable small-aperture fluxes from radio to 
X-ray wavelengths.  This was followed by a study of another 
five similar objects, which have the additional distinction of having
double-peaked broad emission lines (Ho et al. 2000; Ho 2002b).  Figure~7 gives
the latest update from a comprehensive analysis of the SEDs of 150 nearby
type~1 AGNs spanning 4 dex in BH mass ($M_{\rm BH}\approx 10^5-10^9$ \solmass)
and 6.5 dex in Eddington ratio ($L_{\rm bol}/L_{\rm Edd} \approx 10^{-6} -
10^{0.5}$).  Let us focus on two regimes: $L_{\rm bol}/L_{\rm Edd} = 0.1$ to
1, typical of classical, luminous AGNs, and $L_{\rm bol}/L_{\rm Edd} <
10^{-3.0}$, which characterizes most nearby LLAGNs (\S~5.10).  I defer the
discussion of the physical implications until \S~8, but for now list the most
notable features concerning the LLAGN SED, some of which are also apparent in 
the composite LINER SED assembled by Eracleous, Hwang \& Flohic (2008a).  (1) 
The big blue bump is conspicuously absent. (2) Instead, a broad excess is 
shifted to the mid-IR, forming a ``big red bump''; this component is probably 
related to the mid-IR excess previously noted by Lawrence et al. (1985), 
Willner et al. (1985), and Chen \& Halpern (1989), and more recently from 
{\it Spitzer}\ observations (e.g., Willner et al. 2004; Bendo et al. 2006; Gu 
et al. 2007).  (3) As a consequence of this shift, the optical-UV slope is 
exceptionally steep, generally in the range $\alpha_{\rm ou} \approx -1$ to 
$-2.5$, to be compared with $\alpha_{\rm ou} \approx -0.5$ to $-0.7$ for 
luminous AGNs 
(Vanden~Berk et al. 2001; Shang et al. 2005); the X-ray--to--optical ratio is 
large, resulting in \alphaox\ \gax\ $-1$. (4) There is no evidence for a soft 
X-ray excess. (5) Lastly, the overall SED can be considered radio-loud, 
defined here by the convention that the radio-to-optical luminosity ratio 
exceeds a value 

\clearpage
\begin{figure}
\hskip -0.2cm
\psfig{figure=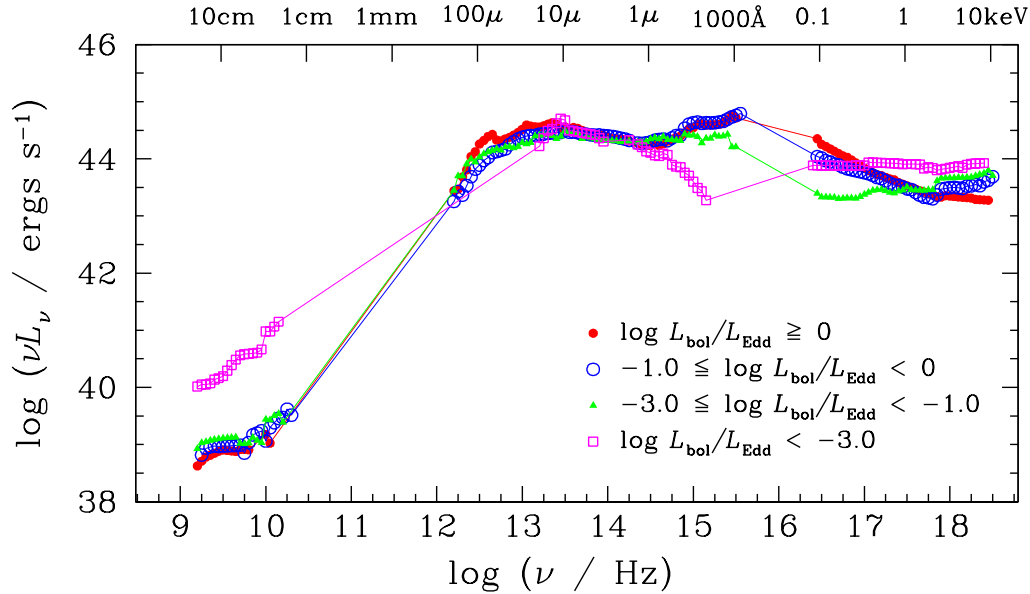,width=14cm,angle=270}
\caption{Composite SEDs for radio-quiet AGNs binned by Eddington
ratio.  The SEDs are normalized at 1 \micron. (Adapted from L.C. Ho, in
preparation.)
}
\end{figure}

\noindent
of $R_o \equiv\,L_{\nu}({\rm 5~GHz})/L_{\nu}(B) = 10$.  
Radio-loudness, in fact, seems to be a property common to essentially 
{\it all}\ nearby weakly active nuclei (Ho 1999b, 2002a; Ho et al. 2000) and a 
substantial fraction of Seyfert nuclei (Ho \& Peng 2001).  Defining 
radio-loudness based on the relative strength of the radio and X-ray emission, 
$R_{\rm X} \equiv\,\nu L_{\nu}({\rm 5~GHz})/ L_{\rm X}$, Terashima \& Wilson 
(2003b) also find that LINERs tend to be radio-loud, here taken to be 
$R_{\rm X} > 10^{-4.5}$.  Moreover, the degree of radio-loudness scales 
inversely with $L_{\rm bol}/L_{\rm Edd}$ (Ho 2002a; Terashima \& Wilson 2003b; 
Wang, Luo \& Ho 2004; Greene, Ho \& Ulvestad 2006; Panessa et al. 2007; 
Sikora, Stawarz \& Lasota 2007; L.C. Ho, in preparation; see Figure~10{\it b}).

In a parallel development, studies of the low-luminosity, often LINER-like 
nuclei of FR~I radio galaxies also support the notion that they lack a UV 
bump.  M84 (Bower et al. 2000) and M87 (Sabra et al. 2003) are two familiar 
examples, but it has been well documented that FR~I nuclei tend to exhibit 
flat \alphaox\ (Donato, Sambruna \& Gliozzi 2004; Balmaverde, Capetti \& 
Grandi 2006; Gliozzi et al. 2008) and steep slopes in the optical (Chiaberge, 
Capetti \& Celotti 1999; Verdoes~Kleijn et al. 2002) and optical-UV (Chiaberge 
et al. 2002).

Finally, I note that the UV spectral slope can be indirectly constrained
from considering the strength of the \heii\ \lamb4686 line.  While this 
line is clearly detected in Pictor~A (Carswell et al. 1984; Filippenko 1985), 
its weakness in NGC~1052 prompted P\'equignot (1984) to deduce that the 
ionizing spectrum must show a sharp cutoff above the He$^+$ ionization limit 
(54.4 eV).  In this respect, NGC~1052 is quite representative of LINERs in 
general.  \heii\ \lamb4686 was not detected convincingly in a {\it single}\ 
case among a sample of 159 LINERs in the entire Palomar survey (Ho, 
Filippenko \& Sargent 1997a).  Starlight contamination surely contributes 
partly to this, but the line has also eluded detection in {\it HST}\ spectra 
(e.g., Ho, Filippenko \& Sargent 1996; Nicholson et al. 1998; Barth et al. 
2001b; Sabra et al. 2003; Sarzi et al. 2005; Shields et al. 2007), which 
indicates that it is truly intrinsically very weak.  To a first 
approximation, the ratio of \heii\ \lamb4686 to H\bet\ reflects the relative 
intensity of the ionizing continuum between 1 and 4 Ryd.  For an ionizing 
spectrum $f_\nu \propto \nu^{\alpha}$, case~B recombination predicts 
\heii\ \lamb4686/H\bet\ = $1.99 \times 4^\alpha$ (Penston \& Fosbury 1978).  
The current observational limits of \heii\ \lamb4686/H\bet\ \lax\ 0.1 thus 
imply $\alpha$ \lax\ $-2$, qualitatively consistent with the evidence from the 
SED studies.  

Maoz (2007) has offered an alternative viewpoint to the one presented above. 
Using a sample of 13 LINERs with variable UV nuclei, he argues that their 
SEDs do not differ appreciably from those of more luminous AGNs, and hence that
LINERs inherently have very similar accretion disks compared to powerful AGNs.  
Maoz does not disagree that LINERs have large X-ray--to-UV flux ratios or that 
they tend to be radio-loud; his data show both trends.  Rather, he contends 
that because LINERs lie on the low-luminosity extrapolation of the well-known 
relation between \alphaox\ and luminosity (Zamorani et al. 1981; Avni \& 
Tananbaum 1982; Strateva et al.  2005) they do not form a distinct population.
And while LINERs do have large values of $R_o$, they nonetheless occupy the 
``radio-quiet'' branch of the $R_o$ versus $L_{\rm bol}/L_{\rm Edd}$ plane 
(Sikora, Stawarz \& Lasota  2007).  In my estimation, the key point is not, 
and has never been, whether LINERs constitute a disjoint class of AGNs, but 
whether they fit into a physically plausible framework in which their 
distinctive SEDs, among other properties, find a natural, coherent 
explanation.  Section 8 attempts to offer such a framework.

It should be noted that Maoz's results strongly depend on his decision to 
exclude all optical and near-IR data from the SEDs, on the grounds that they 
may be confused by starlight.  I think this step is too draconian, as it 
throws away valuable information.  While stellar contamination is certainly a
concern, one can take necessary precautions to try to isolate the nuclear 
emission as much as possible, either through high-resolution imaging (e.g., Ho 
\& Peng 2001; Ravindranath et al. 2001; Peng et al. 2002) or spectral 
decomposition.  In well-studied sources, there is little doubt that the 
optical continuum is truly both featureless and nonstellar (e.g., Halpern \& 
Filippenko 1984; Ho, Filippenko \& Sargent 1996; Ho et al. 2000; Bower et al. 
2000; Sabra et al. 2003).  Given what we know about the nuclear stellar 
population, we cannot assign the featureless continuum to young stars.  In a 
few cases, the nonstellar nature of the nucleus can even be established 
through variability in the optical (Bower et al. 2000; Sabra et al. 2003; 
O'Connell et al. 2005) and mid-IR (Rieke, Lebofsky \& Kemp 1982; Grossan et 
al. 2001; Willner et al. 2004).

While the SEDs of LINERs differ from those of traditional AGNs, it is
important to recognize that they are decidedly {\it nonstellar}\ and approximate
the form predicted for radiatively inefficient accretion flows (RIAFs)
onto BHs, often coupled to a jet (Quataert et al. 1999; Yuan, Markoff \& 
Falcke 2002; Yuan et al. 2002; Fabbiano et al. 2003; Pellegrini et al. 2003b; 
Ptak et al. 2004; Nemmen et al. 2006; Wu, Yuan \& Cao 2007).  They bear 
little resemblance to SEDs characteristic of normal stellar systems.  
Inactive galaxies or starburst systems not strongly affected by dust 
extinction emit the bulk of their radiation in the optical--UV and in the 
thermal IR regions, with only an energetically miniscule contribution from 
X-rays.

\clearpage
\begin{figure}
\psfig{figure=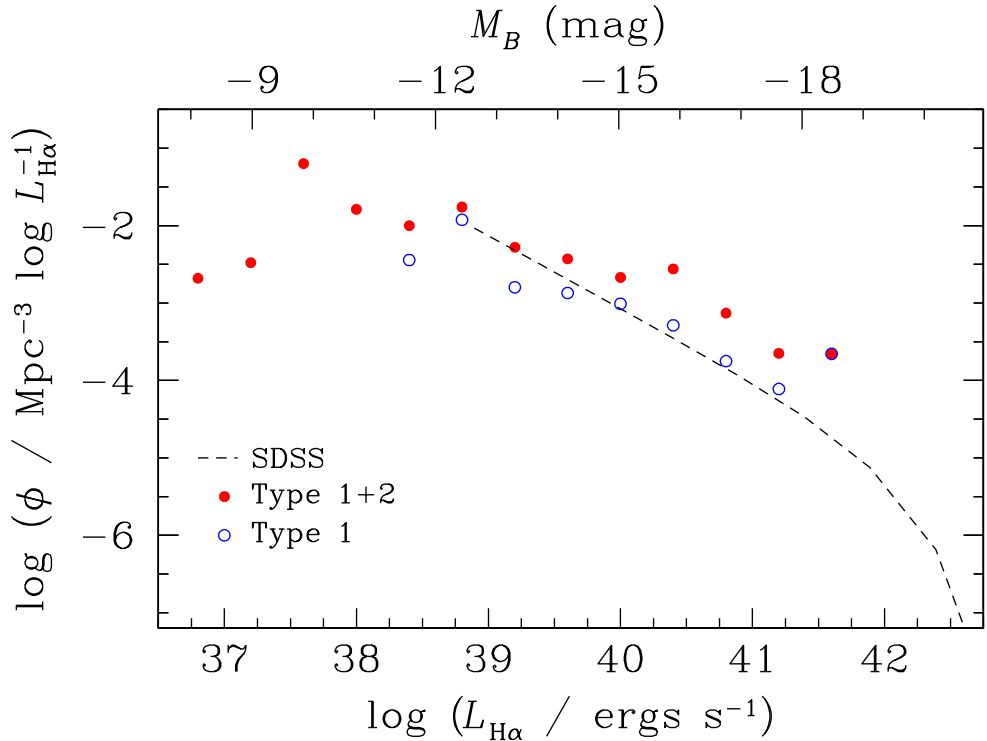,width=13.5cm,angle=270}
\caption{The H\al\ nuclear luminosity function of nearby AGNs derived from the
Palomar survey.  The top axis gives an approximate conversion to absolute
magnitudes in the $B$ band, using the H\al-continuum correlation of Greene \&
Ho (2005b).  The unfilled circles include only type~1 sources, while the
filled circles represent both type~2 and type~1 sources.  The luminosities
have been corrected for extinction, and in the case of type~1 nuclei, they
include both the narrow and broad components of the line.  For
comparison, I show the  $z < 0.35$ luminosity function for SDSS Seyfert
galaxies (types 1 and 2; dashed line; Hao et al. 2005b).  (Adapted from
L.C. Ho, A.V. Filippenko \& W.L.W.  Sargent, in preparation.)
}
\end{figure}

\subsection{Luminosity Function}

Many astrophysical applications of AGN demographics benefit from knowing
the AGN luminosity function, $\Phi(L,z)$.  Whereas $\Phi(L,z)$ has been 
reasonably well charted at high $L$ and high $z$ using quasars, it is very 
poorly known at low $L$ and low $z$.  Indeed, until very recently there has 
been no reliable determination of $\Phi(L,0)$.  The difficulty in determining 
$\Phi(L,0)$ can be ascribed to a number of factors, as discussed in Huchra \& 
Burg (1992).  First and foremost is the challenge of securing a reliable, 
spectroscopically selected sample.  Since nearby AGNs are expected to be faint 
relative to their host galaxies, most of the traditional techniques used to 
identify quasars cannot be applied without introducing large biases.  The 
faintness of nearby AGNs presents another obstacle, namely how to disentangle 
the nuclear emission---the only component relevant to the AGN---from the 
usually much brighter contribution from the host galaxy.  Finally, most 
optical luminosity functions of bright, more distant AGNs are specified in 
terms of the nonstellar optical continuum (usually the $B$ band), whereas 
spectroscopic surveys of nearby galaxies generally only reliably measure 
optical line emission (e.g., H\al) because the featureless nuclear continuum 
is often impossible to detect in ground-based, seeing-limited apertures.

A different strategy can be explored by taking advantage of the fact that
H\al\ luminosities are now available for nearly all of the AGNs in the Palomar
survey.  Figure~8 shows the H\al\ luminosity function for the Palomar sources, 
computed using the $V/V_{\rm max}$ method (L.C. Ho, A.V. Filippenko \& W.L.W. 
Sargent, in preparation).  Two versions are shown, each representing an 
extreme view of what kind of sources should be regarded as {\it bona fide}\ 
AGNs.  The open symbols include only type~1 nuclei, whose AGN status is 
incontrovertible.  This may be regarded as the most conservative assumption 
and a lower bound, since we know that genuine narrow-line AGNs do exist.  The 
filled symbols lump together all sources classified as LINERs, transition 
objects, and Seyferts, both type~1 and type~2.  This represents the most 
optimistic view and an upper bound, if some type~2 sources are in fact AGN 
impostors, although, as I argue in \S~6.5, this is likely to 
be a small effect.  The true space density of local AGNs
lies between these two possibilities.  In either case, the differential
luminosity function can be approximated by a single power law from
$L_{{\rm H}\alpha} \approx 10^{38}$ to $3\times 10^{41}$ \lum, roughly of the
form $\Phi \propto L^{-1.2\pm0.2}$.  The slope seems to flatten below
$L_{{\rm H}\alpha} \approx 10^{38}$ \lum, but the luminosity function is
highly uncertain at the faint end because of density fluctuations in our local
volume.  Nevertheless, it is remarkable that the Palomar luminosity function
formally begins at $L_{{\rm H}\alpha} \approx 6 \times 10^{36}$ \lum,
roughly the luminosity of the Orion nebula (Kennicutt 1984).  In units more 
familiar to the AGN community, this corresponds to an absolute $B$-band 
magnitude of roughly $-8$ (using the H\al-optical continuum conversion of 
Greene \& Ho 2005b), no brighter than a single supergiant star. 

For comparison, I have overlaid the H\al\ luminosity function of $z$ \lax\ 
0.35 Seyfert galaxies derived from the SDSS by Hao et al. (2005b).  The 
Palomar survey reaches $\sim 2$ orders of magnitude fainter in H\al\ 
luminosity than SDSS, but the latter extends a factor of 10 higher at the 
bright end.  Over the region of overlap, the two surveys show reasonably good 
agreement, especially considering the small number statistics of the Palomar 
survey and the fact that Hao et al.'s sample only includes Seyferts.

\subsection{Bolometric Luminosities and Eddington Ratios}

\begin{figure}
\vbox{\hbox{
\psfig{figure=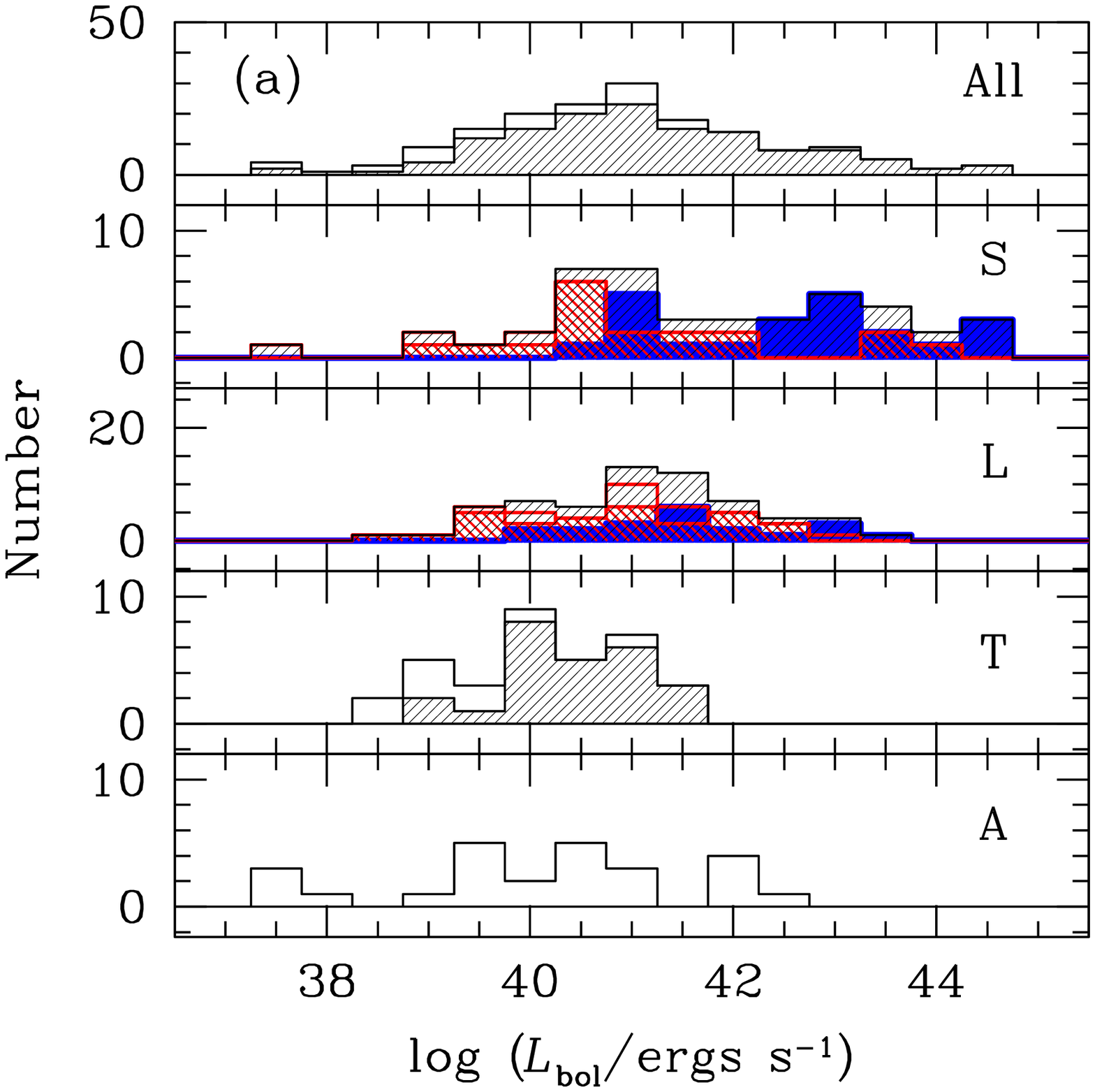,width=6.7cm,angle=0}
\hskip -0.2 cm
\psfig{figure=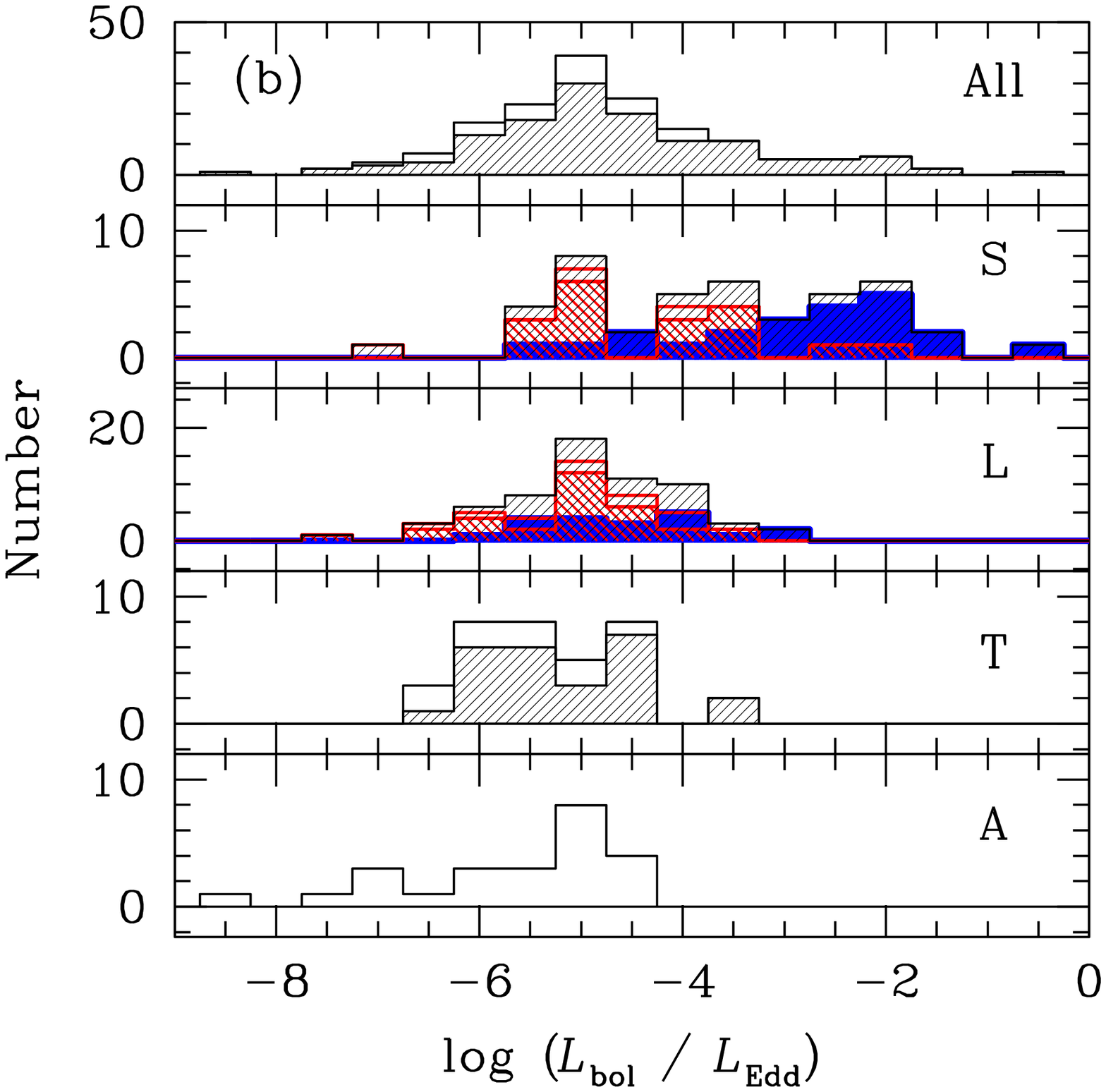,width=6.7cm,angle=0}
}}
\caption{Distribution of ({\it a}) bolometric luminosity, $L_{\rm bol}$, and
({\it b}) ratio of bolometric luminosity to the Eddington luminosity,
$L_{\rm bol}/L_{\rm Edd}$, for all objects, Seyferts (S), LINERs (L),
transition objects (T), and absorption-line nuclei (A).  $L_{\rm bol}$ is
based on the X-ray ($2-10$ keV) luminosity.  The hatched and open histograms
denote detections and upper limits, respectively; type~1 objects are plotted
in blue, type~2 objects in red.  (Adapted from L.C. Ho, in preparation.)
}
\end{figure}

To gain further insight into the physical nature of LLAGNs, it is more
instructive to examine their bolometric luminosities rather than their
luminosities in a specific band or emission line.  Because AGNs emit a very
broad spectrum, their bolometric luminosities ideally should be measured
directly from their full SEDs.  In practice, however, complete SEDs are not 
readily available for most AGNs, and one commonly estimates $L_{\rm bol}$
by applying bolometric corrections derived from a set of well-observed 
calibrators.  As discussed in \S~5.8, the SEDs of LLAGNs differ quite markedly 
from those of conventionally studied AGNs.  Nonetheless, they do exhibit a 
characteristic shape, which enables bolometric corrections to be calculated.  
The usual practice of choosing the optical $B$ band as the reference point 
should be abandoned for LLAGNs, not only because reliable optical continuum 
measurements are scarce but also because the optical/UV region of the SED 
shows the maximal variance with respect to accretion rate (\S~5.8) and depends 
sensitively on extinction.  What is available, by selection, is nuclear 
emission-line fluxes, and upper limits thereon.  Although the H\al\ luminosity 
comprises only a small percentage of the total power, its fractional 
contribution to $L_{\rm bol}$ turns out to be fairly well defined: from the 
SED study of L.C. Ho (in preparation), $L_{\rm bol} \approx 220 
L_{{\rm H}\alpha}$, with an rms scatter of $\sim 0.4$ dex, consistent with the 
calibration given in Greene \& Ho (2005b, 2007a).  Because of the wide range 
of ionization levels among LLAGNs, a bolometric correction based on H\al\ 
should be more stable than one tied to \oiii\ \lamb 5007 (e.g., Heckman et al. 
2005).  Nevertheless, in light of the nonnuclear component of the nebular 
flux in LLAGNs (\S~6.4), the luminosity of the narrow H\al\ line will tend to 
overestimate $L_{\rm bol}$.  I recommend that, whenever possible, $L_{\rm bol}$
should be based on the hard X-ray ($2-10$ keV) luminosity, bearing in mind the 
added complication that the bolometric correction in this band is 
luminosity-dependent.  Making use again of the database from L.C. Ho (in 
preparation), I estimate $L_{\rm bol}/L_{\rm X} \approx 83$, 28, and 16 for 
quasars, luminous Seyferts, and LLAGNs, respectively.

Figure~9 shows the distributions of $L_{\rm bol}$ and their values normalized 
with respect to the Eddington luminosity for Palomar galaxies with 
measurements of $L_{\rm X}$ and central stellar velocity dispersion. The 
$M_{\rm BH} - \sigma$ relation of Tremaine et al. (2002) was used to obtain
\ledd.  Although there is substantial overlap, the four spectral classes 
clearly delineate a luminosity sequence, with $L_{\rm bol}$ decreasing 
systematically as S$\rightarrow$L$\rightarrow$T$\rightarrow$A.  The 
differences become even more pronounced in terms of $L_{\rm bol}/L_{\rm Edd}$, 
with Seyferts having a median value (1.3\e{-4}) 20 times higher than in LINERs 
(5.9\e{-6}), which in turn are higher than transition objects by a factor of 
$\sim 5$.  Among Seyferts and LINERs, type~1 sources are systematically more
luminous than type~2s.  Notably, the vast majority of nearby nuclei have 
highly sub-Eddington luminosities.  The total distribution of Eddington ratios 
is characterized by a prominent peak at $L_{\rm bol}/L_{\rm Edd} \approx 
10^{-5}$ dominated by Seyfert~2s, LINERs, and transition objects, and a 
precipitous drop toward larger Eddington ratios.  Contrary to previous claims 
(Wu \& Cao 2005; Hopkins \& Hernquist 2006) based on the smaller sample of Ho 
(2002a), the distribution of Eddington ratios shows no bimodality.  The 
systematic difference in Eddington ratios between LINERs and Seyferts has been 
noticed before in the Palomar survey (Ho 2002b, 2003, 2005) and in SDSS 
(Kewley et al. 2006), but this is the first time that the more subtle 
differences among the different subclasses can be discerned.

\section{EXCITATION MECHANISMS}

\subsection{Nonstellar Photoionization}

The origin and excitation of the ionized gas in the central regions of nearby 
galaxies has been a longstanding problem (Minkowski \& Osterbrock 1959; 
Osterbrock 1960).  Ever since the early suggestion of Ferland \& Netzer (1983) 
and Halpern \& Steiner (1983), photoionization by a central AGN has surfaced 
as the leading candidate for the excitation mechanism of LINERs.  Given the 
success with which more luminous sources have been explained within this 
framework, and the growing realization that BHs are commonplace, extending 
photoionization models to LINERs is both natural and appealing.  The requisite 
relative strengths of the low-ionization lines in LINERs can be achieved by 
lowering the ionization parameter, commonly defined as $U = Q_{\rm ion}/4\pi 
r^2 c n$, where $Q_{\rm ion}$ is the number of photons s$^{-1}$ capable of 
ionizing hydrogen, $r$ is the distance of the inner face of the cloud from the 
central continuum, $n$ the hydrogen density, and $c$ the speed of light.  While 
Seyfert line ratios can be matched with $\log U \approx -2\pm0.5$ (e.g., 
Ferland \& Netzer 1983; Stasi\'nska 1984; Ho, Shields \& Filippenko 1993), 
LINERs require $\log U \approx -3.5\pm0.5$ (Ferland \& Netzer 1983; Halpern \& 
Steiner 1983; P\'equignot 1984; Binette 1985; Ho, Filippenko \& Sargent 1993; 
Groves, Dopita \& Sutherland 2004).  

An issue that has not been properly addressed is which of the primary variables 
--- $Q_{\rm ion}$, $r$, or $n$ --- conspire to reduce $U$ in LINERs to the 
degree required by the models.  The answer seems to be all three. The dominant 
factor comes from the luminosity, as LINERs emit an order of magnitude less 
ionizing luminosity than do Seyferts: $\langle L_{{\rm H}\alpha} \rangle = 3 
\times 10^{39}$ \lum\ versus $29 \times 10^{39}$ \lum\ (Ho, Filippenko \& 
Sargent 2003).  Given the gas-poor environments of LINERs (see \S~5.6), we can 
now say with some certainty that the reduction in ionizing luminosity is {\it 
intrinsic}\ and not due to obscuration (as proposed by Halpern \& Steiner 
1983).  But this is unlikely to be the end of the story.  The electron density 
of LINERs ($\langle n_e \rangle \approx 280$ \cc), at least as probed by the 
relatively low-density tracer \sii\ \lamb\lamb6716, 6731, is $\sim 50$\% lower 
than in Seyferts ($\langle n_e \rangle \approx 470$ \cc).  Very little is 
known about the detailed morphology and spatial distribution of the NLR in 
LINERs, or for that matter in low-luminosity Seyferts.  Whereas the NLR in 
luminous Seyferts span $\sim 50-1000$ pc in radius, scaling roughly as 
$L^{0.5}$ (Bennert et al. 2006), LINERs seem to be significantly more compact.  
This is perhaps not surprising, if the NLR size-luminosity relation extends to 
LLAGNs.  At typical ground-based resolution, narrow-band imaging studies find 
that the ionized gas in LINERs tends to be quite centrally peaked, with typical 
dimensions of $r$ \lax\ $50-100$ pc (Keel 1983a; Pogge 1989).  In the 
instances where narrow-band images or slit spatial profiles are available from 
\hst\ (Bower et al. 2000; Pogge et al. 2000; Cappellari et al. 2001; 
Verdoes~Kleijn et al.  2002; Gonz\'alez Delgado et al. 2004; Walsh et al. 2008),
the line-emitting gas appears even more concentrated still, 
with scales \lax\ tens pc, although some of it clearly extends to scales of at 
least $\sim 200$ pc (Shields et al. 2007).  The covering factor is high, on 
average $\sim 0.3$ for the LINER nuclei of the radio galaxies studied by 
Capetti, Verdoes~Kleijn \& Chiaberge (2005).  With very few exceptions (e.g., 
NGC~1052; Pogge et al. 2000), extended, elongated structures analogous to 
classical ionization cones in Seyferts do not exist in LINERs.  Interestingly, 
both of these trends (smaller $n_e$ and $r$) would naively drive $U$ in the 
opposite direction needed to explain LINERs.  On the other hand, we know that 
the NLRs of AGNs in general, and perhaps of LINERs especially, contain a wide 
range of densities not probed by \sii\ and that this material is highly 
stratified radially (Wilson 1979; Pelat, Alloin \& Fosbury 1981; Carswell et 
al. 1984; Filippenko \& Halpern 1984; P\'equignot 1984; Binette 1985; 
Filippenko 1985; Ho, Filippenko \& Sargent 1993, 1996; Barth et al.  2001b; 
Laor 2003; Shields et al.  2007; Walsh et al. 2008).  The 
effective ionization parameter, therefore, depends on the detailed spatial 
distribution of the gas.  

Whereas basic single-zone photoionization models can match many of the strong 
lines, it is well-known that more complex, multi-component models, especially
ones that incorporate a range of densities, are required to achieve a 
satisfactory fit (Binette 1985; Gabel et al. 2000; Sabra et al. 2003).  A 
notorious deficiency of single-zone models has been their inability to 
reproduce the high values of the temperature-sensitive ratio 
\oiii\ \lamb4363/\oiii\ \lamb5007.  However, because of the different critical 
densities of these two transitions, they need not originate cospatially, making
\oiii\ \lamb4363/\oiii\ \lamb5007 no longer a valid thermometer.  The 
discovery that these two lines indeed have different line widths removed one 
of the principal objections to photoionization models (Filippenko \& Halpern 
1984; Filippenko 1985).  

The observed weakness of \heii\ \lamb4686/H\bet\ (see also \S~5.8) has also 
been a thorny problem.  P\'equignot (1984) achieved a consistent fit to the 
spectrum of NGC~1052 by invoking a modified ionizing spectrum consisting of an 
80,000 K blackbody coupled with an X-ray tail extending to higher energies.  
The main difficulty with this proposal is that the observed SEDs of LINERs do 
not have a blackbody component peaked in the UV, nor is P\'equignot's 
specific model unique because Gabel et al. (2000) achieved an equally good---if 
not better---fit using a simple power-law continuum with $\alpha = -1.2$.  
Since we now have ample evidence that the SEDs of LINERs are {\it not}\ the 
same as those of more luminous AGNs, future photoionization calculations 
should adopt empirically motivated input spectra.  Important steps in these 
directions have been taken (e.g., Nicholson et al. 1998; Gabel et al. 2000; 
Nagao et al. 2002; Lewis, Eracleous \& Sambruna 2003), but much more can be 
done.  One fruitful avenue to pursue is to incorporate the {\it full}\ observed
SED, which is noteworthy not only because of its hard ionizing spectrum but, 
due to its radio-loudness, also because it presents a copious supply of 
relativistic, synchrotron-emitting particles, which can dramatically alter the 
excitation of the NLR (Aldrovandi \& P\'equignot 1973; Ferland \& Mushotzky 
1984; Gruenwald \& Viegas-Aldrovandi 1987). ``Cosmic ray heating'' boosts the 
strengths of low-ionization lines such as \nii\ \lamb\lamb6548, 6583 and 
\sii\ \lamb\lamb6716, 6731 (Viegas-Aldrovandi \& Gruenwald 1990), which are 
normally underpredicted (e.g., Ho, Filippenko \& Sargent 1993; Lewis, 
Eracleous \& Sambruna 2003), and thereby help to constrain alternative 
solutions that invoke selective abundance enhancement of N and S 
(Storchi-Bergmann \& Pastoriza 1990) or dust grain depletion (Gabel et al. 
2000).  

\subsection{Contribution from Fast Shocks}

Despite the natural appeal of AGN photoionization, alternative excitation
mechanisms for LINERs have been advanced.  Collisional ionization by shocks
has been a popular contender from the outset (Burbidge, Gould \& Pottasch 1963;
Osterbrock 1971; Osterbrock \& Dufour 1973; Koski \& Osterbrock 1976; 
Danziger, Fosbury \& Penston 1977; Fosbury et al. 1978; Ford \& Butcher 1979; 
Heckman 1980b; Baldwin, Phillips \& Terlevich 1981).  Shocks continue to be 
invoked (Bonatto, Bica \& Alloin 1989; Dopita \& Sutherland 1995; 
Alonso-Herrero et al. 2000; Sugai \& Malkan 2000) even after concerns over the 
\oiii\ \lamb 4363 temperature problem had been dispelled as a result of either
revised measurements (Keel \& Miller 1983; Rose \& Tripicco 1984) or 
complications arising from density stratification (Filippenko \& Halpern 1984).
Dopita \& Sutherland (1995) showed that the diffuse radiation field 
generated by fast ($\upsilon\, \approx 150-500$ \kms) shocks can reproduce the 
optical narrow emission lines seen in both LINERs and Seyferts.  In their 
models, LINER-like spectra are realized under conditions in which the 
precursor \hii\ region of the shock is absent, as might be the case in 
gas-poor environments.  The postshock cooling zone attains a much higher 
equilibrium electron temperature than a photoionized plasma; consequently, a 
robust prediction of shock models is that shocked gas should produce a 
higher excitation spectrum, most readily discernible in the UV, than 
photoionized gas.  In all the cases studied so far, however, the UV spectra 
are inconsistent with the fast-shock scenario because the observed intensities 
of high-excitation lines such as \civ\ \lamb1549 and \heii\ \lamb1640 are much 
weaker than predicted (Barth \etal 1996, 1997; Maoz \etal 1998; Nicholson 
\etal 1998; Gabel et al. 2000).  Dopita \etal (1997) used the spectrum of the 
circumnuclear {\it disk} of M87 to advance the view that LINERs are 
shock-excited.  This argument is misleading because their analysis deliberately 
avoids the nucleus.  Sabra et al. (2003) demonstrate that the UV--optical 
spectrum of the {\it nucleus}\ of M87 is best explained by a multi-component 
photoionization model.

Analysis of the emission-line profiles of the Palomar nuclei further casts 
doubt on the viability of the fast-shock scenario (Ho, Filippenko \& Sargent 
2003).  The velocity dispersions of the nuclear gas generally fall short of 
the values required for fast-shock excitation to be important.  Furthermore, 
the close similarity between the velocity field of LINERs and Seyferts as 
deduced from their line profiles contradicts the basic premise that shocks 
are primarily responsible for the spectral differences between the two classes 
of objects.  For a given bulge potential, LINERs, if anything, have {\it 
smaller}, not larger, gas velocity dispersions than Seyferts (L.C. Ho, in 
preparation).  And as discussed in \S~5.2, the incidence of extended radio 
jets, the most likely source of kinetic energy injection into the NLR, is 
actually lower in LINERs than in Seyferts, again contrary to naive expectations.

Notwithstanding these complications, it is inconceivable that mechanical 
heating, especially by lower velocity ($\sim 50-100$ \kms) shocks, does not 
play some role in the overall excitation budget of LINERs.  The velocities 
of the line-emitting gas are, after all, highly supersonic, turbulent, and 
most likely pressure-dominated (L.C. Ho, in preparation).  The trick is to 
figure out what is the balance between shocks and photoionization, and what 
physical insights can be gained from knowing the answer. It would be 
worthwhile to revisit composite shock plus photoionization models, such as 
those developed by Viegas-Aldrovandi \& Gruenwald (1990) and Contini (1997) 
with the latter component maximally constrained by observations so that 
robust, quantitative estimates can be placed on the former.  Such an approach 
might yield meaningful measurements of the amount of mechanical energy 
deposited into the host galaxy by AGN feedback.

\subsection{Contribution from Stellar Photoionization}

Another widely discussed class of models invokes hot stars formed in a
short-duration burst of star formation to supply the primary ionizing
photons.  Ordinary O-type stars with effective temperatures typical of those
found in giant \hii\ regions in galactic disks do not produce sufficiently
strong low-ionization lines to account for the spectra of LINERs.  The
physical conditions in the centers of galaxies, on the other hand, may be more
favorable for generating LINER-like spectra.  For example, Terlevich \&
Melnick (1985) postulate that the high-metallicity environment of galactic
nuclei may be particularly conducive to forming very hot, $T \approx (1-2)
\times 10^5$ K, luminous Wolf-Rayet stars whose ionizing spectrum would
effectively mimic the power-law continuum of an AGN. The models of Filippenko
\& Terlevich (1992) and Shields (1992) appeal to less extreme conditions.
These authors show that photoionization by ordinary O stars, albeit of 
somewhat higher effective temperature than normal (but see Schulz \& Fritsch 
1994), embedded in an environment with high density and low ionization 
parameter can explain the spectral properties of transition objects.  Barth \& 
Shields (2000) extended this work by modeling the ionizing source not as single
O-type stars but as a more realistic evolving young star cluster.  They
confirm that young, massive stars can indeed generate optical emission-line
spectra that match those of transition objects, and, under some plausible
conditions, even those of {\it bona fide} LINERs.  But there is an important 
caveat: the star cluster must be formed in an instantaneous burst, and its age 
must coincide with the brief phase ($\sim$3$-$5 Myr after the burst) during 
which sufficient Wolf-Rayet stars are present to supply the extreme-UV photons 
necessary to boost the low-ionization lines.  The necessity of a sizable 
population of Wolf-Rayet stars is also emphasized in the study by Gabel 
\& Bruhweiler (2002).  As discussed in Ho, Filippenko \& Sargent (2003), the 
main difficulty with this scenario, and indeed with all models that appeal to 
young or intermediate-age stars (e.g., Engelbracht et al.  1998; 
Alonso-Herrero et al. 2000; Taniguchi, Shioya \& Murayama 2000), is that the 
nuclear stellar population of the host galaxies of the majority of nearby 
AGNs, irrespective of spectral class, is demonstrably {\it old}\ (\S~4.2).  
Stellar absorption indices indicative of young or intermediate-age stars are 
seldom seen, and the telltale emission features of Wolf-Rayet stars are 
notably absent, both in ground-based and {\it HST}\ spectra.  Sarzi et al. 
(2005) find that young stars can account for at most a few percent of the blue 
light within the central few parsecs of nearby LLAGNs, in most cases incapable 
of providing enough ionizing photons to account for the observed H\al\ 
emission.  Post-starburst scenarios face another serious dilemma: if most 
LLAGNs, which constitute the majority of nearby AGNs and a large percentage of 
all galaxies, are described by this scenario, then where are their precursors?
They do not exist.  These empirical facts seriously undermine the viability of 
starburst or post-starburst models for LLAGNs.

Evolved, low-mass stars, on the other hand, probably contribute at some level 
to the ionization.  This idea was advocated by Binette et al. (1994), who 
proposed that post-asymptotic giant branch (post-AGB) stars, which can attain 
effective temperatures as high as $\sim 10^5$ K, might be responsible for 
photoionizing the extended ionized gas often observed in elliptical galaxies.  
The emission-line spectrum of these nebulae, in fact, tend to be of relatively 
low ionization (Demoulin-Ulrich, Butcher \& Boksenberg 1984; Phillips et al. 
1986).   Invoking evolved stars has the obvious appeal of not violating the 
stellar population constraints discussed above.  This mechanism, however, 
cannot be the dominant contributor to compact LINERs.  The line emission tends 
to be very centrally concentrated (\S~6.1), much more so than the underlying 
stellar density profile.  Moreover, the line strengths in most LLAGNs are 
simply too high.  The calculations of Binette et al. predict H\al\ equivalent 
widths of EW $\approx$ 1 \AA, whereas the LINERs and transition objects in the 
Palomar survey have an average EW 3$-$4 times higher (Ho, Filippenko \& 
Sargent 2003), with over 70\% of the sample having EW $>$ 1 \AA.  

To obtain a quantitative estimate of the contribution of post-AGB stars to the 
ionization budget of the weaker emission-line nuclei, I convert the nuclear 
stellar magnitudes ($m_{44}$) given in Ho, Filippenko \& Sargent (1997a) to 
stellar masses assuming a mass-to-light ratio of $M/L_B = 8(M/L_B)_\odot$ and 
that post-AGB stars have a specific ionization rate of $Q_{\rm ion} = 7.3 
\times 10^{40} (M/M_\odot)\, {\rm s}^{-1}$ (Binette et al.  1994).  Within the 
$100 \times 200$ pc aperture of the Palomar spectra, the integrated stellar 
mass is $\sim 10^7$ to $10^{10}$ \solmass, with a median 
value of $2\times10^9$ \solmass, which corresponds to an ionizing photon rate 
of $Q_{\rm ion} = 1.5 \times 10^{50}\,{\rm s}^{-1}$.  These estimates depend 
on the choice of the stellar initial mass function, and, most importantly, on 
detailed processes during the late stages of stellar evolution that are still 
not fully understood (see, e.g., O'Connell 1999).   Nevertheless, taking the
fiducial estimates as a rough guide, the predicted values of $Q_{\rm ion}$ can 
be compared to the actually observed, extinction-corrected H\al\ luminosity.  
Assuming complete reprocessing of the ionizing continuum and that on average it 
takes 2.2 Lyman continuum photons to produce one H\al\ photon, I estimate that 
the nebular line emission in roughly one-third of the Palomar sources can be 
powered by photoionization from post-AGB stars.  The fraction is higher for 
LINERs (39\%) than Seyferts (16\%), being most prevalent for LINER~2s (44\%) 
and transition objects (33\%).  Eracleous, Hwang \& Flohic (2008b) performed a 
similar analysis for a handful of LINERs with central stellar luminosity 
profiles available from \hst, concluding also that post-AGB stars can 
alleviate the ionizing photon deficit in some objects. 

\begin{figure}
\vbox{\hbox{
\psfig{figure=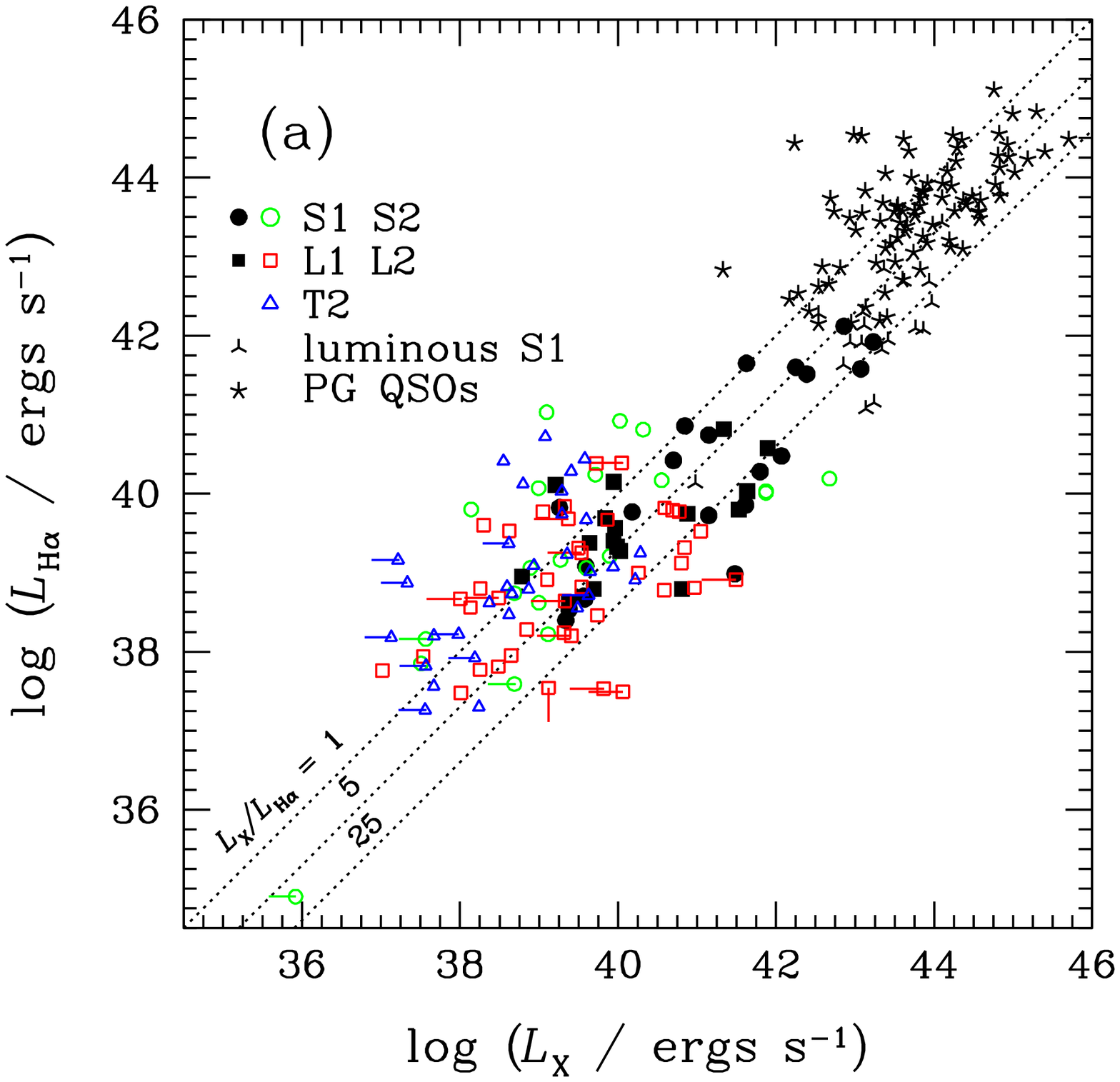,width=6.7cm,angle=0}
\hskip -0.2 cm
\psfig{figure=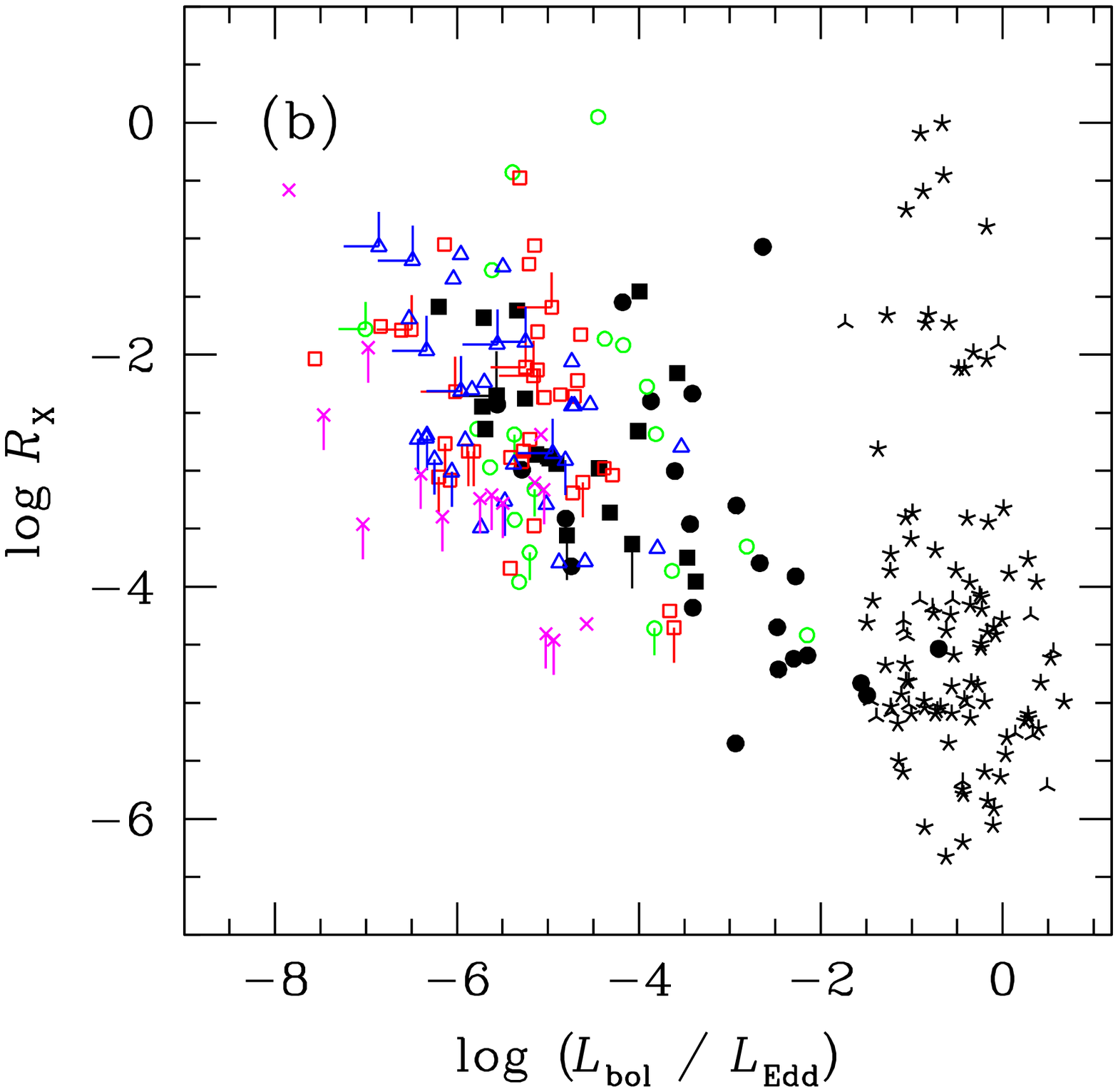,width=6.7cm,angle=0}
}}
\caption{({\it a}) Correlation between (total) H\al\ luminosity and X-ray
($2-10$ keV) luminosity.  The dotted lines mark $L_{\rm X}/L_{{\rm H}\alpha}$
= 1, 5, and 25.  ({\it b}) Distribution of radio-loudless parameter versus
Eddington ratio, with the bolometric luminosity estimated assuming
$L_{\rm bol} = 16  L_{\rm X}$.  
}
\end{figure}

\subsection{Energy Budget}

In luminous AGNs, important confirmation of the basic photoionization
paradigm comes from the strong empirical scaling and correlated variability 
between line flux and the strength of the ionizing continuum.  Although very 
little information exists in terms of line-continuum variability for LLAGNs,
enough X-ray observations have now been amassed to examine the correlation 
between optical line luminosity and X-ray luminosity.  The X-ray band
is only indirectly coupled to the bulk of the Lyman continuum, but 
in LLAGNs, it offers the most reliable probe of the high-energy spectrum.  
Studies in the soft X-ray band suggest that LLAGNs roughly follow the 
extrapolation of the $L_{{\rm H}\alpha}-L_{\rm X}$ correlation established for 
higher luminosity sources (Koratkar et al. 1995; Roberts \& Warwick 2000; 
Halderson et al. 2001).  Intriguingly, no clear differences could be discerned 
between LINERs and Seyferts, confirming preliminary evidence presented in 
Halpern \& Steiner (1983).  A more complex picture, however, emerges at higher 
energies ($2-10$ keV).  Terashima, Ho \& Ptak (2000) and Terashima et al. 
(2000) note that LINER~2s, unlike LINER~1s, suffer from a deficit of ionizing 
photons: the X-ray emission of the nucleus, when extrapolated to the UV, 
underpredicts the observed H\al\ luminosity by a factor of $\sim 10-100$.  This 
trend persists in more recent {\it Chandra}\ observations (Terashima \& Wilson 
2003b; Flohic et al. 2006; Eracleous, Hwang \& Flohic 2008b), one that seems 
to be especially endemic to transition objects (Ho et al. 2001; Filho et al. 
2004).  

Figure~10{\it a}\ shows an update of the $L_{{\rm H}\alpha}-L_{\rm X}$ relation 
for all Palomar sources with high-resolution (\lax\ 5\asec) X-ray measurements.
For comparison, I also include $z < 0.5$ Palomar-Green (PG; Schmidt \& Green 
1983) quasars and luminous Seyfert~1s with 
well-determined SEDs (Figure~7; L.C. Ho, in preparation).  All broad-line 
sources follow an approximately linear relation over nearly 7 orders of 
magnitude in luminosity. In detail, $L_{{\rm H}\alpha}\propto L_{\rm X}^{1.1}$,
such that quasars and luminous Seyferts have a median ratio 
$L_{\rm X}/L_{{\rm H}\alpha} \approx 2$, to be compared with 
$L_{\rm X}/L_{{\rm H}\alpha} \approx 7$ for Palomar Seyfert~1s.  Ignoring the 
possible effect of a luminosity-dependent covering factor, this can be 
interpreted as the consequence of a decrease in the ratio of UV radiation to 
X-rays with decreasing luminosity, reflecting a pattern familiar from samples 
of bright AGNs (e.g., Strateva et al. 2005), now extended down to lower 
luminosities by $\sim 3$ orders of magnitude.  Since LINER~1s are weaker than 
Seyfert~1s and their SEDs lack a UV bump, it is surprising that they actually 
have a somewhat {\it lower}\ $L_{\rm X}/L_{{\rm H}\alpha}$ ratio ($\sim 5$) 
than Seyfert~1s.  This suggests that at least some of the H\al\ emission in 
LINER~1s, presumably in the narrow component, is not powered by AGN 
photoionization.  In fact, this turns out to be a property symptomatic of 
{\it all}\ type~2 LLAGNs (Table~1).  The most extreme manifestation can be 
seen among transition objects, with $L_{\rm X}/L_{{\rm H}\alpha} \approx 0.4$, 
but both LINER~2s and Seyfert~2s also exhibit an ionizing photon deficit.  
For conditions typical of LLAGNs, Eracleous, Hwang \& Flohic (2008b) use 
photoionization calculations to infer $L_{\rm ion} = 18 L_{{\rm H}\alpha}\,
f_c^{-1}$, where $L_{\rm ion}$ is the ionizing luminosity between 1 Ryd and 10 
keV and $f_c$ is the covering factor of the line-emitting gas.  Since 
$L_{\rm ion} \approx 3 L_{\rm X}$ for a power-law spectrum with $\alpha = 
-0.1$ to $-0.9$, $L_{\rm X}/L_{{\rm H}\alpha} \approx 6 f_c^{-1}$.  It is 
clear that most narrow-line LLAGNs violate this energy balance condition, even 
for the optimistic assumption of $f_c = 1$.

There are several possible solutions to the ionization deficit problem. (1) The 
X-rays could be highly obscured, perhaps even Compton-thick.  In light of the 
evidence given in \S\S~5.3, 5.6, I consider this solution to be untenable for 
LINER~2s; highly absorbed sources do exist (e.g., NGC~4261; Zezas et al. 
2005), but they are in the minority.  Moreover, many of the X-ray measurements 
have already been corrected for absorption.  The situation is more complex for 
Seyfert~2s.  Some of the sources with low $L_{\rm X}/L_{{\rm H}\alpha}$ indeed 
show direct evidence for Compton thickness from their X-ray spectra (Cappi et 
al. 2006).  Others, however, are too faint for spectral analysis, and for 
these, their status as Compton-thick sources was based on the observed ratio of 
2$-$10 keV flux to \oiii\ \lamb5007 flux (Bassani et al. 1999; Panessa \& 
Bassani 2002).  Applying an average correction factor of 60 to the X-ray 
luminosity would bring the Seyfert~2s into agreement with the Seyfert~1s on the 
$L_{{\rm H}\alpha}-L_{\rm X}$ relation (Panessa et al. 2006). But this 
procedure {\it assumes}\ that the low values of $L_{\rm X}/L_{{\rm H}\alpha}$ 
are due to a reduction of the X-rays by absorption rather than an enhancement 
of H\al\ (see below).  (2) The SED could be drastically different, 
specifically in having a much more prominent UV component.  This proposition 
can be promptly dismissed because the SEDs of LLAGNs generically lack a UV 
bump.  There is certainly no indication that type~2 sources are preferentially 
brighter in the UV; in the case of LINERs, type~2 sources, if anything, tend 
to be redder than type~1 sources (Maoz et al. 2005).  (3) Lastly, and most 
plausibly, a significant fraction of the ionization for the narrow-line gas 
comes from nonnuclear sources.  As discussed in connection with the preceding 
two subsections, young, massive stars and fast shocks are generally not viable 
options.  There are a number of candidate sources of ``extra'' ionization, 
including hot, evolved stars, turbulent mixing layers, diffuse X-ray emitting 
plasma, low-mass XRBs, cosmic ray heating, and mechanical heating 
from radio jets.  As all of these sources probably contribute at some level, 
efforts to single out any dominant mechanism may be hopelessly challenging.
Nevertheless, as discussed in \S~6.3, post-AGB stars appear especially 
promising.  Taking the calculations of Binette et al. (1994) as a guide, the 
stellar mass within the central $100-200$ pc region generates sufficient Lyman 
continuum photons to account for the H\al\ emission in $\sim 30\%-40\%$ of the 
LINER~2s and transition objects.  This estimate is crude and admittedly 
optimistic, as it assumes a covering factor of unity for the NLR, but it 
serves as a useful illustration of the types of effects that should be 
included in any complete treatment of the energy budget problem in LLAGNs.

\subsection{The Nature of Transition Objects and a Unified View of LLAGNs}

The physical origin of transition nuclei continues to be a thorny, unresolved 
problem.  In standard line-ratio diagrams (Figure~3), transition nuclei are 
empirically defined to be those sources that lie sandwiched between the loci 
of ``normal'' \hii\ regions and LINERs.  This motivated Ho, Filippenko \& 
Sargent (1993) to propose that transition objects may be composite systems 
consisting of a LINER nucleus plus an \hii\ region component.  The latter 
could arise from neighboring circumnuclear \hii\ regions or from \hii\ regions 
randomly projected along the line of sight.  A similar argument, based on 
decomposition of line profiles, was made by V\'eron, Gon\c{c}alves \& 
V\'eron-Cetty (1997).  If transition objects truly are LINERs sprinkled with a 
frosting of star formation, one would expect that their host galaxies should 
be similar to those of LINERs, modulo minor differences due to extra 
contaminating star formation.  The study of Ho, Filippenko \& Sargent (2003) 
tentatively supports this picture.  The host galaxies of transition nuclei 
exhibit systematically higher levels of recent star formation, as indicated by 
their far-IR emission and broad-band optical colors, compared to LINERs of 
matched morphological types.  Moreover, the hosts of transition nuclei tend to 
be slightly more inclined than LINERs.  Thus, all else being equal, 
transition-type spectra seem to be found precisely in those galaxies whose 
nuclei have a higher probability of being contaminated by extra-nuclear 
emission from star-forming regions.

This story, however, has some holes.  If spatial blending of circumnuclear 
\hii\ regions is sufficient to transform a regular LINER into a transition 
object, the LINER nucleus should reveal itself unambiguously in spectra taken 
with angular resolution sufficiently high to isolate it.  This test was 
performed by Barth, Ho \& Filippenko (2003), who obtained {\it HST}/STIS 
spectra, taken with a 0\farcs2-wide slit, of a well-defined subsample of 15 
transition objects selected from the Palomar catalog.  To their surprise, the 
small-aperture spectra of the nuclei, for the most part, look very similar to 
the ground-based spectra; they are {\it not} more LINER-like.  Shields et al. 
(2007) reached the same conclusion from their STIS study of Palomar nuclei, 
which included six transition objects, showing that even at \hst\ 
resolution these objects do not reveal the expected excitation gradients. 

The ``masqueraded-LINER'' hypothesis can be further tested by searching for
compact radio and X-ray cores using high-resolution images.  Recall that this
is a highly effective method to filter out weak AGNs (\S\S~5.2, 5.3).  Filho,
Barthel \& Ho (2000, 2002a; Filho et al. 2004) have systematically surveyed 
the full sample of Palomar transition objects using the VLA at 8.4~GHz.  They 
find that $\sim$25\% of the population contains arcsecond-scale radio cores.  
These cores appear to be largely nonstellar in nature.  The brighter subset of
these sources that are amenable to follow-up Very Long Baseline Interferometry 
(VLBI) observations (Filho et al.  2004) all reveal more compact 
(milliarcsecond-scale) cores with flat radio
spectra and high brightness temperatures (\gax\ $10^7$ K).  These radio 
statistics are hard to interpret, however, in the absence of a control sample 
of other LLAGNs surveyed to the same depth, resolution, and wavelength.  The 
Nagar, Falcke \& Wilson (2005) 15~GHz survey satisfies these criteria.  As 
Table~1 shows, the frequency of radio cores in transition objects is roughly 
half of that in Seyfert~2s and LINER~2s.  On the other hand, the detection 
rate of X-ray cores is actually remarkably high---74\%---identical to that of 
LINER~2s and similar to that of Seyfert~2s.  This observation strongly 
suggests that the majority of transition objects indeed {\it do}\ harbor AGNs.

In light of these recent developments, the basic picture for the physical 
nature of transition objects needs to be revised.  Inspection of the 
statistical properties in Table~1 offers the following clues, which help not
only to explain transition objects but provide a unified view to relate the 
different classes of LLAGNs.  

\begin{enumerate}

{\item 
Seyferts, LINERs, and transition objects define a sequence of decreasing 
accretion rate.  This is most evident from $L_{\rm X}$ and 
$L_{\rm bol}/L_{\rm Edd}$, but it is also seen in $L_{{\rm H}\alpha}$ 
and $P_{\rm rad}$.
}

{\item 
As noted in \S~5.10, type~1 sources have significantly higher luminosities and 
Eddington ratios than type~2 systems.  The basic premise of the conventional 
orientation-based unification scenario does not hold for LLAGNs.  The 
systematic reduction in accretion rate along the sequence 
S$\rightarrow$L$\rightarrow$T also provides a viable explanation for the 
systematic decrease in the detection rate of broad H\al\ emission, especially 
the precipitous drop among transition sources ($f_b$ in Table~1).
}

{\item 
Transition objects appear to be anomalously strong in their H\al\ emission.
In light of the \hst\ evidence for a distributed source of ionization, I 
suspect that a significant fraction of the H\al\ emission in these objects in 
fact is {\it not}\ photoionized by the central AGN.  This leads to misleading 
values of $L_{\rm X}/L_{{\rm H}\alpha}$ and $R_o$ (which is based on 
$L_{{\rm H}\alpha}$).  For this class either the X-ray or radio core provides 
a cleaner measure of the AGN power.  
}

{\item
The loose inverse correlation between radio-loudness and accretion rate, best 
seen by comparing $R_{\rm X}$ versus either $L_{\rm X}$ or 
$L_{\rm X}/L_{\rm Edd}$, mirrors the trends found by Ho (2002a) and 
Terashima \& Wilson (2003b).
}

{\item
Focusing on the type~2 sources, note that LINER~2s are very similar to 
Seyfert~2s, the former being $\sim 1/3-1/2$ as strong as the latter in terms 
of H\al\ luminosity and radio power.  The two groups have almost identical 
$L_{\rm X}$ and $L_{\rm X}/L_{\rm Edd}$, although this may be an artifact of 
incomplete absorption correction for Seyfert~2s, some of which are highly
absorbed (Panessa et al. 2006).  In the same vein, I propose that transition 
objects represent the next step in the luminosity sequence. Judging by their 
X-ray luminosity, radio power, $L_{\rm X}/L_{\rm Edd}$, and radio detection 
fraction, the AGN component in transition objects is $\sim$1/4 to 1/2 as 
strong as that in LINER~2s. 
}
\end{enumerate}

According to the picture just outlined, most, if not all, type~2 sources are 
genuinely accretion powered.  Using the accretion rate as the metric for 
the level of AGN activity, Seyfert~1s rank at the top of the scale, followed 
by Seyfert~2s, LINER~1s, LINER~2s, and finally ending with transition objects. 
This scenario, which in broad-brush terms explains a wide range of data 
summarized in Table~1, has the virtue of simplicity.  It is also physically 
appealing, given the broad spectrum of accretion rates anticipated in nearby 
galaxies.

There is, however, one loose end that needs to be tied.  What powers the 
spatially extended, ``excess'' optical line emission in transition objects?  
For the reasons explained before, the source of the ionization is unlikely to 
be shock heating or photoionization by hot, massive stars, notwithstanding the 
success with which such models have been applied to some individual objects
(Engelbracht et al. 1998; Barth \& Shields 2000; Gabel \& Bruhweiler 2002).  
Shields et al. (2007) suggest two candidates for a spatially extended source 
of ionization: hot, evolved stars and turbulent mixing layers in the 
interstellar medium (Begelman \& Fabian 1990).  In \S~6.3, I showed that
the stellar mass in the central 100--200 pc indeed seems to provide enough 
post-AGB stars to account for the correct level of H\al\ emission in a 
significant fraction of the transition objects.  I would like to suggest two 
other sources, ones that have the advantage of being empirically well 
motivated by recent observations.  These processes probably operate in all 
nuclear environments all the time, maintaining a ``baseline'' level of weak 
optical line emission that is only noticeable after the AGN has subsided to a 
very low level.  

As discussed in \S\S~5.3, 5.4, the X-ray morphology of the central few hundred 
parsecs of galaxies can be quite complex.  The nucleus, if present, is often 
encircled by other point sources, mostly XRBs (Fabbiano 2006).
With X-ray luminosities ranging from $\sim 10^{37}$ to $10^{39}$ \lum\ (Flohic 
et al. 2006), XRBs individually or collectively can outshine the nucleus 
itself (Ho et al. 2001; Eracleous et al. 2002; Ho, Terashima \& Ulvestad 2003; 
see Figure~5).  The discrete sources themselves are embedded in extended 
emission, consisting of an optically thin thermal plasma with $kT \approx 0.5$ 
keV and a spectrally harder diffuse component, which contributes a luminosity 
of $\sim (5-9)\times 10^{38}$ \lum\ in the 0.5$-$10 keV band (Flohic et al. 
2006).  The hard diffuse component most likely represents the cumulative 
emission from faint, unresolved low-mass XRBs, although this interpretation 
seems somewhat at odds with the spectrum derived by Flohic et al. (power 
law with $\alpha = -0.3$ to $-0.5$).  Low-mass XRBs typically can be fit by a
thermal bremsstrahlung model with $kT = 5-10$ keV or a power law with $\alpha
= -0.6$ to $-0.9$ (e.g., Makishima et al. 1989).  High-mass XRBs would provide 
a better match to the observed spectrum, but in view of what we know about the 
stellar populations (\S~4.2), they are probably untenable.  A possible 
solution is to invoke a multi-temperature plasma (M. Eracleous, private 
communications); a hotter component ($kT \approx$ few keV), when added to the 
cooler $kT = 0.5$ keV component, would presumably permit a significant 
contribution from low-mass XRBs without violating the spectral constraints.
We can estimate the expected X-ray output from low-mass XRBs from the 
correlation between optical and X-ray luminosity established for normal 
galaxies (Fabbiano \& Trinchieri 1985).  Using again the nuclear stellar 
magnitudes from the Palomar survey, I obtain a median $L_{\rm X}$($2-10$ keV) 
= $(3\pm1)\times10^{38}$ \lum\ within the central 2\asec$\times$4\asec\ 
aperture, which falls within the ballpark of the value measured by Flohic et 
al. (2006).  The combination of hot gas and XRB emission, therefore, supplies 
$\sim10^{39}$ \lum\ in X-rays, comparable to the amount coming from the 
nucleus alone for Seyfert~2s and LINER~2s, and double the amount from 
transition nuclei (Table~1).  Voit \& Donahue (1997) suggest that hot plasma 
additionally may transfer heat conductively to the line-emitting gas in 
LINERs, in a manner analogous to the situation in cooling flow filaments in 
galaxy clusters.

Lastly, cosmic ray heating (\S~6.1) by the central radio core will further 
enhance the optical line luminosity (Ferland \& Mushotzky 1984).  The very 
source of the fast particles, namely compact radio jets, itself probably 
injects an additional source of mechanical heating, although this is more 
difficult to model concretely.  Both processes---photoionization by off-nuclear
X-rays and cosmic ray heating---have a convenient virtue: they will tend to 
produce low-ionization spectra and therefore provide a natural match to the 
spectral characteristics of nearby LLAGNs.  

\section{IMPLICATIONS FOR BLACK HOLE DEMOGRAPHICS}

As summarized by Kormendy (2004), spatially resolved kinematical observations
have convincingly measured BH masses in a sizable number of nearby galaxies, 
to the point that important inferences on their demographics can be drawn 
(Richstone 2004).  Following an argument first due to So\l tan (1982), 
comparison of the integrated radiation density from quasars to the integrated
mass density in local BHs shows that BHs have grown mostly via a radiatively 
efficient mode of accretion during their bright AGN phase (e.g., Yu \& 
Tremaine 2002).  Nearby galaxies, therefore, should be home to AGN relics.  

The LLAGNs summarized in this review provide an important confirmation of this
basic prediction.  Not only are weakly accreting BHs found in great abundance 
in the local Universe, but they are found where prevailing wisdom says that 
they should be found, namely in the centers of galaxies that contain bulges.
Among E, S0, and Sb galaxies, the AGN detection rate is $\sim 50$\%, 
increasing to over 70\% among Sa galaxies.  Since sensitivity and confusion 
impact the detection rates, these statistics are not inconsistent with the 
notion that BHs are ubiquitous in essentially {\it all}\ E$-$Sb galaxies.  

Notably, the incidence of AGNs plummets for galaxies with Hubble types 
Sc and later, precisely at the point where classical ($r^{1/4}$ profile) 
bulges effectively cease to exist and secular dissipation processes begin 
to kick in (Kormendy \& Kennicutt 2004).  Although the AGN fraction is low for 
late-type spirals, it is also not zero.  Careful scrutiny of this minority 
population addresses two important questions: (1) Are there
central (nonstellar) BHs with masses below $10^6$ \solmass?  (2) Must central 
BHs always be encased in a bulge?

\begin{figure}
\vbox{\hbox{
\psfig{figure=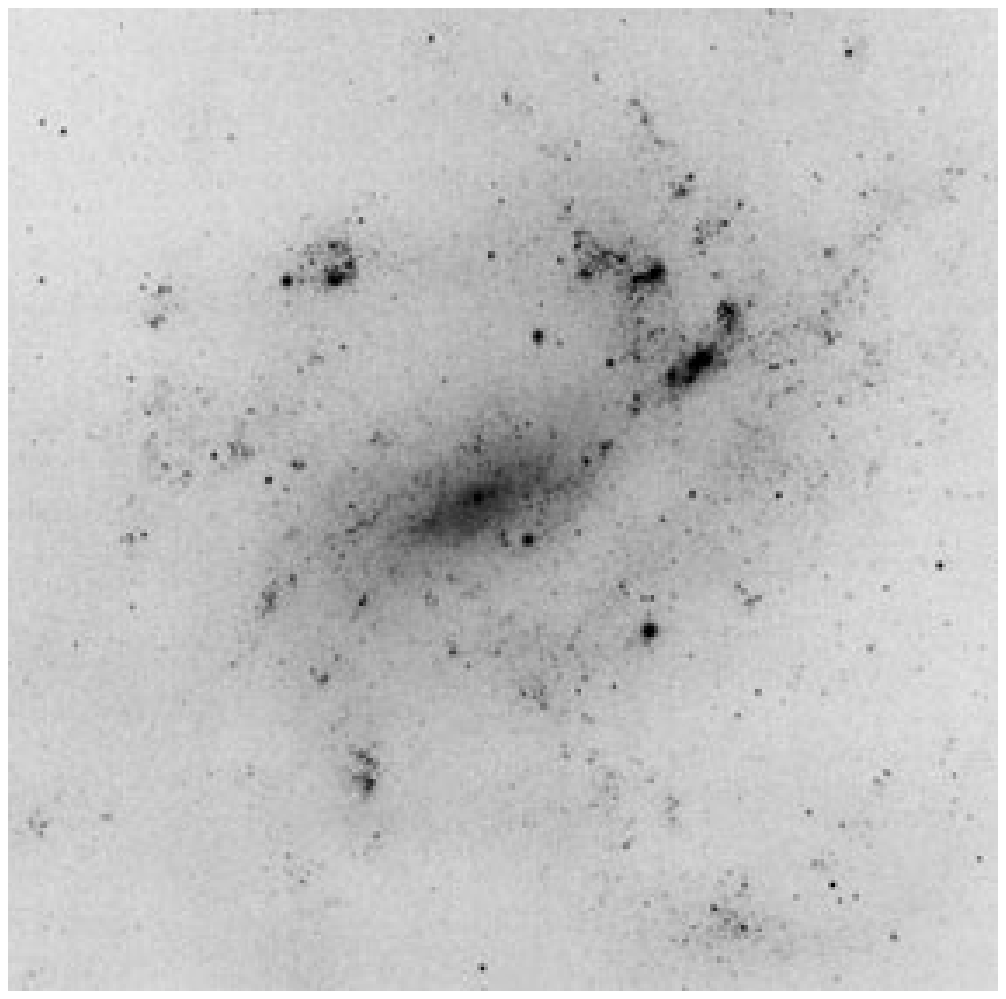,width=0.465\columnwidth,angle=0}
\hskip +0.2 cm
\psfig{figure=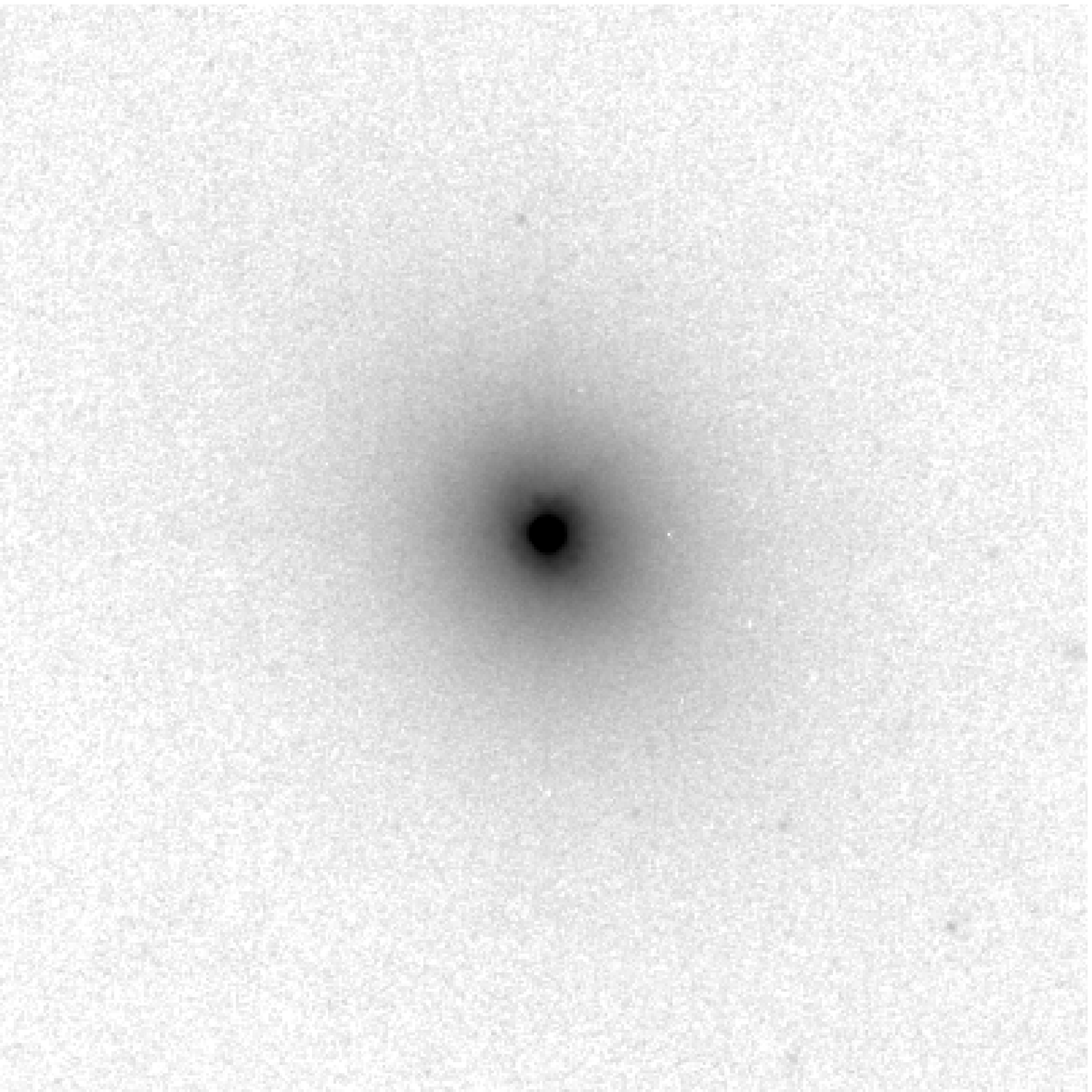,width=0.465\columnwidth,angle=0}
}}
\caption{
Two examples of AGNs in late-type galaxies.  The {\it left}\ panel shows an
optical image of NGC 4395, adapted from the {\it Carnegie Atlas of Galaxies}\
(Sandage \& Bedke 1994); the image is $\sim$15\amin\ (17 kpc) on a side.
The {\it right}\ panel shows an {\it HST}\ $I$-band image of POX 52, adapted 
from C.E. Thornton, A.J. Barth, L.C. Ho, R.E. Rutledge, J.E. Greene (in 
preparation); the image is $\sim$11\asec\ (5 kpc) on a side.
}
\end{figure}

Two remarkable galaxies give the clearest testimony that low-mass BHs do, in 
fact, exist.  Within the Palomar survey, the nearby ($\sim$4 Mpc) galaxy 
NGC~4395 contains all the usual attributes of a self-respecting AGN: broad 
optical and UV emission lines (Filippenko \& Sargent 1989; Filippenko, Ho \& 
Sargent 1993), a compact radio core (Ho \& Ulvestad 2001) of high brightness 
temperature (Wrobel, Fassnacht \& Ho 2001; Wrobel \& Ho 2006), and rapidly 
variable hard X-ray emission (Shih, Iwasawa \& Fabian 2003; Moran et al. 2005).
Contrary to expectations, however, NGC~4395 is an extremely late-type
(Sdm) spiral (Figure~11, {\it left}), whose central stellar velocity
dispersion does not exceed $\sim 30$ \kms\ (Filippenko \& Ho 2003).  If
NGC~4395 obeys the $M_{\rm BH}-\sigma$ relation, its central BH should have a
mass \lax\ $10^5$ \solmass.  This limit agrees surprisingly well with the value
of \mbh\ estimated from its broad H\bet\ line width or X-ray variability
properties ($\sim 10^4-10^5$ \solmass; Filippenko \& Ho 2003), or from 
reverberation mapping (3.6\e{5} \solmass; Peterson et al. 2005).  POX~52 
(Figure~11, {\it right}) presents another interesting case.  As first noted by 
Kunth, Sargent \& Bothun (1987), the presence of a Seyfert-like nucleus in 
POX~52 is unusual because of the low luminosity of the host galaxy.  Barth et 
al. (2004) show that POX~52 bears a close spectroscopic resemblance to 
NGC~4395.  Based on the broad profile of H\bet, these authors derive a virial 
BH mass of 1.6\e{5} \solmass\ for POX~52, again surprisingly close to the 
value of 1.3\e{5} \solmass\ predicted from the $M_{\rm BH} - \sigma$ relation 
given the measured central stellar velocity dispersion of 35 \kms.  

The two objects highlighted above demonstrate that the mass spectrum of nuclear
BHs indeed does indeed extend below $10^6$ \solmass, providing great leverage 
for anchoring the $M_{\rm BH} - \sigma$ at the low end.  Furthermore, they 
shed light on the variety of environments in which nuclear BHs can form, 
providing much-needed empirical clues to the conditions that fostered the 
formation of the seeds for supermassive BHs.  NGC~4395 has little or no bulge, 
but it does have a compact, central star cluster in which the BH is embedded 
(Filippenko \& Ho 2003).  Interestingly, G1, a massive star cluster in M31 
that to date contains the best direct detection of an intermediate-mass 
($\sim 20,000$ \solmass) BH (Gebhardt, Rich \& Ho 2002, 2005; Ulvestad, Greene
\& Ho 2007), is thought to be the stripped nucleus of a once small galaxy. The 
same holds for the Galactic cluster $\omega$ Cen, for which Noyola, Gebhardt 
\& Bergmann (2008) report a central dark mass of $5\times10^4$ \solmass.
POX~52 is equally striking.  Deep images reveal POX~52 to be most akin to a 
spheroidal galaxy (Barth et al. 2004; C.E. Thornton, A.J. Barth, L.C. 
Ho, R.E. Rutledge, J.E. Greene, in preparation), to date an unprecedented 
morphology for an AGN host galaxy.   This is quite unexpected because 
spheroidal galaxies, while technically hot stellar systems, bear little 
physical resemblance to classical bulges.  Spheroidals occupy a distinct 
locus on the fundamental plane (e.g., Geha, Guhathakurta \& van~der~Marel 
2002; Kormendy et al. 2008), and they may originate from harassment 
and tidal stripping of late-type disk galaxies (e.g., Moore et al. 1996).  
Thus, like NGC~4395, POX~52 stands as testimony that a classical bulge is not 
a prerequisite for the formation of central BH.

But how common are such objects?  AGNs hosted in high-surface brightness,
late-type spirals appear to be quite rare in the nearby Universe.  Within the
comprehensive Palomar survey, NGC~4395 emerges as a unique case of an 
unambiguous broad-line AGN hosted in a late-type system.  The majority of 
late-type spirals do possess compact, photometrically distinct nuclei (B\"oker 
et al. 2002), morphologically not dissimilar from that in NGC~4395, but, with 
few exceptions (Shields et al. 2008), these nuclei are compact star clusters 
with no compelling evidence for an accompanying accreting central BH (Walcher 
et al. 2006).  Nuclear star clusters do not appear to directly impact a 
galaxy's ability to host an AGN (Seth et al. 2008).  Several serendipitous 
cases of AGNs in late-type galaxies have recently been found from analysis of 
{\it Spitzer}\ mid-IR spectra (Satyapal et al. 2007, 2008), as well as a 
number of AGN candidates from inspection of {\it Chandra}\ images (Desroches 
\&  Ho 2008).  Among earlier Hubble types, Gallo et al. (2008) report the 
detection of X-ray nuclei in two low-luminosity early-type galaxies.

To assess the true incidence of AGNs like NGC~4395 and POX~52 requires a 
spectroscopic survey much larger than Palomar.  Greene \& Ho (2007b) 
performed a systematic spectral analysis of over 500,000 SDSS spectra with $z < 
0.35$ to search for broad-line AGNs, producing not only a detailed BH mass 
function for low-redshift AGNs (Greene \& Ho 2007a) but also a comprehensive 
catalog of $\sim 200$ low-mass ($M_{\rm BH} < 10^6$ \solmass) objects.  Not 
much is known yet about the host galaxies, except that on average they are 
about 1 mag fainter than $L^*$.   {\it HST}\ imaging of the initial sample of 
19 objects discovered by Greene \& Ho (2004) reveal that the host galaxies are 
either mid- to late-type spirals (although none seems as late-type as 
NGC~4395) or compact, spheroidal-looking systems not unlike POX~52 (Greene, Ho 
\& Barth 2008).  When projected onto the galaxy fundamental plane, the 
``bulge'' component in some systems resides within the locus of spheroidal
galaxies.   Follow-up high-dispersion spectroscopy shows that these 
objects approximately follow the $M_{\rm BH} - \sigma$ relation (Barth, Greene 
\& Ho 2005).

\section{IMPLICATIONS FOR ACCRETION PHYSICS}

\subsection{Why Are LLAGNs So Dim?}

This review highlights a class of galactic nuclei that are extraordinary 
for being so ordinary.  At their most extreme manifestation, LLAGNs emit a 
billion times less light than the most powerful known quasars.  When quasars 
were first discovered, the challenge then was to explain their tremendous 
luminosities.  Ironically, more than four decades later, the problem has been 
reversed: the challenge now is to explain how dead quasars can remain so 
dormant.  The luminosity deficit problem was noted by Fabian \& Canizares 
(1988), who drew attention to the fact that elliptical galaxies, despite being 
suffused with a ready supply of hot gas capable of undergoing spherical 
accretion, have very dim nuclei.  We can no longer speculate that these 
systems lack supermassive BHs, for we now know that they are there, at least 
in galaxies with bulges.  And as I have shown in this review, the problem is 
by no means confined to ellipticals either.

Explanations of the luminosity paradox fall in several categories.

\begin{itemize}

{\item
{\underbar{\it Obscuration}} \ \ \ This trivial possibility can be summarily 
dismissed as a general solution in light of the evidence presented in 
\S\S~5.3, 5.6.
}

{\item
{\underbar{\it Low accretion rate}} \ \ \ A more plausible strategy is to 
starve the BH.  Present-day massive galaxies, after all, should be relatively 
gas-poor, especially in their central regions, which are largely devoid of 
significant ongoing star formation.  This argument quickly falls apart when 
one realizes just how little material is needed to light up the nuclei.  The 
bolometric luminosities of nearby nuclei span $\sim 10^{38}-10^{44}$ \lum, 
with a median value of $L_{\rm bol} = 3\times 10^{40}$ \lum\ and half of the 
sample lying between $3\times 10^{39}$ and $3\times 10^{41}$ \lum.  For a 
canonical radiative efficiency of $\eta = 0.1$, the required accretion rate is 
merely $\dot M=L_{\rm bol}/\eta c^2=5\times10^{-6\pm1}$ \solmass\ \peryr.  
This is a pitifully miniscule amount, in comparison with the amount of fuel 
actually available to be accreted.  Galactic nuclei unavoidably receive fuel
from two sources: ordinary mass loss from evolved stars and gravitational 
capture of gas from the hot interstellar medium.

During the normal course of stellar evolution, evolved stars return a
significant fraction of their mass to the ISM through mass loss.  For a
Salpeter stellar initial mass function with a lower-mass cutoff of 0.1
\solmass, an upper-mass cutoff of 100 \solmass, solar metallicities, and an
age of 15 Gyr, Padovani \& Matteucci (1993) estimate that

\begin{displaymath}
\dot M_*\,\approx\,3\times10^{-11} \, \left(\frac{L}{L_{\odot, V}}\right)
\,\,\,\,\,\,\, M_{\odot}\,{\rm yr}^{-1}.  
\end{displaymath}

\noindent
This result is consistent, within a factor of $\sim$2, with the work of Ciotti 
et al. (1991) and Jungwiert, Combes \& Palous (2001). {\it HST}\ images reveal
that galaxies contain central density concentrations, either in the form of 
nuclear cusps or photometrically distinct, compact stellar nuclei. The
cusp profiles continue to rise to the resolution limit of \hst\ (0\farcs1),
which is $r \approx 10$ pc at a distance of 20 Mpc, where 
$\rho\, \approx\, 10-10^3$ $L_{\odot, V}$ pc$^{-3}$ for the ``core''
ellipticals and $\rho\, \approx\, 10^2-10^4$ $L_{\odot, V}$ pc$^{-3}$ for the
``power-law'' ellipticals and bulges of early-type spirals and S0s (e.g.,
Faber et al. 1997).   Within a spherical region of $r = 10$ pc, the diffuse
cores have $L \approx 4\times 10^4 - 4\times 10^6\, L_{\odot, V}$, which
yields $\dot M_*\,\approx\,1\times10^{-6} - 1\times10^{-4}$ \solmass\
\peryr; for the denser power-law cusps, $L \approx 4\times 10^5 -
4\times 10^7\, L_{\odot, V}$, or $\dot M_*\,\approx\,1\times10^{-5} -
1\times10^{-3}$ \solmass\ \peryr.  Centrally dominant nuclear star clusters,
present in a large fraction of disk galaxies, typically have luminosities
$L \approx 10^7$ \solum\ (Carollo et al. 1997; B\"oker et al. 2002), and hence 
$\dot M_*\,\approx\,10^{-3}$ \solmass\ \peryr.

Diffuse, hot gas in the central regions of galaxies holds another potential
fuel reservoir.  Low-angular momentum gas sufficiently close to a BH can
accrete spherically in the manner described by Bondi (1952). To estimate the
Bondi accretion rate, one needs to know the gas density and temperature at the 
accretion radius, $R_a \approx GM_{\rm BH}/c_s^2$, where $c_s \approx 0.1 
T^{1/2}$ \kms\ is the sound speed of the gas at temperature $T$.  The mass 
accretion rate follows from the continuity equation, $\dot M_{\rm B}\,=\,4\pi 
R_a^2 \rho_a c_s$, where $\rho_a$ is the gas density at $R_a$.   Expressed in 
terms of typical observed parameters (see below),

\begin{displaymath}
\dot M_{\rm B}\,\approx\,7.3\times 10^{-4}\,
\left(\frac{M_{\rm BH}}{10^8\, M_{\odot}}\right)^2\,
\left(\frac{n}{0.1\, {\rm cm}^{-3}}\right)\,
\left(\frac{200\, {\rm km\,s}^{-1}}{c_s}\right)^3\,
\,M_{\odot}\,{\rm yr}^{-1}.
\end{displaymath}

{\it Chandra}\ observations with sufficient resolution to resolve $R_a$ find 
that the diffuse gas in the central regions of elliptical galaxies typically 
has temperatures of $kT\,\approx\,0.3-1$ keV and densities of $n\,\approx\,
0.1-0.5$ \cc\ (e.g., Di~Matteo et al. 2001, 2003; Loewenstein et al. 2001;
Pellegrini 2005).  Our knowledge of the hot gas 
content in the central regions of the bulges of spiral and S0 galaxies is more 
fragmentary.  {\it Chandra}\ has so far resolved the hot gas in the centers of
a handful of bulges (Milky Way: Baganoff et al. 2003; M31: Garcia et al. 2005; 
M81: Swartz et al. 2002; NGC~1291: Irwin, Sarazin \& Bregman 2002; NGC~1553: 
Blanton, Sarazin \& Irwin 2001; Sombrero: Pellegrini et al. 2003a).  These 
studies suggest that bulges typically have gas temperatures of $kT\,\approx\,
0.3-0.6$ keV.  Information on gas densities is sketchier, but judging from the 
work on M81 and the Sombrero, a fiducial value might be $n\,\approx\,0.1$ \cc.

If, for simplicity, we assume that the hot gas in most bulges is characterized
by $n\,=\,0.1$ \cc\ and $kT\,=\,0.3$ keV, then $\dot M_{\rm B}\,\approx\,
10^{-5}-10^{-3}$ \solmass\ \peryr\ for \mbh\ = $10^7-10^8$ \solmass.  In
elliptical galaxies \mbh\ $\approx\,10^8-10^9$ \solmass, and for $n\,=\,0.2$
\cc\ and $kT\,=\,0.5$ keV, $\dot M_{\rm B}\,\approx\, 10^{-4}-10^{-2}$
\solmass\ \peryr.  We note that these estimates of the Bondi accretion rates, 
which fall within the range given in recent compilations (e.g., Donato, 
Sambruna \& Gliozzi 2004; Pellegrini 2005; Soria et al. 2006), are probably 
lower limits because the actual densities near $R_a$ are likely to be higher 
than we assumed.  For the above fiducial temperatures and BH masses, 
$R_a\,\approx\,1-10$ pc for bulges and $\sim 10-100$ pc for ellipticals, 
roughly an order of magnitude smaller than the typical linear resolution 
achieved by {\it Chandra}\ for nearby galaxies.  In well-resolved cases, the 
gas temperature profile generally remains constant to within $\sim 50$\% 
whereas the density typically increases by a factor of a few toward the center 
(e.g., Milky Way: Baganoff et al. 2003; M31: Garcia et al. 2005; M87: 
Di~Matteo et al. 2003; NGC~1316: Kim \& Fabbiano 2003).

Although the above estimates are very rough, and they are valid only in a 
statistical sense, one cannot escape the conclusion that in general in the 
central few parsecs of nearby galaxies $\dot M_{\rm B} + \dot M_* \gg \dot M$.
Although meager, the joint contributions from stellar mass loss and Bondi 
accretion, if converted to radiation with $\eta = 0.1$, would generate nuclei 
far more luminous than actually observed.  The net accretion from the Bondi 
flow would be considerably reduced if the gas possesses some angular momentum 
at large radii (Proga \& Begelman 2003)---as inevitably it must---but even so 
it seems likely that the BH still has plenty of food at its disposal.  LLAGNs 
are by no means fuel-starved.  Moreover, the above estimates have erred on 
the conservative side.  For example, I have assumed that all of the stars are 
evolved, although in reality most nuclei have composite populations and hence 
larger mass loss rates.  Furthermore, I have neglected additional dissipation 
from larger scales (e.g., due to nuclear bars or spirals), as well as 
discrete, episodic events such as stellar tidal disruptions, which can provide 
a significant source of fuel, especially for BHs with masses \lax\ $10^{7}$ 
\solmass\ (Milosavljevi\'c, Merritt \& Ho 2006).  All of these additional 
sources will only exacerbate the fuel surplus crisis.
}

{\item
{\underbar{\it Low radiative efficiency}} \ \ \ If accretion does proceed 
at a reasonable fraction of the supply rate, then one has no option but to 
conclude that the radiative efficiency is much less than $\eta = 0.1$, the 
standard value for an optically thick, geometrically thin disk.  This type of 
argument has been invoked to explain the apparent conflict between the nuclear 
luminosities and Bondi accretion rates in many early-type galaxies (e.g., 
Fabian \& Rees 1995; Reynolds et al. 1996; Di~Matteo \& Fabian 1997; Mahadevan 
1997; Di~Matteo et al. 2001, 2003; Loewenstein et 
al. 2001; Ho, Terashima \& Ulvestad 2003; Pellegrini et al. 2003a; Donato, 
Sambruna \& Gliozzi 2004;  Evans et al. 2006).  Accretion flows with low 
radiative efficiency, of which the most popular version is the ADAF (see 
reviews in Narayan 2002; Yuan 2007), arise when the accreting medium is 
sufficiently tenuous that its cooling time exceeds the accretion timescale.  
RIAFs are predicted to exist for accretion rates below a critical threshold of 
$\dot M_{\rm crit} \approx 0.3\alpha^2 \dot M_{\rm Edd} \approx 0.01 \dot 
M_{\rm Edd}$, where the Shakura \& Sunyaev (1973) viscosity parameter is taken 
to be $\alpha \approx 0.1-0.3$ and $\dot M_{\rm Edd} \equiv L_{\rm Edd}/\eta 
c^2 = 0.22 \left(\eta/0.1\right) \left(M_{\rm BH}/10^8\, M_{\odot}\right)$ 
\solmass\ \peryr.  LLAGNs lie comfortably below this threshold.
}

{\item
{\underbar{\it Inefficient accretion/jet feedback}} \ \ \ Precisely how low 
$\eta$ can become depends on how much of the native fuel supply actually gets 
accreted.  In the presence of some rotation in the ambient medium, numerical 
simulations find that the amount of material accreted is much less than is 
available at large radii (e.g., Stone \& Pringle 2001; Igumenshchev, Narayan 
\& Abramowicz 2003).  Since the gravitational binding energy in a RIAF cannot 
be radiated efficiently, it must be lost by nonradiative means (Blandford \& 
Begelman 1999), either through convective transport of energy and angular 
momentum to large radii or through a global outflowing wind (see review by 
Quataert 2003).  The net effect of either process is to flatten the density 
profile near the center and to dramatically reduce the accretion rate.  At 
very low accretion rates \lax\ $(10^{-5}-10^{-6}) \,\dot M_{\rm Edd}$, such as 
in Sgr~A$^*$ and some giant elliptical galaxies, electron heat conduction will 
further suppress the accretion rate (Johnson \& Quataert 2007).  While these 
effects will ease the burden of invoking extremely small and perhaps 
physically unrealistic radiative efficiencies, it is important to note that 
these more recent models are {\it still}\ radiatively inefficient.

Whether the outflows generated in RIAFs can lead directly to relativistic jets 
is unclear, but what observations have established is the tendency for lowly 
accreting systems to become increasingly jet-dominated.  We see this not only 
in FR~I radio galaxies (Chiaberge, Capetti \& Celotti 1999; Donato, Sambruna 
\& Gliozzi 2004; Kharb \& Shastri 2004; Chiaberge, Capetti \& Macchetto 2005; 
Balmaverde \& Capetti 2006; Wu, Yuan \& Cao 2007), but it is also reflected in 
the nuclear properties of more run-of-the-mill LLAGNs (\S\S~5.3, 5.8), as well 
as in nearly quiescent nuclei (Pellegrini et al. 2007; Wrobel, Terashima \& Ho 
2008).  Detailed analysis of some sources, in fact, indicates that most of the 
accretion power is not radiated but instead channeled into the kinetic energy 
of relativistic jets (M87: Di~Matteo et al. 2003; IC~1459: Fabbiano et al. 
2003; IC~4296: Pellegrini et al. 2003b).  By analogy with the situation in 
cooling flows in galaxy clusters, the propensity for LLAGNs to become 
radio-loud opens up the possibility that the kinetic energy from small-scale 
jets or collimated outflows provides a major source of ``feedback'' into the 
circumnuclear environment, perhaps to the extent that accretion can be 
significantly interrupted or curtailed (Binney \& Tabor 1995; Di~Matteo et al. 
2003; Pellegrini et al. 2003a; Omma et al. 2004).  Indeed, calculations by 
K\"ording, Jester \& Fender (2008; see also Heinz, Merloni \& Schwab 2007) 
show that the total kinetic energy injected by LLAGN jets is very substantial.
Given the prominent hard X-ray component in LLAGN spectra, inverse-Compton 
scattering of the hard photons might also provide another avenue to heat the 
ambient medium (Ciotti \& Ostriker 2001).  Either form of energy 
injection---mechanical or radiative---can lead to unsteady, intermittent 
accretion with a short duty cycle.
}

{\item
{\underbar{\it Subluminous disk}} \ \ \  A thin disk can be tolerated if it 
can be made extremely subluminous during periods of intermittent activity 
(Shields \& Wheeler 1978).  This situation would arise if accretion disks in 
AGNs undergo the thermal-viscous ionization instability (Lin \& Shields 1986; 
Siemiginowska, Czerny \& Kostyunin 1996), in which case they spend most of 
their time in quiescence, punctuated by brief episodes of intense outbursts.  
Menou \& Quataert (2001) questioned the applicability of the ionization 
instability in AGNs, but they suggested that low-luminosity systems (with 
$\dot M$ \lax\ $10^{-3}$ \solmass\ \peryr) may contain disks in which mass 
accumulates in a stable, nonaccreting ``dead zone.''  Others have managed 
to stall accretion by condensing the hot flow into an inner cold, inert disk 
(Nayakshin 2003; Tan \& Blackman 2005; Jolley \& Kuncic 2007), which may form 
naturally from Compton cooling of the corona (Liu et al. 2007).  Finally, 
Merloni \& Fabian (2002) proposed that LLAGNs do contain a cold thin disk, but 
because of their low mass accretion rates, they liberate a large fraction of 
their gravitational energy in a strongly magnetized, unbound corona.  Since a 
cold disk component is present in all these models, they face a serious, and 
in my opinion insurmountable, challenge because LLAGNs generally do not show 
fluorescent Fe~K\al\ emission or reflection features in their X-ray spectra.  
The Merloni \& Fabian model may be spared of this criticism, as the disk may 
be highly ionized, but it does predict strong and rapid X-ray variability, 
which is generally {\it not}\ observed in LLAGNs (\S~5.3; Ptak et al. 1998).
}
\end{itemize}

\subsection{The Disk-Jet Connection}

As the mass accretion rate drops and the radiative efficiency declines, an 
increasing fraction of the accretion power gets channeled into a relativistic 
jet whose energy release is mainly kinetic rather than radiative.  The 
principal evidence for the growing importance of jets in LLAGNs comes from the 
broad-band SEDs, which invariably are prominent in the radio, with the degree 
of radio-loudness rising systematically (albeit with significant scatter) with 
decreasing Eddington ratio (\S~5.8; Figure~10{\it b}).  Where available, VLBI 
imaging on milliarcsecond scales reveals unresolved cores with nonthermal 
brightness temperatures and a flat or slightly inverted spectrum---classical 
signposts of a relativistic jet (Blandford \& K\"onigl 1979).  Detailed 
modeling of the SEDs of individual sources often shows that the accretion flow 
itself does not produce enough radio emission to match the data: that extra 
``something else'' is most plausibly attributed to the jet component (Quataert 
et al. 1999; Ulvestad \& Ho 2001b; Fabbiano et al.  2003; Pellegrini et al.  
2003b; Anderson, Ulvestad \& Ho 2004; Ptak et al. 2004; Wu \& Cao 2005; Nemmen 
et al. 2006; Wu, Yuan \& Cao 2007).  Moreover, RIAF models predict radio 
spectral indices of $\alpha \approx +0.4$ (Mahadevan 1997), whereas the 
observed values more typically fall in the range $\alpha \approx -0.2$ to 
$+0.2$.

The jet may contribute substantially outside of the radio band, especially 
in the optical and X-rays.  Some advocate that the jet, in fact, accounts 
for most or even all of the emission across the broad-band SED.  For example, 
Yuan et al. (2002) successfully fitted the multiwavelength data of NGC~4258 
with effectively a jet-only model.  In their picture, a radiative shock at the 
base of the jet gives rise to synchrotron emission in the near-IR and optical 
regions, whose self-Compton component then explains the X-rays; the 
flat-spectrum radio emission comes from further out in the jet.  Similar 
models have been devised for the Galactic Center source Sgr~A$^*$ (Falcke \& 
Markoff 2000; Yuan, Markoff \& Falcke 2002).  The gross similarity between the 
SEDs of some FR~I nuclei and BL Lac objects, which are jet-dominated sources 
but otherwise also low-accretion rate systems (Wang, Staubert \& Ho 2002), has 
also been noted (e.g., Bower et al. 2000; Capetti et al. 2000; Chiaberge 
et al.  2003; Meisenheimer et al. 2007).  

Statistical samples that are larger but more limited in spectral coverage have 
come from combining radio data with high-resolution optical or X-ray 
observations. 
Studies that specifically target radio galaxies, particularly FR~I sources 
and weak-line FR~IIs, report that the core radio power scales tightly with the 
optical and/or X-ray continuum luminosity, a finding often taken to support a 
common nonthermal, jet origin for the broad-band emission (Worrall \& 
Birkinshaw 1994; Canosa et al. 1999; Chiaberge, Capetti \& Celotti 1999, 2000; 
Capetti et al. 2002; Verdoes~Kleijn et al. 2002; Donato, Sambruna \& Gliozzi 
2004; Balmaverde \& Capetti 2006; Balmaverde, Capetti \& Grandi 2006; Evans 
et al. 2006; for a counterargument, see Rinn, Sambruna \& Gliozzi 2005 and 
Gliozzi et al. 2008).  A similar radio-optical correlation, after correcting 
for Doppler boosting, is also seen among BL Lac objects (Giroletti et al. 
2006), strengthening the case that FR~I radio galaxies and BL Lac objects are 
intrinsically the same but misoriented siblings.  Many FR~II systems, on the 
other hand, especially those with broad lines, deviate systematically from the 
baseline FR~I correlations, by exhibiting stronger optical and 
X-ray emission for a given level of radio emission (Chiaberge, Capetti \& 
Celotti 2000, 2002; Varano et al. 2004).  In concordance with the frequent 
detection of X-ray absorption and Fe~K$\alpha$ emission (Evans et al. 2006), 
this suggests that FR~IIs have higher accretion rates and a much more dominant 
accretion flow component, relative to the jet, than FR~Is.

Any attempt to explain the broad-band spectrum of LLAGNs with either just a
RIAF or just a jet runs the risk of oversimplification.  Clearly both are 
required. The trick is to figure out a reliable way to divvy up the two 
contributions to the SED.  We cannot deny that there is a jet, because we see 
it directly in the radio at a strength far greater than can be attributed to 
the RIAF. The jet emission must contribute at some level outside of the radio 
band.  At the same time, the jet cannot exist in isolation; it is anchored to 
and fed by some kind of accretion flow, of which a promising configuration is 
a vertically thick RIAF (Livio, Ogilvie \& Pringle 1999; Meier 1999).  An 
outstanding problem is that the interpretation of the data is not unique.  
Because many of the model parameters are poorly constrained and the broad-band 
data remain largely fragmentary and incomplete, the SEDs often can be fit with 
pure jet models, pure accretion flow models, or some combination of the two.  
The recent detection of high levels of polarization in the optical nuclei of 
FR~Is (Capetti et al. 2007) strongly points toward a synchrotron origin in the 
jet for the optical continuum, but even this observation cannot be considered 
definitive, because a RIAF can also produce nonthermal flares (e.g., in 
Sgr~A$^*$; Quataert 2003).

The so-called BH fundamental plane---a nonlinear correlation among radio 
luminosity, X-ray luminosity, and BH mass---offers a promising framework to 
unify accreting BHs over a wide range in mass and accretion rates.  Merloni, 
Heinz \& Di~Matteo (2003) first demonstrated that the correlation between 
$L_{\rm rad}$ and $L_{\rm X}$ tightens considerably after including \mbh\ as a 
third variable.  Combining observational material for several Galactic stellar 
BHs and a large sample of nearby LLAGNs, they find that

\begin{displaymath}
\log L_{\rm rad} = 0.60 \log L_{\rm X} + 0.78 \log M_{\rm BH}.
\end{displaymath}

\noindent
This empirical correlation agrees well with the theoretical relations between 
radio flux, BH mass, and accretion rate derived from the scale-invariant 
disk-jet model of Heinz \& Sunyaev (2003).  The BH fundamental plane, however,
appears to be a very blunt tool.  In an independent analysis, Falcke, 
K\"ording \& Markoff (2004) obtained a similar empirical relation, but unlike 
Merloni, Heinz \& Di~Matteo these authors explained the scaling coefficients 
entirely in terms of a jet-dominated model.  Moreover, as emphasized by 
K\"ording, Falcke \& Corbel (2006), objects with very different emission 
processes, including luminous quasars and BL Lac objects, sit on the same 
correlation, albeit with larger scatter.

\begin{figure}
\vbox{\hbox{
\hskip -0.2 cm
\psfig{figure=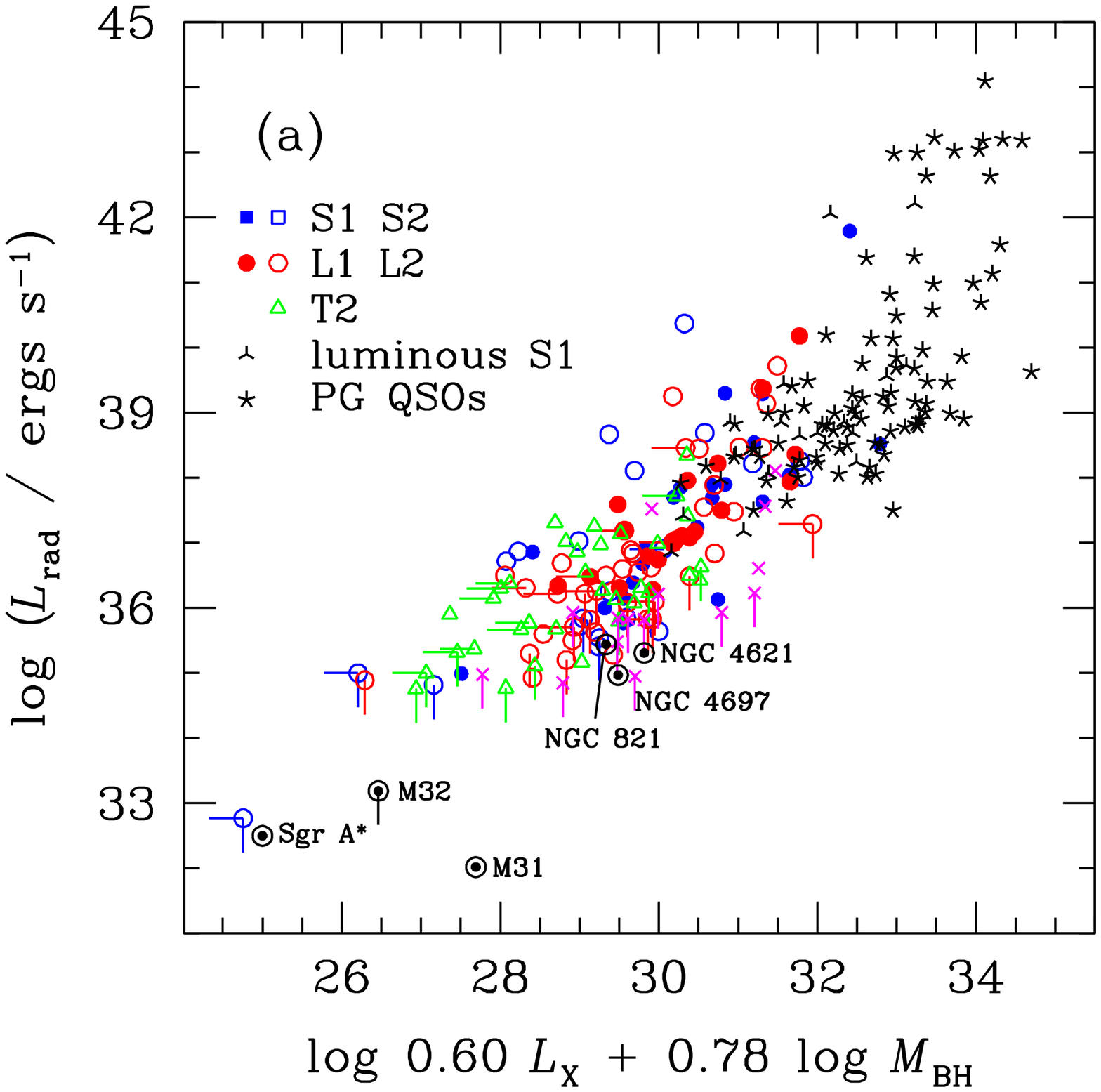,width=6.7cm,angle=0}
\hskip -0.0 cm
\psfig{figure=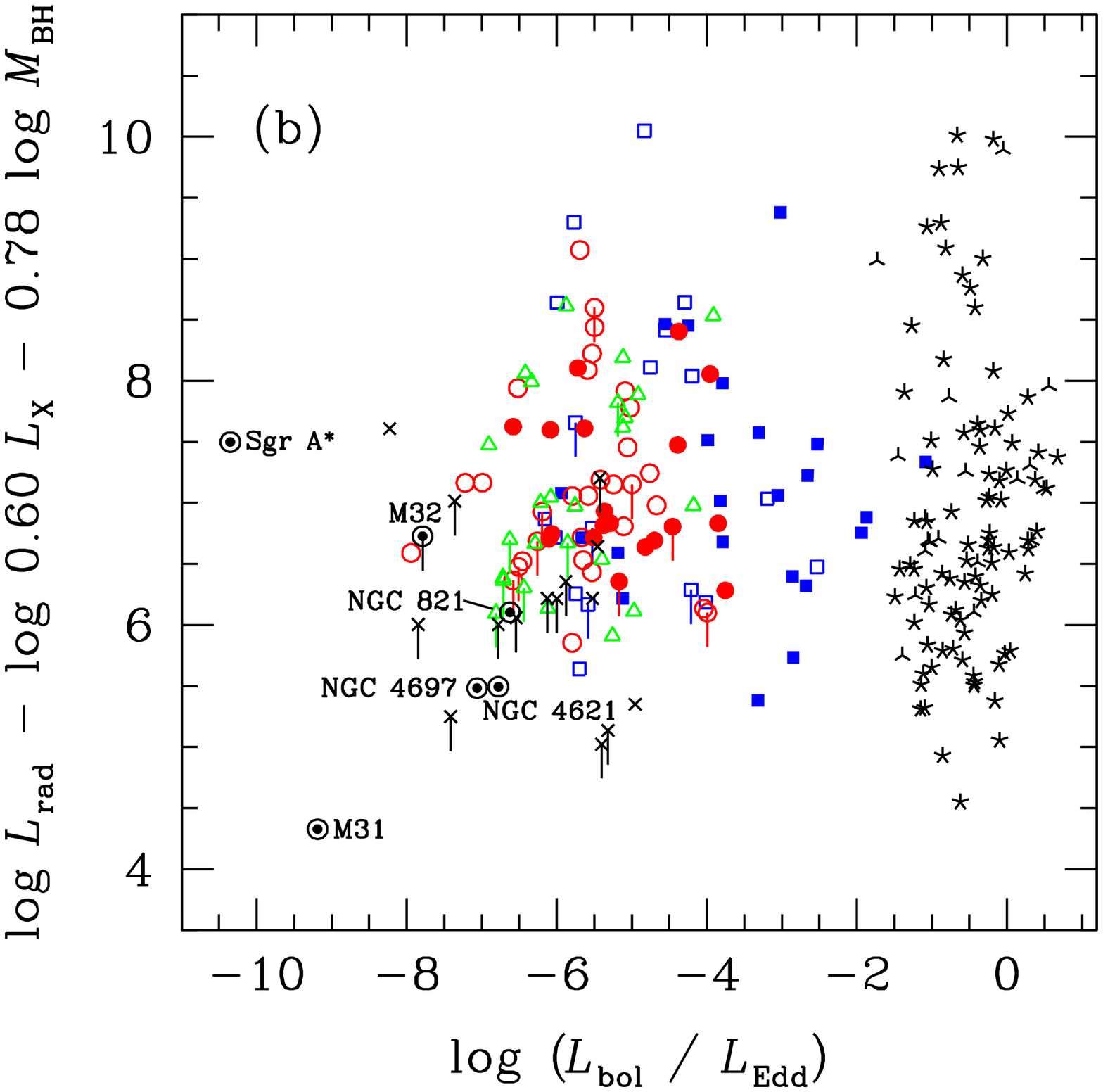,width=6.7cm,angle=0}
}}
\caption{({\it a}) Fundamental plane correlation among core radio luminosity,
X-ray luminosity, and BH mass.  ({\it b}) Deviations from the fundamental
plane as a function of Eddington ratio.  
}
\end{figure}

I illustrate this point in Figure~12{\it a}, which includes all Palomar LLAGNs 
with suitable data, along with the collection of high-luminosity sources from 
L.C. Ho (in preparation).  With the exception of a handful of radio-loud 
quasars, the vast 
majority of the objects fall on a well-defined swath spanning $\sim 10$ orders 
of magnitude in luminosity.  There are no obvious differences among the various
subclasses of LLAGNs, except that the type~1 sources appear more tightly 
correlated.  Plotting the residuals of the fundamental plane relation versus 
the Eddington ratio reveals two interesting points (Figure~12{\it b}).  First, 
although the intrinsic scatter of the relation is quite large, it markedly 
increases for objects with high Eddington ratios, at $L_{\rm bol}/L_{\rm Edd} 
\approx 10^{-1\pm1}$, as already noted by Maccarone, Gallo \& Fender (2003) 
and Merloni, Heinz \&  Di~Matteo (2003).  The scatter flares up because the 
radio-loud quasars lie offset above the relation and the radio-quiet quasars 
on average lie offset below the relation.  At the opposite extreme, sources 
with $L_{\rm bol}/L_{\rm Edd}$ \lax\ $10^{-6.5}$ may also show a systematic 
downturn, in possible agreement with the proposal by Yuan \& Cui (2005) that 
below a critical threshold, $L_{\rm X} \approx 10^{-5.5} L_{\rm Edd}$, both 
the radio {\it and}\ the X-rays should be dominated by emission from the jet.  
M31 (Garcia et al. 2005), NGC~821 (Pellegrini et al. 2007), and NGC~4621 and 
NGC~4697 (Wrobel, Terashima \& Ho 2008) seem to conform to Yuan \& Cui's 
prediction, but M32 and especially Sgr~A$^*$ clearly do not.  Additional deep 
radio and X-ray observations of ultra-low-luminosity nuclei would be very 
valuable to clarify the situation in this regime.

If, as surmised, the relative proportions between jet and disk output depend 
on accretion rate, with the bulk of the radiated power, even in the X-rays, 
originating from the former in the lowest accretion rate systems, two 
important consequences ensue. With respect to the microphysics of RIAFs, it 
implies that the radiative efficiencies are even lower than previously 
inferred under the assumption that the X-rays emanate solely from the accretion 
flow.  On a more global, environmental scale, shifting the emphasis from 
the disk to the jet changes the balance between kinetic versus radiative 
output, with important implications for prescriptions of AGN feedback in models
of galaxy formation because BHs spend most of their lives in a low-state.  
From empirical and theoretical considerations (Heinz, Merloni \& Schwab 2007; 
K\"ording, Jester \& Fender 2008), the jet carries a substantial fraction 
of the accreted rest mass energy: 
$P_{\rm jet} \approx 0.2 \eta \dot M c^2 \approx 7.2\times10^{36} 
(L_{\rm rad}/10^{30}\,{\rm ergs\,\,s}^{-1})^{12/17}$ \lum.  In 
fact, the total kinetic energy injected by LLAGN jets is comparable to or 
perhaps even greater than the contribution from supernovae.  At low 
redshifts, radiative feedback from quasars, which is commonly assumed to 
operate with an efficiency of $\sim 5$\%, may 

\clearpage
\begin{figure}
\psfig{figure=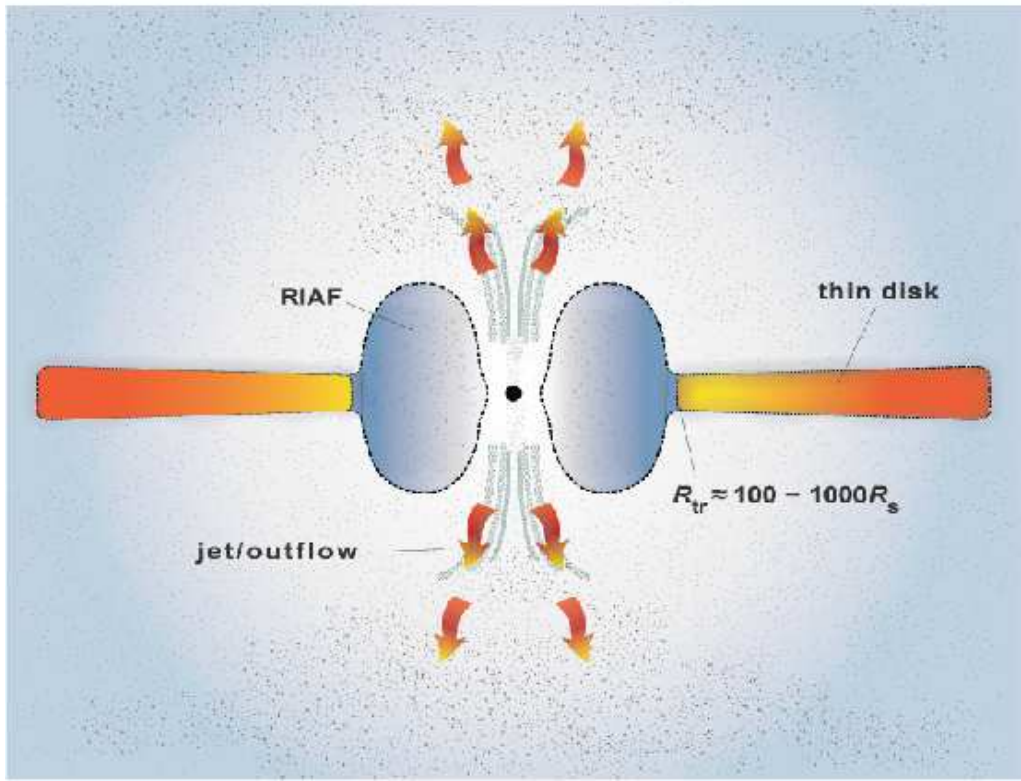,width=13.5cm,angle=0}
\caption{
A cartoon of the central engine of LLAGNs, consisting of three
components: an inner, radiatively inefficient accretion flow (RIAF), an outer,
truncated thin disk, and a jet or outflow.  (Courtesy of S. Ho.)
}
\end{figure}

\noindent
be less important then 
jet-driven feedback from LLAGNs (K\"ording, Jester \& Fender 2008).

\subsection{The Central Engine of LLAGNs}

The preceding sections argue that the weak nuclear activity seen in the 
majority of nearby galaxies traces low-level BH accretion akin to the more 
familiar form observed in powerful AGNs.  However, multiple lines of evidence 
indicate that LLAGNs are not simply scaled-down versions of their more luminous 
cousins. They are qualitatively different.  From the somewhat fragmentary clues 
presented in this review, we can piece together a schematic view of the 
structure of the central engine in LLAGNs (Ho 2002b, 2003, 2005).  As sketched 
in Figure~13, it has three components.

\begin{enumerate}
\item{{\underbar{\it Radiatively inefficient accretion flow}} \ \ \ 
In the present-day Universe, and especially in the centers of big bulges, the 
amount of material available for accretion is small, resulting in mass 
accretion rates that fall far below $10^{-2} \dot M_{\rm Edd}$.  In such a 
regime, the low-density, tenuous material is optically thin and cannot cool
efficiently.  Rather than settling into a classical optically thick, 
geometrically thin, radiatively efficient disk---the normal configuration for
luminous AGNs---the accretion flow puffs up into a hot, quasi-spherical, 
radiatively inefficient distribution, whose dynamics may be dominated by 
advection, convection, or outflows.  This is an area of active ongoing 
theoretical research.  In the interest of brevity, I will gloss over the 
technical details and simply follow Quataert (2003) by calling these RIAFs.  
The existence of RIAFs, or conversely the absence of a standard thin disk 
extending all the way to small radii (a few Schwarzschild radii $R_{\rm S}$), 
is suggested by the feeble luminosities of LLAGNs, by their low Eddington 
ratios, and especially by their low inferred radiative efficiencies.  The great disparity between the available fuel supply and the actually observed accretion 
luminosity demands that the radiative efficiency of the accretion flow be much 
less than $\eta = 0.1$ (\S~8.1).   Additional support for RIAFs comes from 
considerations of the SED, particularly the absence of the big blue bump, a 
classical signature of the thin disk, and the preponderance of intrinsically 
hard X-ray spectra.
}

\item{{\underbar{\it Truncated thin disk}} \ \ \ 
Beyond a transition radius $R_{\rm tr} \approx 100-1000 R_{\rm S}$, the RIAF
switches to a truncated optically thick, geometrically thin disk.  The 
observational evidence for this component comes in three forms.  First, the 
SEDs of some well-studied LLAGNs require a truncated thin disk to explain the 
big red bump---the prominent mid-IR peak and the steep fall-off of the 
spectrum in the optical--UV region (\S~5.8).  The thermal disk emission is 
cool (red) not only because of a low accretion rate (Lawrence 2005) but also 
because the inner radius of the disk does not extend all the way in to a few 
$R_{\rm S}$ as in luminous AGNs.  Second, the very same truncated disk 
structure employed to model the SED simultaneously accounts for the 
relativistically broadened, double-peaked emission-line profiles observed in 
some sources (\S~5.5).  Indeed, in the case of NGC~1097 (Nemmen et al. 2006), 
the transition radius derived from modeling the SED ($R_{\rm tr} = 225 
R_{\rm S}$) agrees remarkably well with the inner radius of the disk obtained 
from fitting the double-peaked broad H\al\ profile.  Ho et al. (2000) suggested 
that double-peaked broad emission lines are commonplace in LLAGNs. By 
implication, the truncated disk configuration inferred from this special class 
of line profiles must be commonplace too.  Lastly, the striking absence of 
broad Fe~K\al\ emission in the X-ray spectra of LLAGNs (\S~5.3), a feature 
commonly attributed to X-ray fluorescence off of a cold accretion disk 
extending inward to a few $R_{\rm S}$ in bright Seyfert~1 nuclei (e.g., Nandra 
et al. 1997b, 2007), strongly suggests that in low-luminosity sources such a 
structure is either absent or truncated interior to some radius, such that it 
subtends a significantly smaller solid angle.  Similar lines of reasoning have 
been advanced for broad-line radio galaxies that show weak Fe~K\al\ emission 
and weak Compton reflection (Wo\'zniak et al. 1998; Eracleous, Sambruna \& 
Mushotzky 2000; Lewis et al. 2005), although these characteristics can be 
mimicked by an ionized but otherwise untruncated disk (Ballantyne, Ross \& 
Fabian 2002).
}

\item{{\underbar{\it Jet/outflow}} \ \ \ The empirical connection between 
LLAGNs and jets has been established unequivocally from radio observations.  
Not only are the SEDs of LLAGNs generically radio-loud, but the strength of the
radio emission generally cannot be fit without recourse to a jet component, 
which in many cases can be seen directly from VLBI-scale radio images.  From a 
theoretical point of view, jets may share a close physical connection with 
RIAFs. As emphasized by Narayan \& Yi (1995) and Blandford \& Begelman (1999), 
RIAFs have a strong tendency to drive bipolar outflows due to the high thermal 
energy content of the hot gas.  Whether such outflows can develop into highly 
collimated, relativistic ejections remains to be seen, but they at least 
provide a promising starting point.  RIAFs may be additionally conducive to 
jet formation because its vertically thick structure enhances the large-scale 
poloidal component of the magnetic field, which plays a critical role in 
launching jets (Livio, Ogilvie \& Pringle 1999; Meier 1999; Ballantyne \& 
Fabian 2005; Ballantyne 2007).  It is interesting to recall that the original 
motivation for ion-supported tori (Rees et al. 1982), an early incarnation of 
RIAFs, was to explain the low luminosity of radio galaxies. Rees et al.  
postulated that the puffed-up structure of the ion torus may help facilitate 
the collimation of the jet.
}
\end{enumerate}

The above-described three-component structure has been applied to model the 
broad-band spectrum of a number of LLAGNs, including NGC~4258 (Lasota et al. 
1996; Gammie, Narayan \& Blandford 1999), M81 and NGC~4579 (Quataert et al. 
1999), NGC~3998 (Ptak et al. 2004), and NGC~1097 (Nemmen et al. 2006).  For the 
handful of LLAGNs with available estimates of the transition radii, 
$R_{\rm tr}$ seems to scale roughly inversely with $L_{\rm bol}/L_{\rm Edd}$ 
(Yuan \& Narayan 2004).  This trend may be in agreement with models 
for disk evaporation (Liu \& Meyer-Hofmeister 2001).  As the latter authors
note, however, disks attain their maximum evaporation efficiency at 
$R_{\rm tr} \approx 300 R_{\rm S}$, making sources such as M81 and NGC~4579, 
both with $R_{\rm tr} \approx 100 R_{\rm S}$ (Quataert et al. 1999), difficult 
to explain.  At a qualitative level, at least, the general idea that the thin
disk recedes to larger and larger radii as the accretion rate drops is 
probably correct.  In an analysis of 33 PG quasars with Fe~K\al\ emission 
detected in {\it XMM-Newton}\ spectra, Inoue, Terashima \& Ho (2007) find 
that the iron line profile varies systematically with Eddington ratio. 
Specifically, the Fe~K\al\ profile becomes narrower with decreasing 
$L_{\rm bol}/L_{\rm Edd}$, a result that can be interpreted as a systematic 
increase in the inner radius of the accretion disk at low accretion rates.

The basic schematic proposed in Figure~13 is hardly new.  To my knowledge, a 
hybrid model consisting of a RIAF---then called an ion-supported torus---plus 
a truncated thin disk was most clearly articulated in a prescient paper by 
Chen \& Halpern (1989) in their description of Arp~102B, later elaborated by 
Eracleous \& Halpern (1994) in the general context of double-peaked broad-line 
radio galaxies.  Chen \& Halpern identified the 25 \micron\ peak in the SED 
with the turnover frequency of the synchrotron peak from the RIAF, whose 
elevated structure illuminates an outer thin disk that emits the double-peaked 
broad optical lines. The overall weakness of the UV continuum in Arp~102B 
(Halpern et al.  1996) further corroborates a truncated thin disk structure 
and potentially provides an explanation for the low-ionization state of the 
emission-line spectrum.  As for the jet component, it was assumed to be 
present, at least implicitly, insofar as the double-peaked broad-line AGNs 
were thought to reside preferentially in radio-loud AGNs, and the very concept 
of ion-supported tori was invented with reference to radio galaxies (Rees et 
al. 1982).

Recent developments add important refinements and modifications to Chen \& 
Halpern's original picture.  First, the mid-IR peak in most objects is 
dominated by thermal emission from the truncated thin disk rather than by the 
synchrotron peak of the RIAF.  Second, the jet component, which was once 
regarded as somewhat incidental, has emerged as a natural and perhaps inevitable
outgrowth of the inner accretion flow itself.  Third, although the original 
model was invented to explain a small minority of the AGN population 
(double-peaked radio-loud sources), now we have good reason to believe that 
similar physical conditions prevail in LINERs as a class (Ho et al. 2000), 
and, by extension, in the majority of nearby accreting BHs.

The physical picture outlined above for LLAGNs shares strong similarities with 
that developed for X-ray binaries in their hard state (see Maccarone, Fender 
\& Ho 2005 and references therein), suggesting that the basic architecture
of the central engine around accreting BHs---across 10 orders of magnitude in 
mass---is essentially scale-invariant (Meier 2001; Maccarone, Gallo \& 
Fender 2003; Merloni, Heinz \& Di~Matteo 2003; Falcke, K\"ording \& Markoff 
2004; Ho 2005; K\"ording, Jester \& Fender 2006).

\section{CONCLUDING REMARKS}

The topics covered in this article are both very old and very new.  It has been 
known for over three decades that a large segment of the galaxy population 
exhibits signs of unusual activity in their nuclei, and for nearly as long 
people have puzzled over the physical origin of this activity.  Over time the 
observational material at optical wavelengths has improved markedly, 
especially with the completion of the Palomar survey, but the debate has only 
intensified.  Given their abundance, most of the controversy has centered, not 
surprisingly, on LINERs.  While the nonstellar nature of a sizable fraction of 
LINERs is now incontrovertible (e.g., those with broad H\al\ emission), the 
AGN content in the majority of the class remains unsettled.  Determining the 
physical origin of these systems is more than of mere phenomenological 
interest.  Because LINERs are so numerous---being the dominant constituent of 
the local LLAGN population and a sizable fraction of all galaxies---they have 
repercussions on virtually every issue related to AGN and BH demographics.  

The main source of contention stems from the fact that Mother Nature knows too 
many ways of generating nebular conditions that qualitatively look similar to 
the low-ionization characteristics of LINERs at optical wavelengths.  A 
dizzying array of excitation mechanisms has been proposed to explain LINERs, 
ranging from variants of conventional AGN photoionization, to shocks of 
various flavors, to interstellar processes such as cooling flows and turbulent 
mixing layers, to stellar-based photoionization by populations both young and 
old, prosaic and exotic.  This field has suffered not from a shortage of 
ideas, but from too many.  As a consequence, whenever LINERs are discussed, it 
is customary to end with the pessimistic mantra that they are a mixed-bag, 
heterogeneous class of objects, a statement with somewhat dismissive 
connotations that is often taken to mean that we have no idea what they are 
and that they are too messy to deal with.  This is an unfair characterization 
of the progress that has been made, and I think that there is good reason to 
sound a more positive note.  

As summarized in this review, a number of developments during the last few 
years shed considerable light on the physical origin of LLAGNs in general and 
LINERs in particular.  The key advances have come from the broader 
perspective afforded by observations outside of the traditional optical 
window, especially in the radio and X-rays, although important 
insights can also be credited to optical and UV data taken with \hst.  In all 
instances, high angular resolution has been a critical factor to disentangle 
the weak nuclear emission from the blinding host galaxy background.  

Other developments have been instrumental in forging a coherent view of nuclear
activity in the nearby Universe.  On the theoretical side, rapid advances
in the study of radiatively inefficient accretion flows, originally primarily
motivated by applications to X-ray binaries and to the Galactic Center source
Sgr~A$^*$, has led to a growing appreciation that they are also relevant to
LLAGNs in general.  Many investigators have sharpened the physical analogy 
between the spectral states of X-ray binaries and certain classes of AGNs, an 
effort that has resulted in a more holistic picture of BH accretion, 
especially as it concerns the evolution of the accretion flow in response to 
variations in mass accretion rate and the mechanism for generating jets or 
outflows.  Meanwhile, the dynamical detection of supermassive BHs, their 
ubiquity, and the discovery of scaling relations between BHs and their host 
galaxies have given a major boom to studies of AGNs in all their multi-faceted 
manifestations.  More than ever, in the grand scheme of things, AGNs are no 
longer viewed as rare and exotic oddities but as natural episodes during the 
life cycle of galaxies during which their BHs accrete, grow, and shine.  The 
impact of BH growth and AGN feedback have emerged forcefully as major new 
themes in galaxy formation.  LLAGNs gain an even greater prominence within 
this context.  Although the bulk of the mass density of BHs was accreted in a 
luminous, radiatively efficient mode, it behooves us to understand how BHs 
spend most of their lives.   The detection of supermassive BHs has also 
fundamentally altered the character of the discourse on LLAGNs.  We can now 
shift our attention from the question of {\it whether}\ LLAGNs contain 
BHs---an implicit or explicit motivation for much of the past discussion on 
the nature of these sources---to {\it why}\ these BHs have the properties that 
they do.  Among other things, LLAGNs can be used as an effective platform for 
exploring accretion physics in highly sub-Eddington systems and for 
investigating physical processes in the circumnuclear regions of galaxies that 
are normally masked by brighter nuclei.

The following is a list of the ``top ten'' results from this paper.

\begin{enumerate}

\item{Approximately 2/3 of local E--Sb galaxies exhibit weak nuclear activity 
incompatible with normal stellar processes; in contrast, only about 15\% of 
Sc--Sm galaxies show AGN activity (\S 3). }

\item{The vast majority of LINERs, and, by implication, most nearby weakly 
active nuclei, are genuine, accretion-powered AGNs (\S\S 6.1, 6.5).}

\item{The ubiquity of LLAGNs in galaxies with bulges strongly supports the
current paradigm derived from dynamical studies that all bulges contain BHs.
However, the detection of AGNs in some bulgeless, even dwarf, galaxies proves
that bulges are not necessary for the formation of central BHs (\S 7).}

\item{The luminosity function of nearby LLAGNs follows $\Phi \propto
L^{-1.2\pm0.2}$ from $L_{{\rm H}\alpha} \approx 3\times 10^{41}$ to $10^{38}$
\lum, below which it appears to flatten down to $L_{{\rm H}\alpha} \approx 6 
\times 10^{36}$ \lum\ or $M_B \approx -8$ mag (\S 5.9).}

\item{Stellar photoionization by young or intermediate-stars and shock heating
can be ruled out as the excitation mechanisms for LLAGNs (\S\S 6.2, 6.3).}

\item{Despite the overall success of AGN photoionization models, many LLAGNs,
especially type~2 sources, have a shortage of ionizing photons.  The energy 
deficit problem could be solved with cosmic ray heating and extra ionization 
from evolved (post-AGB) stars, diffuse thermal plasma, and the cumulative 
X-ray emission from low-mass X-ray binaries (\S 6.4).}

\item{Variations in the mass accretion rate give rise to the different classes 
of emission-line nuclei.  LINERs are the low-luminosity, low-accretion rate 
extension of Seyferts, followed by transition nuclei, and ending with 
absorption-line nuclei at the end of the BH starvation sequence (\S 6.5).}

\item{LLAGNs are not simply scaled-down versions of powerful AGNs.  Their 
central engines undergo fundamental changes when the accretion rate drops to 
extremely sub-Eddington values.  In this regime, the BLR and obscuring torus 
disappear (\S\S 5.5, 5.6).  LLAGNs do not follow the standard AGN unification 
model.}

\item{Below a characteristic luminosity of $\sim$1\% Eddington, the
canonical optically thick, geometrically thin accretion disk transforms
into a three-component structure consisting of an inner vertically thick and 
radiatively inefficient accretion flow, a truncated outer thin disk, and a jet 
or outflow (\S\S 5.8, 8.3).}

\item{At the lowest accretion rates, an increasing fraction of the accretion
energy gets channeled into a relativistic jet.  The emitted energy is mainly
kinetic rather than radiative.  Since radiation and kinetic jets interact
differently with the surrounding gas, this has important implications for
AGN feedback into galaxy formation (\S 8.2).}

\end{enumerate}

\vskip 0.2cm
My research is supported by the Carnegie Institution of Washington and by NASA 
grants from the Space Telescope Science Institute (operated by AURA, Inc., 
under NASA contract NAS5-26555).  I would like to recognize my collaborators 
who have contributed to the work covered in this review, especially A.J. Barth,
M. Eracleous, M.E. Filho, A.V. Filippenko, J.E. Greene, D. Maoz, E.C. Moran, 
C.Y.  Peng, A. Ptak, E. Quataert, H.-W. Rix, W.L.W. Sargent, M.  Sarzi, J.C. 
Shields, Y. Terashima, J.S. Ulvestad, and J.M.  Wrobel.  Several of them
(A.J. Barth, M. Eracleous, J.E. Greene, M. Sarzi, J.C. Shields, Y. Terashima) 
read an early draft of the manuscript and provided useful feedback that helped 
to improve it.  Some of the concepts expressed in the review were sharpened 
after correspondence with M. Eracleous, G. Ferland, J.C. Shields, and Y. 
Terashima.  I thank A.J. Barth, A.V. Filippenko, and W.L.W. 
Sargent for permission to cite material in advance of publication, H.M.L.G. 
Flohic and M. Eracleous for providing the images for Figure 5, and Salvador Ho 
for drafting Figure 13.  I am grateful to J. Kormendy for his steadfast 
encouragement, wise counsel, and meticulous editing.

\vskip 0.5cm

%\begin{thebibliography}{99}

\noindent {LITERATURE CITED}

\frenchspacing

\vskip 0.5cm

\nhi 
Aldrovandi SMV, P\'equignot D. 1973. {\it Astron. Astrophys.} 26:33

\nhi 
Allen MG, Dopita, MA, Tsvetanov ZI. 1998. {\it Ap. J.} 493:571

\nhi 
Alonso-Herrero A, Rieke MJ, Rieke GH, Ruiz M. 1997. {\it Ap. J.} 482:74

\nhi 
Alonso-Herrero A, Rieke MJ, Rieke GH, Shields JC. 2000. {\it Ap. J.} 530:688

\nhi 
Anderson JM, Ulvestad JS. 2005. {\it Ap. J.} 627:674

\nhi 
Anderson JM, Ulvestad JS, Ho LC. 2004. {\it Ap. J.} 603:42

\nhi 
Antonucci R. 1993.  {\it Annu. Rev. Astron. Astrophys.} 31:473

\nhi 
Armus L, Heckman TM, Miley GK. 1990. {\it Ap. J.} 364:471

\nhi 
Atkinson JW, Collett JL, Marconi A, Axon DJ, Alonso-Herrero A, et al. 2005. 
{\it MNRAS} 359:504

\nhi 
Avni Y, Tananbaum H. 1982. {\it Ap. J.} 262:L17

\nhi 
Awaki H, Mushotzky R, Tsuru T, Fabian AC, Fukazawa Y, et al. 1994. {\it PASJ} 
46:L65

\nhi 
Baganoff FK, Maeda Y, Morris M, Bautz MW, Brandt WN, et al. 2003. {\it Ap. J.} 
591:891

\nhi 
Baldwin JA, Phillips MM, Terlevich R. 1981. {\it PASP} 93:5

\nhi 
Ballantyne DR. 2007. {\it Mod. Phys. Lett. A} 22:2397

\nhi 
Ballantyne DR, Fabian AC. 2005. {\it Ap. J.} 622:L97

\nhi 
Ballantyne DR, Ross RR, Fabian AC. 2002. {\it MNRAS} 332:L45

\nhi 
Balmaverde B, Capetti A. 2006.  {\it Astron. Astrophys.} 447:97

\nhi 
Balmaverde B, Capetti A, Grandi P. 2006. {\it Astron. Astrophys.} 451:35

\nhi
Barger AJ, Cowie LL, Mushotzky RF, Richards EA. 2001. {\it Astron. J.} 121:662

\nhi 
Barth AJ. 2004. In {\it Carnegie Observatories Astrophysics Series, Vol 1:
Coevolution of Black Holes and Galaxies}, ed. LC Ho, p. 21. Cambridge:
Cambridge Univ. Press

\nhi 
Barth AJ, Filippenko AV, Moran EC. 1999a. {\it Ap. J.} 515:L61

\nhi 
Barth AJ, Filippenko AV, Moran EC. 1999b. {\it Ap. J.} 525:673

\nhi 
Barth AJ, Greene JE, Ho LC. 2005.  {\it Ap. J.} 619:L151

\nhi 
Barth AJ, Ho LC, Filippenko AV. 2003.  In {\it Active Galactic Nuclei: from 
Central Engine to Host Galaxy}, ed. S Collin, F Combes, I Shlosman, p. 387.
San Francisco: ASP

\nhi 
Barth AJ, Ho LC, Filippenko AV, Rix H-W, Sargent WLW 2001a. {\it Ap. J.} 546:205

\nhi 
Barth AJ, Ho LC, Filippenko AV, Sargent WLW. 1998. {\it Ap. J.} 496:133

\nhi 
Barth AJ, Ho LC, Rutledge RE, Sargent WLW. 2004. {\it Ap. J.} 607:90

\nhi 
Barth AJ, Reichert GA, Filippenko AV, Ho LC, Shields JC, Mushotzky RF, 
Puchnarewicz EM. 1996. {\it Astron. J.} 112:1829

\nhi 
Barth AJ, Reichert GA, Ho LC, Shields JC, Filippenko AV, Puchnarewicz EM. 
1997. {\it Astron. J.} 114:2313

\nhi 
Barth AJ, Sarzi M, Rix H-W, Ho LC, Filippenko AV, Sargent WLW. 2001b. 
{\it Ap. J.} 555:685

\nhi 
Barth AJ, Shields JC. 2000. {\it PASP} 112:753

\nhi 
Barth AJ, Tran H, Brotherton MS, Filippenko AV, Ho LC, et al. 1999. 
{\it Astron. J.} 118:1609

\nhi
Bassani L, Dadina M, Maiolino R, Salvati M, Risaliti G, et al.
1999.  {\it Ap.~J.~Suppl.} 121:473

\nhi 
Begelman MC, Fabian AC. 1990. {\it MNRAS} 244:36P

\nhi 
Bendo GJ, Dale DA, Draine BT, Engelbracht CW, Kennicutt RC Jr, et al. 2006. 
{\it Ap. J.} 645:134

\nhi 
Bendo GJ, Joseph RD. 2004. {\it Astron. J.} 127:3338

\nhi 
Bennert N, Jungwiert B, Komossa S, Haas M, Chini R. 2006. 
{\it Astron. Astrophys.} 456:953

\nhi 
Bianchi S, Corral A, Panessa F, Barcons X, Matt G, et al. 2008. {\it MNRAS} 
385:195

\nhi 
Bietenholz MF, Bartel N, Rupen NP. 2000. {\it Ap. J.} 532:895

\nhi 
Binette L. 1985. {\it Astron. Astrophys.} 143:334

\nhi 
Binette L, Magris CG, Stasi\'nska G, Bruzual AG. 1994. 
{\it Astron. Astrophys.} 292:13

\nhi 
Binney J, Tabor G. 1995. {\it MNRAS} 276:663

\nhi 
Blandford RD, Begelman MC. 1999. {\it MNRAS} 303:L1

\nhi 
Blandford RD, K\"onigl A. 1979. {\it Ap. J.} 232:34

\nhi 
Blanton EL, Sarazin CL, Irwin JA. 2001. {\it Ap. J.} 552:106

\nhi 
B\"ohringer H, Belsole E, Kennea J, Matsushita K, Molendi S, et al. 2001. 
{\it Astron. Astrophys.} 365::181

\nhi 
Boisson C, Joly M, Moultaka J, Pelat D, Serote-Roos M. 2000.  
{\it Astron. Astrophys.} 357:850

\nhi 
B\"{o}ker T, van~der~Marel RP, Laine S, Rix H-W, Sarzi M, Ho LC, Shields JC. 
2002. {\it Astron. J.} 123:1389

\nhi 
Bonatto C, Bica E, Alloin D. 1989. {\it Astron. Astrophys.} 226:23

\nhi 
Bondi H. 1952. {\it MNRAS} 112:195

\nhi 
Bower GA, Green RF, Quillen AC, Danks A, Gull T, et al. 2000. {\it Ap. J.} 
534:189

\nhi 
Bower GA, Wilson AS, Heckman TM, Richstone DO. 1996. {\it Astron. J.} 111:1901

\nhi 
Burbidge EM, Burbidge G. 1962. {\it Ap. J.} 135:694

\nhi 
Burbidge EM, Burbidge G. 1965. {\it Ap. J.} 142:634

\nhi 
Burbidge G, Gould RJ, Pottasch SR. 1963. {\it Ap. J.} 138:945

\nhi 
Canosa CM, Worrall DM, Hardcastle MJ, Birkinshaw M. 1999. {\it MNRAS} 310:30

\nhi 
Capetti A, Axon DJ, Chiaberge M, Sparks WB, Macchetto FD, et al.
2007. {\it Astron. Astrophys.} 471:137

\nhi 
Capetti A, Celotti A, Chiaberge M, de Ruiter HR, Fanti R, et al.
2002. {\it Astron. Astrophys.} 383:104

\nhi 
Capetti A, Trussoni E, Celotti A, Feretti L, Chiaberge M. 2000. {\it MNRAS} 
318:493

\nhi 
Capetti A, Verdoes Kleijn G, Chiaberge M. 2005. {\it Astron. Astrophys.} 
439:935

\nhi 
Cappellari M, Bertola F, Burstein D, Buson LM, Greggio L, Renzini A. 2001. 
{\it Ap. J.} 551:197

\nhi 
Cappi M, Panessa F, Bassani L, Dadina M, Dicocco G, et al. 2006. 
{\it Astron. Astrophys.} 446:459

\nhi 
Carollo CM, Stiavelli M, de Zeeuw PT, Mack, J. 1997. {\it Astron. J.} 114:2366

\nhi 
Carrillo R, Masegosa J, Dultzin-Hacyan D, Ordo\~{n}ez R. 1999.  {\it Rev. Mex. 
Astron. Astrof.} 35:187

\nhi 
Carswell RF, Baldwin JA, Atwood B, Phillips MM. 1984. {\it Ap. J.} 286:464

\nhi 
Chary R, Becklin EE, Evans AS, Neugebauer G, Scoville NZ, et al.
2000. {\it Ap. J.} 531:756

\nhi 
Chen K, Halpern JP. 1989. {\it Ap. J.} 344:115

\nhi 
Chiaberge M, Capetti A, Celotti A. 1999.  {\it Astron. Astrophys.} 349:77

\nhi 
Chiaberge M, Capetti A, Celotti A. 2000.  {\it Astron. Astrophys.} 355:873

\nhi 
Chiaberge M, Capetti A, Celotti A. 2002.  {\it Astron. Astrophys.} 394:791

\nhi 
Chiaberge M, Capetti A, Macchetto FD. 2005. {\it Ap. J.} 625:716

\nhi 
Chiaberge M, Gilli R, Capetti A, Macchetto FD. 2003. {\it Ap. J.} 597:166

\nhi 
Chiaberge M, Macchetto FD, Sparks W, Capetti A, Allen MG, Martel AR. 2002.
{\it Ap. J.} 571:247

\nhi 
Cid Fernandes R Jr, Golz\'alez Delgado RM, Schmitt H, Storchi-Bergmann T, Martins LP, et al. 2004. {\it Ap. J.} 605:105

\nhi 
Ciotti L, D'Ercole A, Pellegrini S, Renzini A. 1991. {\it Ap. J.} 376:380

\nhi 
Ciotti L, Ostriker JP. 2001. {\it Ap. J.} 551:131

\nhi 
Colbert EJM, Mushotzky RF. 1999. {\it Ap. J.} 519:89

\nhi 
Colina L, Golz\'alez Delgado RM, Mas-Hesse JM, Leitherer C. 2002. {\it Ap. J.} 
579:545

\nhi 
Collins JA, Rand RJ. 2001. {\it Ap. J.} 551:57

\nhi 
Combes F. 2003. In {\it Active Galactic Nuclei: from Central Engine to Host 
Galaxy}, ed. S Collin,  F Combes, I Shlosman, p. 411. San Francisco: ASP

\nhi 
Constantin A, Vogeley MS. 2006. {\it Ap. J.} 650:727

\nhi 
Contini M. 1997. {\it Astron. Astrophys.} 323:71

\nhi 
Corbett EA, Kewley L, Appleton PN, Charmandaris V, Dopita MA, et al.
2003. {\it Ap. J.} 583:670

\nhi 
Costero R, Osterbrock DE. 1977. {\it Ap. J.} 211:675

\nhi 
Dale DA, Smith JDT, Armus L, Buckalew BA, Helou G, et al. 2006. {\it Ap. J.} 
646:161

\nhi 
Danziger IJ, Fosbury RAE, Penston MV. 1977. {\it MNRAS} 179:41P

\nhi 
Decarli R, Gavazzi G, Arosio I, Cortese L, Boselli A, et al. 2007. {\it MNRAS} 
381:136

\nhi 
Demoulin-Ulrich M-H, Butcher HR, Boksenberg A. 1984. {\it Ap. J.} 285:527

\nhi 
Desroches L-B, Ho LC. 2008. {\it Ap. J.}, submitted

\nhi 
Dewangan GC, Griffiths RE, Di Matteo T, Schurch NJ. 2004. {\it Ap. J.} 607:788

\nhi 
D\'\i az AI, Pagel BEJ, Wilson IRG. 1985. {\it MNRAS} 212:737

\nhi 
D\'\i az AI, Pagel BEJ, Terlevich E. 1985. {\it MNRAS} 214:41P

\nhi 
Di Matteo T, Allen SW, Fabian AC, Wilson AS, Young AJ. 2003. {\it Ap. J.} 
582:133

\nhi 
Di Matteo T, Fabian AC. 1997. {\it MNRAS} 286:L50

\nhi 
Di Matteo T, Johnstone RM, Allen SW, Fabian AC. 2001. {\it Ap. J.} 550:L19

\nhi 
Disney MJ, Cromwell RH. 1971. {\it Ap. J.} 164:L35

\nhi 
Doi A, Kameno S, Kohno K, Nakanishi K, Inoue M. 2005. {\it MNRAS} 363:692

\nhi 
Donato D, Sambruna RM, Gliozzi M. 2004. {\it Ap. J.} 617:915

\nhi 
Done C., Gierli\'nski M, Sobolewska M, Schurch N.  2007. In {\it The Central 
Engine of Active Galactic Nuclei}, ed.  LC Ho, J-M Wang, p. 121. San 
Francisco: ASP

\nhi 
Dong X-B, Wang T, Yuan W, Shan H, Zhou H, et al. 2007. {\it Ap. J.} 657:700

\nhi 
Dopita MA, Koratkar AP, Allen MG, Tsvetanov ZI, Ford HC, et al.
1997. {\it Ap. J.} 490:202

\nhi 
Dopita MA, Sutherland RS. 1995. {\it Ap. J.} 455:468

\nhi 
Dudik RP, Satyapal S, Gliozzi M, Sambruna RM. 2005. {\it Ap. J.} 620:113

\nhi 
Elitzur M, Shlosman I. 2006. {\it Ap. J.} 648:L101

\nhi 
Elvis M, Wilkes BJ, McDowell JC, Green RF, Bechtold J, et al. 1994. 
{\it Ap.~J.~Suppl.} 95:1

\nhi 
Engelbracht CW, Rieke MJ, Rieke GH, Kelley DM, Achtermann JM. 1998. 
{\it Ap. J.} 505:639

\nhi 
Eracleous M, Halpern JP. 1994. {\it Ap.~J.~Suppl.} 90:1

\nhi 
Eracleous M, Halpern JP. 2001. {\it Ap.~J.} 554:240

\nhi 
Eracleous M, Hwang JA, Flohic HMLG. 2008a. {\it Ap.~J.} submitted

\nhi 
Eracleous M, Hwang JA, Flohic HMLG. 2008b. {\it Ap.~J.} submitted

\nhi 
Eracleous M, Sambruna R, Mushotzky RF. 2000. {\it Ap. J.} 537:654

\nhi 
Eracleous M, Shields JC, Chartas G, Moran EC. 2002. {\it Ap.~J.} 565:108

\nhi 
Evans DA, Worrall DM, Hardcastle MJ, Kraft RP, Birkinshaw M. 2006. 
{\it Ap.~J.} 642:96

\nhi 
Fabbiano G. 2006. {\it Annu. Rev. Astron. Astrophys.} 44:323

\nhi 
Fabbiano G, Baldi A, Pellegrini S, Siemiginowska A, Elvis M, et al.
2004. {\it Ap.~J.} 616:730

\nhi 
Fabbiano G, Elvis M, Markoff S, Siemiginowska A, Pellegrini S, et al.
2003. {\it Ap.~J.} 588:175

\nhi 
Fabbiano G, Trinchieri G. 1985. {\it Ap. J.} 296:430

\nhi 
Faber SM, Tremaine S, Ajhar EA, Byun Y-I, Dressler A, et al. 1997. 
{\it Astron. J.} 114:1365

\nhi 
Fabian AC, Arnaud KA, Nulsen PEJ, Mushotzky RF. 1986. {\it Ap. J.} 305:9

\nhi 
Fabian AC, Canizares CR. 1988. {\it Nature} 333:829

\nhi 
Fabian AC, Rees MJ. 1995. {\it MNRAS} 277:L55

\nhi 
Falcke H, K\"ording E, Markoff S. 2004. {\it Astron. Astrophys.} 414:895

\nhi 
Falcke H, Markoff S. 2000. {\it Astron. Astrophys.} 362:113

\nhi 
Falcke H, Nagar NM, Wilson AS, Ulvestad JS. 2000. {\it Ap. J.} 542:197

\nhi 
Fanaroff BL, Riley JM. 1974. {\it MNRAS} 167:31P

\nhi 
Ferrarese L, Merritt D. 2000. {\it Ap. J.} 539:L9

\nhi 
Ferland GJ, Mushotzky RF. 1984. {\it Ap. J.} 286:42

\nhi 
Ferland GJ, Netzer H. 1983. {\it Ap. J.} 264:105

\nhi 
Filho ME, Barthel PD, Ho LC. 2000. {\it Ap.~J.~Suppl.} 129:93

\nhi 
Filho ME, Barthel PD, Ho LC. 2002a. {\it Ap.~J.~Suppl.} 142:223

\nhi 
Filho ME, Barthel PD, Ho LC. 2002b. {\it Astron. Astrophys.} 385:425

\nhi 
Filho ME, Barthel PD, Ho LC. 2006. {\it Astron. Astrophys.} 451:71

\nhi 
Filho ME, Fraternali F, Nagar NM, Barthel PD, Markoff S, et al. 2004. 
{\it Astron. Astrophys.} 418:429

\nhi 
Filippenko AV. 1985. {\it Ap. J.} 289:475

\nhi 
Filippenko AV. 1996. In {\it The Physics of LINERs in View of Recent 
Observations}, ed.  M Eracleous, A Koratkar, C Leitherer, LC Ho, p. 17. 
San Francisco: ASP

\nhi 
Filippenko AV, Halpern JP. 1984. {\it Ap. J.} 285:458
 
\nhi 
Filippenko AV, Ho LC. 2003. {\it Ap. J.} 588:L13

\nhi 
Filippenko AV, Ho LC, Sargent WLW. 1993. {\it Ap. J.} 410:L75

\nhi 
Filippenko AV, Sargent WLW. 1985. {\it Ap.~J.~Suppl.} 57:503 
 
\nhi 
Filippenko AV, Sargent WLW. 1988. {\it Ap. J.} 324:134

\nhi 
Filippenko AV, Sargent WLW. 1989. {\it Ap. J.} 342:L11

\nhi 
Filippenko AV, Terlevich R. 1992. {\it Ap. J.} 397:L79
 
\nhi 
Fiore F, ed. 2006. {\it AGNs and Galaxy Evolution}. {\it MmSAI} 77

\nhi 
Fiore F, Pellegrini S, Matt G, Antonelli LA, Comastri A, et al. 2001. 
{\it Ap. J.} 556:150

\nhi 
Flohic HMLG, Eracleous M, Chartas G, Shields JC, Moran EC. 2006. {\it Ap. J.} 
647:140

\nhi 
Ford HC, Butcher H. 1979. {\it Ap.~J.~Suppl.} 41:147

\nhi 
Fosbury RAE, Melbold U, Goss WM, Dopita MA. 1978. {\it MNRAS} 183:549

\nhi 
Fosbury RAE, Melbold U, Goss WM, van Woerden H. 1977. {\it MNRAS} 179:89

\nhi 
Fukazawa Y, Iyomoto N, Kubota A, Matsumoto Y, Makishima K. 2001. 
{\it Astron. Astrophys.} 374:73

\nhi 
Gabel JR, Bruhweiler FC. 2002. {\it Astron. J.} 124:737

\nhi 
Gabel JR, Bruhweiler FC, Crenshaw DM, Kraemer SB, Miskey CL. 2000. 
{\it Ap. J.} 532:883

\nhi 
Gallimore JF, Axon DJ, O'Dea CP, Baum SA, Pedlar A. 2006. {\it Astron. J.} 
132:546

\nhi 
Gallo E, Treu T, Jacob J, Woo J-H, Marshall RJ, Antonucci R. 2008. {\it Ap. J.}
 in press

\nhi 
Gammie CF, Narayan R, Blandford RD. 1999. {\it Ap. J.} 516:177

\nhi 
Garcia MR, Williams BF, Yuan F, Kong AKH, Primini FA, et al.
2005. {\it Ap. J.} 632:1042

\nhi 
Gebhardt K, Bender R, Bower G, Dressler A, Faber SM, et al. 2000. 
{\it Ap. J.} 539:L13

\nhi 
Gebhardt K, Rich RM, Ho LC. 2002. {\it Ap. J.} 578:L41

\nhi 
Gebhardt K, Rich RM, Ho LC. 2005. {\it Ap. J.} 634:1093

\nhi 
Geha M, Guhathakurta P, van~der~Marel RP. 2002. {\it Astron. J.} 124:3073

\nhi 
Georgantopoulos I, Panessa F, Akylas A, Zezas A, Cappi M, Comastri A. 2002. 
{\it Astron. Astrophys.} 386:60

\nhi 
George IM, Fabian AC. 1991. {\it MNRAS} 249:352

\nhi 
Ghosh H, Pogge RW, Mathur S, Martini P, Shields JC. 2007. {\it Ap. J.} 656:105

\nhi 
Giroletti M, Giovannini G, Taylor GB, Falomo R. 2006. {\it Ap. J.} 646:801

\nhi 
Gliozzi M, Foschini L, Sambruna RM, Tavecchio F. 2008. 
{\it Astron. Astrophys.} 478:723

\nhi 
Gliozzi M, Sambruna RM, Brandt WN, Mushotzky RF, Eracleous M. 2004. 
{\it Astron. Astrophys.} 413:139

\nhi 
Gliozzi M, Sambruna RM, Foschini L. 2007. {\it Ap. J.} 662:878

\nhi 
Gonz\'alez Delgado RM, Cid Fernandes R, Per\'ez E, Martins LP, 
Storchi-Bergmann T, et al. 2004. {\it Ap. J.} 605:127

\nhi 
Gonz\'alez-Mart\'\i n O, Masegosa J, M\'arquez I, Guerrero MA, Dultzin-Hacyan 
D. 2006. {\it Astron. Astrophys.} 460:45

\nhi 
Granato GL, De Zotti G, Silva L, Bressan A, Danese L. 2004. {\it Ap. J.} 600:580

\nhi 
Grandi SA, Osterbrock DE. 1978. {\it Ap. J.} 220:783

\nhi 
Greene JE, Ho LC. 2004. {\it Ap. J.} 610:722

\nhi 
Greene JE, Ho LC. 2005a. {\it Ap. J.} 627:721

\nhi 
Greene JE, Ho LC. 2005b. {\it Ap. J.} 630:122

\nhi 
Greene JE, Ho LC. 2006. {\it Ap. J.} 641:L21

\nhi 
Greene JE, Ho LC. 2007a. {\it Ap. J.} 667:131

\nhi 
Greene JE, Ho LC. 2007b. {\it Ap. J.} 670:92

\nhi 
Greene JE, Ho LC, Barth AJ. 2008. {\it Ap. J.} submitted

\nhi 
Greene JE, Ho LC, Ulvestad JS. 2006. {\it Ap. J.} 636:56

\nhi 
Gronwall C, Jangren A, Salzer JJ, Werk JK, Ciardullo R. 2004. {\it Astron. J.}
128:644

\nhi 
Grossan B, Gorjian V, Werner M, Ressler M. 2001. {\it Ap. J.} 563:687

\nhi 
Groves BA, Dopita MA, Sutherland RS. 2004. {\it Ap.~J.~Suppl.} 153:75

\nhi 
Groves B, Heckman T, Kauffmann G. 2006. {\it MNRAS} 371:1559

\nhi 
Gruenwald RB, Viegas-Aldrovandi SM. 1987. {\it Astron. Astrophys. Suppl.} 70:143

\nhi 
Gu Q-S, Huang J-S, Wilson G, Fazio GG. 2007. {\it Ap.~J.} 671:L105

\nhi 
Guainazzi M, Oosterbroek T, Antonelli LA, Matt G. 2000.  
{\it Astron. Astrophys.} 364:L80

\nhi 
Haas M, M\"uller SAH, Bertoldi F, Chini R, Egner S, et al. 2004. 
{\it Astron. Astrophys.} 424:531

\nhi 
Halderson EL, Moran EC, Filippenko AV, Ho LC. 2001. {\it Astron. J.} 122:637

\nhi 
Hall PB, Yee HKC, Lin H, Morris SL, Patton DR, et al. 2000. {\it Astron. J.} 
120:2220

\nhi 
Halpern JP, Eracleous M. 1994. {\it Ap. J.} 433:L17

\nhi 
Halpern JP, Eracleous M, Filippenko AV, Chen K. 1996. {\it Ap. J.} 464:704

\nhi 
Halpern JP, Filippenko AV. 1984. {\it Ap. J.} 285:475

\nhi 
Halpern JP, Steiner JE. 1983. {\it Ap. J.} 269:L37

\nhi 
Hao L, Strauss MA, Fan XH, Tremonti CA, Schlegel DJ, et al. 2005a. 
{\it Astron. J.} 129:1783

\nhi 
Hao L, Strauss MA, Tremonti CA, Schlegel DJ, Heckman TM, et al. 2005b. 
{\it Astron. J.} 129:1795

\nhi 
Hawkins MRS. 2004. {\it Astron. Astrophys.} 424:519

\nhi 
Heckman TM. 1980a. {\it Astron. Astrophys.} 87:142

\nhi 
Heckman TM. 1980b. {\it Astron. Astrophys.} 87:152
 
\nhi 
Heckman TM, Balick B, Crane PC. 1980. {\it Astron. Astrophys. Suppl. } 40:295

\nhi 
Heckman TM, Baum SA, van Breugel WJM, McCarthy P. 1989. {\it Ap. J.} 338:48

\nhi 
Heckman TM, Ptak A, Hornschemeier A, Kauffmann G. 2005. {\it Ap. J.} 634:161

\nhi 
Heinz S, Merloni A, Schwab J. 2007. {\it Ap. J.} 658:L9

\nhi 
Heinz S, Sunyaev R. 2003. {\it MNRAS} 343:L59

\nhi 
Heller CH, Shlosman I. 1994. {\it Ap. J.} 424:84

\nhi 
Hernquist L. 1989. {\it Nature}. 340;687

\nhi 
Ho LC. 1996. In {\it The Physics of LINERs in View of Recent Observations}, 
ed.  M Eracleous, A Koratkar, C Leitherer, LC Ho, p. 103. San Francisco: ASP

\nhi 
Ho LC. 1999a. {\it Ap. J.} 510:631

\nhi 
Ho LC. 1999b. {\it Ap. J.} 516:672

\nhi 
Ho LC. 2002a. {\it Ap. J.} 564:120

\nhi 
Ho LC. 2002b. In {\it Issues in Unification of AGNs}, ed. R Maiolino, 
A Marconi, N Nagar, p. 165. San Francisco: ASP

\nhi 
Ho LC. 2003. In {\it Active Galactic Nuclei: from Central Engine to Host 
Galaxy}, ed. S Collin,  F Combes, I Shlosman, p. 379. San Francisco: ASP

\nhi 
Ho LC, ed. 2004a. {\it Carnegie Observatories Astrophysics Series, Vol. 1: 
Coevolution of Black Holes and Galaxies} Cambridge: Cambridge Univ. Press

\nhi 
Ho LC. 2004b. In {\it Carnegie Observatories Astrophysics Series, Vol. 1: 
Coevolution of Black Holes and Galaxies}, ed. LC Ho, p. 293. Cambridge: 
Cambridge Univ. Press

\nhi 
Ho LC. 2005. In {\it From X-ray Binaries to Quasars: Black Hole Accretion on 
All Mass Scales}, ed. TJ Maccarone, RP Fender, LC Ho, p. 219. Dordrecht: Kluwer

\nhi 
Ho LC, Feigelson ED, Townsley LK, Sambruna RM, Garmire GP, et al. 2001. 
{\it Ap. J.} 549:L51

\nhi 
Ho LC, Filippenko AV. 1993. {\it Astrophys. Space Sci.} 205:19

\nhi 
Ho LC, Filippenko AV, Sargent WLW. 1993. {\it Ap. J.} 417:63
 
\nhi 
Ho LC, Filippenko AV, Sargent WLW. 1995. {\it Ap.~J.~Suppl.} 98:477

\nhi 
Ho LC, Filippenko AV, Sargent WLW. 1996. {\it Ap. J.} 462:183  %M81

\nhi 
Ho LC, Filippenko AV, Sargent WLW. 1997a. {\it Ap.~J.~Suppl.} 112:315

\nhi 
Ho LC, Filippenko AV, Sargent WLW. 1997b. {\it Ap. J.} 487:568 %stats

\nhi 
Ho LC, Filippenko AV, Sargent WLW. 1997c. {\it Ap. J.} 487:579 %HII

\nhi 
Ho LC, Filippenko AV, Sargent WLW. 1997d. {\it Ap. J.} 487:591 %bars

\nhi 
Ho LC, Filippenko AV, Sargent WLW. 2003. {\it Ap. J.} 583:159 %HFSVI

\nhi 
Ho LC, Filippenko AV, Sargent WLW, Peng CY. 1997e. {\it Ap.~J.~Suppl.} 
112:391

\nhi 
Ho LC, Peng CY. 2001. {\it Ap. J.} 555:650

\nhi 
Ho LC, Ptak A, Terashima Y, Kunieda H, Serlemitsos PJ, et al.
1999a. {\it Ap. J.} 525:168

\nhi 
Ho LC, Rudnick G, Rix H-W, Shields JC, McIntosh DH, et al.
2000. {\it Ap. J.} 541:120

\nhi 
Ho LC, Sarzi M, Rix H-W, Shields JC, Rudnick G, et al. 2002. {\it PASP} 114:137

\nhi 
Ho LC, Shields JC, Filippenko AV. 1993. {\it Ap. J.} 410:567

\nhi 
Ho LC, Terashima Y, Ulvestad JS. 2003. {\it Ap. J.} 589:783

\nhi 
Ho LC, Ulvestad JS. 2001. {\it Ap.~J.~Suppl.} 133:77

\nhi 
Ho LC, Van Dyk SD, Pooley GG, Sramek RA, Weiler KW. 1999b. {\it Astron. J} 
118:843

\nhi 
Hopkins PF, Hernquist L. 2006. {\it Ap.~J.~Suppl.} 166:1

\nhi 
Hopkins PF, Hernquist L, Cox TJ, Di Matteo T,  Martini P, et al.
2006.  {\it Ap.~J.~Suppl.} 163:1

\nhi 
Huchra JP, Burg R. 1992. {\it Ap. J.} 393:90

\nhi 
Humason ML, Mayall NU, Sandage AR. 1956. {\it Astron. J.} 61:97

\nhi 
Igumenshchev IV, Narayan R, Abramowicz MA. 2003. {\it Ap. J.} 592:1042

\nhi 
Inoue H, Terashima Y, Ho LC. 2007. {\it Ap. J.} 662:860

\nhi 
Irwin JA, Sarazin CL, Bregman JN. 2002. {\it Ap. J.} 570:152

\nhi 
Ishisaki Y, Makishima K, Iyomoto N, Hayashida K, Inoue H, et al. 1996. 
{\it PASJ} 48:237

\nhi 
Iyomoto N, Fukazawa Y, Nakai N, Ishihara Y. 2001. {\it Ap. J.} 561:L69

\nhi 
Iyomoto N, Makishima K, Fukazawa Y, Tashiro M, Ishisaki Y, et al.
1996. {\it PASJ}, 48:231

\nhi 
Iyomoto N, Makishima K, Fukazawa Y, Tashiro M, Ishisaki Y. 1997. {\it PASJ}
49:425

\nhi 
Iyomoto N, Makishima K, Matsushita K, Fukazawa Y, Tashiro M, Ohashi T. 1998a. 
{\it Ap. J.} 503:168

\nhi 
Iyomoto N, Makishima K, Tashiro M, Inoue S, Kaneda H, et al.
1998b. {\it Ap. J.} 503:L31

\nhi 
Jolley EJD, Kuncic Z. 2007. {\it Ap\&SS} 310:327

\nhi 
Johnson BM, Quataert E. 2007. {\it Ap. J.} 660:1273

\nhi 
Jungwiert B, Combes F, Palous J. 2001. {\it Astron. Astrophys.} 376:85

\nhi 
Kauffmann G, Heckman TM, Tremonti C, Brinchmann J, Charlot S, et al. 2003. 
{\it MNRAS} 346:1055

\nhi 
Kauffmann G, White SDM, Heckman TM, Me\'nard B, Brinchmann J, et al.
2004. {\it MNRAS} 353:713

\nhi 
Keel WC. 1983a. {\it Ap. J.} 268:632
 
\nhi 
Keel WC. 1983b. {\it Ap. J.} 269:466
 
\nhi 
Keel WC. 1983c. {\it Ap.~J.~Suppl.} 52:229

\nhi 
Keel WC, Miller JS. 1983. {\it Ap. J.} 266:L89

\nhi 
Kennicutt RC Jr. 1984. {\it Ap. J.} 287:116

\nhi 
Kewley LJ, Groves B, Kauffmann G, Heckman T. 2006. {\it MNRAS} 372:961

\nhi 
Kewley LJ, Heisler CA, Dopita MA, Lumsden S. 2001. {\it Ap.~J.~Suppl.} 132:37

\nhi 
Khachikian EY, Weedman DW. 1974. {\it Ap. J.} 192:581

\nhi 
Kharb P, Shastri P. 2004. {\it Astron. Astrophys.} 425:825

\nhi 
Kim D-C, Sanders DB, Veilleux S, Mazzarella JM, Soifer BT. 1995. 
{\it Ap.~J.~Suppl.} 98:129

\nhi 
Kim D-W, Fabbiano G. 2003. {\it Ap.~J.} 586:826

\nhi 
Kinkhabwala A, Sako M, Behar E, Kahn SM, Paerels F, et al. 2002. 
{\it Ap. J.} 575:732

\nhi 
Kirhakos S, Phillips MM. 1989. {\it PASP} 101:949

\nhi 
Komossa S, B\"ohringer H, Huchra JP. 1999. {\it Astron. Astrophys.} 349:88

\nhi 
Koratkar AP, Deustua S, Heckman TM, Filippenko AV, Ho LC, Rao M. 1995. 
{\it Ap. J.} 440:132

\nhi 
K\"ording E, Falcke H, Corbel S. 2006. {\it Astron. Astrophys.} 456:439

\nhi 
K\"ording E, Jester S, Fender R. 2006. {\it MNRAS} 372:1366

\nhi 
K\"ording E, Jester S, Fender R. 2008. {\it MNRAS} 383:277

\nhi 
Kormendy J. 1993. In {\it The Nearest Active Galaxies}, ed. J Beckman,
L Colina, H Netzer, p. 197. (Madrid: CSIC)

\nhi 
Kormendy J. 2004. In {\it Carnegie Observatories Astrophysics Series, Vol 1:
Coevolution of Black Holes and Galaxies}, ed. LC Ho, p. 1. Cambridge:
Cambridge Univ. Press

\nhi 
Kormendy J, Fisher DB, Cornell ME, Bender R. 2008. {\it Ap.~J.~Suppl.} submitted

\nhi 
Kormendy J, Kennicutt RC. 2004. {\it Annu. Rev. Astron. Astrophys.} 42:603

\nhi 
Kormendy J, Richstone DO. 1995. {\it Annu. Rev. Astron. Astrophys.} 33:581

\nhi 
Koski AT, Osterbrock DE. 1976. {\it Ap. J.} 203:L49

\nhi 
Krips M, Eckart A, Krichbaum TP, Pott J-U, Leon S, et al. 2007. {\it Astron. 
Astrophys.} 464:553

\nhi 
Krolik JH. 1998. {\it Active Galactic Nuclei}. Princeton: Princeton Univ. Press

\nhi 
Kukula MJ, Pedlar A, Baum SA, O'Dea CP. 1995. {\it MNRAS} 276:1262

\nhi 
Kunth D, Sargent WLW, Bothun GD. 1987. {\it Astron. J.} 92:29

\nhi 
Laor A. 2003. {\it Ap. J.} 590:86

\nhi 
Larkin JE, Armus L, Knop RA, Soifer BT, Matthews K. 1998. {\it Ap.~J.~Suppl.}
114:59

\nhi 
Lasota J-P, Abramowicz MA, Chen X, Krolik J, Narayan R, Yi I. 1996. 
{\it Ap. J.} 462:142

\nhi 
Laurikainen E, Salo H, Buta R. 2004. {\it Ap. J.} 607:103

\nhi 
Lawrence A. 2005. {\it MNRAS} 363:57

\nhi 
Lawrence A, Elvis M. 1982. {\it Ap. J.} 256:410

\nhi 
Lawrence A, Ward M, Elvis M, Fabbiano G, Willner SP, et al. 1985. 
{\it Ap. J.} 291:117

\nhi
Lewis KT, Eracleous M, Gliozzi M, Sambruna RM, Mushotzky RF. 2005. {\it Ap. J.} 
622:816

\nhi
Lewis KT, Eracleous M, Sambruna RM. 2003. {\it Ap. J.} 593:115

\nhi
Li C, Kauffmann G, Heckman TM, White SDM, Jing YP. 2008. {\it MNRAS} in press

\nhi
Lightman AP, White TR. 1988. {\it Ap. J.} 335:57

\nhi
Lin DC, Shields GA. 1986. {\it Ap. J.} 305:28

\nhi
Lira P, Lawrence A, Johnson RA. 2000. {\it MNRAS} 319:17

\nhi
Liu BF, Meyer-Hofmeister E. 2001. {\it Astron. Astrophys.} 372:386

\nhi
Liu BF, Taam RE, Meyer-Hofmeister E, Meyer F. 2007. {\it Ap. J.} 671:695

\nhi
Livio M, Ogilvie GI, Pringle JE. 1999. {\it Ap. J.} 512:100

\nhi
Loewenstein M, Mushotzky RF, Angelini L, Arnaud KA, Quataert E. 2001. 
{\it Ap. J.} 555:L21

\nhi
Maccarone TJ, Fender RP, Ho LC, eds. 2005. {\it From X-ray Binaries to 
Quasars: Black Hole Accretion on All Mass Scales} Dordrecht: Kluwer

\nhi
Maccarone TJ, Gallo E, Fender RP. 2003. {\it MNRAS} 345:L19

\nhi
Magorrian J, Tremaine S, Richstone D, Bender R, Bower G, et al. 1998. 
{\it Astron. J.} 115:2285

\nhi
Mahadevan R. 1997. {\it Ap. J.} 477:585

\nhi
Maia MAG, Machado RS, Willmer CNA. 2003. {\it Astron. J.} 126:1750

\nhi
Maiolino R, Rieke GH. 1995.  {\it Ap. J.} 454:95

\nhi
Makishima K, Fujimoto R, Ishisaki Y, Kii T, Loewenstein M, et al. 1994. 
{\it PASJ} 46:L77

\nhi
Makishima K, Ohashi T, Hayashida K, Inoue H, Koyama K, et al. 1989. {\it PASJ}
41:697

\nhi
Malkan MA, Sargent WLW. 1982. {\it Ap. J.} 254:22

\nhi
Maoz D. 2007. {\it MNRAS} 377:1696

\nhi
Maoz D, Filippenko AV, Ho LC, Macchetto FD, Rix H-W, Schneider DP. 1996.
{\it Ap.~J.~Suppl.} 107:215

\nhi
Maoz D, Filippenko AV, Ho LC, Rix H-W, Bahcall JN, et al. 1995. 
{\it Ap. J.} 440:91

\nhi
Maoz D, Koratkar AP, Shields JC, Ho LC, Filippenko AV, Sternberg A. 1998. 
{\it Astron. J.} 116:55

\nhi
Maoz D, Nagar NM, Falcke H, Wilson AS. 2005. {\it Ap. J.} 625:699

\nhi
Matsumoto Y, Fukazawa Y, Nakazawa K, Iyomoto N, Makishima K. 2001. {\it PASJ} 
53:475

\nhi
Meier DL. 1999. {\it Ap. J.} 522:753

\nhi
Meier DL. 2001. {\it Ap. J.} 548:L9

\nhi
Meisenheimer K, Tristram KRW, Jaffe W, Israel F, Neumayer N, et al. 2007. 
{\it Astron. Astrophys.} 471:453

\nhi
Menou K, Quataert E. 2001. {\it Ap. J.} 552:204

\nhi
Merloni A, Fabian AC. 2002. {\it MNRAS} 332:165

\nhi
Merloni A, Heinz S, Di Matteo T. 2003. {\it MNRAS} 345:1057

\nhi
Merloni A, Nayakshin S, Sunyaev R, eds. 2005. {\it Growing Black Holes: 
Accretion in a Cosmological Context} Berlin: Springer-Verlag

\nhi
Middleton M, Done C, Schurch N. 2008. {\it MNRAS} 383:1501

\nhi
Miller CJ, Nichol RC, Gomez PL, Hopkins AM, Bernardi M. 2003. {\it Ap. J.} 
597:142

\nhi
Milosavljevi\'c M, Merritt D, Ho LC. 2006. {\it Ap. J.} 652:120

\nhi
Minkowski R, Osterbrock DE. 1959. {\it Ap. J.} 129:583

\nhi
Miniutti G, Ponti G, Greene JE, Ho LC, Fabian AC, Iwasawa K. 2008. {\it MNRAS} 
in press

\nhi
Moore B, Katz N, Lake G, Dressler A, Oemler A. 1996. {\it Nature} 379:613

\nhi
Moran EC, Eracleous M, Leighly KM, Chartas G, Filippenko AV, et al.
2005. {\it Astron. J.} 129:2108

\nhi
M\"{u}ller SAH, Haas M, Siebenmorgen R, Klaas U, Meisenheimer K, et al.
2004.  {\it Astron. Astrophys.} 426:L29

\nhi
Muno MP, Baganoff FK, Bautz MW, Feigelson ED, Garmire GP, et al. 2004. 
{\it Ap. J.} 613:326

\nhi
Murray N, Chiang J. 1997. {\it Ap. J.} 474:91

\nhi
Mushotzky RF. 1993. In {\it The Nearest Active Galaxies}, ed. J Beckman,
L Colina, H Netzer, p. 47. Madrid: CSIC Press

\nhi
Mushotzky RF, Wandel A. 1989. {\it Ap. J.} 339:674

\nhi
Nagao T, Murayama T, Shioya Y, Taniguchi Y. 2002. {\it Ap. J.} 567:73

\nhi
Nagar NM, Falcke H, Wilson AS. 2005. {\it Ap. J.} 435:521

\nhi
Nagar NM, Falcke H, Wilson AS, Ho LC. 2000. {\it Ap. J.} 542:186

\nhi
Nagar NM, Falcke H, Wilson AS, Ulvestad JS. 2002. {\it Astron. Astrophys.} 
392:53

\nhi
Nagar NM, Wilson AS, Falcke H. 2001. {\it Ap. J.} 559:L87

\nhi
Nandra K, George IM, Mushotzky RF, Turner TJ, Yaqoob T. 1997a. {\it Ap. J.} 
476:70

\nhi
Nandra K, George IM, Mushotzky RF, Turner TJ, Yaqoob T. 1997b. {\it Ap. J.} 
477:602

\nhi
Nandra K, O'Neill PM, George IM, Reeves JN. 2007. {\it MNRAS} 382:194

\nhi
Narayan R. 2002. In {\it Lighthouses of the Universe}, ed. M Gilfanov et al.,
p. 405. Berlin: Springer

\nhi
Narayan R, Yi I. 1995. {\it Ap. J.} 444:231

\nhi
Nayakshin S. 2003. {\it Astron. Nachr. Suppl.} 324:3

\nhi
Nelson CH, Whittle M. 1996. {\it Ap. J.} 465:96

\nhi
Nemmen RS, Storchi-Bergmann T, Yuan F, Eracleous M, Terashima Y, Wilson AS. 
2006. {\it Ap. J.} 643:652

\nhi
Nicastro F. 2000. {\it Ap. J.} 530:L65

\nhi
Nicholson KL, Reichert GA, Mason KO, Puchnarewicz EM, Ho LC, et al.
1998. {\it MNRAS} 300:893

\nhi
Noyola E, Gebhardt K, Bergmann M. 2008. {\it Ap. J.} in press

\nhi
O'Connell RW. 1999. {\it Annu. Rev. Astron. Astrophys.} 37:603

\nhi
O'Connell RW, Martin JR, Crane JD, Burstein D, Bohlin RC, et al.
2005. {\it Ap. J.} 635:305

\nhi
Omma H, Binney J, Bryan G, Slyz A. 2004. {\it MNRAS} 348:1105

\nhi 
Osterbrock DE. 1960. {\it Ap. J.} 132:325

\nhi 
Osterbrock DE. 1971. In {\it Nuclei of Galaxies}, ed. DJK O'Connell, p. 151.
Amsterdam: North Holland

\nhi 
Osterbrock DE, Dufour RJ. 1973. {\it Ap. J.} 185:441

\nhi 
Osterbrock DE, Miller JS. 1975. {\it Ap. J.} 197:535

\nhi 
Osterbrock DE, Shaw RA. 1988. {\it Ap. J.} 327:89

\nhi
Padovani P, Matteucci F. 1993. {\it Ap. J.} 416:26

\nhi
Page MJ, Breeveld AA, Soria R, Wu K, Branduardi-Raymont G, et al.
2003. {\it Astron. Astrophys.} 400:145

\nhi
Page MJ, Soria R, Zane S, Wu K, Starling R. 2004. {\it Astron. Astrophys.} 
422:77

\nhi
Panessa F, Barcons X, Bassani L, Cappi M, Carrera FJ, et al.
2007. {\it Astron. Astrophys.} 467:519

\nhi 
Panessa F, Bassani L. 2002. {\it Astron. Astrophys.} 394:435

\nhi 
Panessa F, Bassani L, Cappi M, Dadina M, Barcons X, et al.
2006. {\it Astron. Astrophys.} 455:173

\nhi 
Pappa A, Georgantopoulos I, Stewart GC, Zezas AL. 2001. {\it MNRAS} 326:995

\nhi 
Peimbert M, Torres-Peimbert S. 1981. {\it Ap. J.} 245:845

\nhi 
Pelat D, Alloin D, Fosbury RAE. 1981. {\it MNRAS} 195:787

\nhi 
Pellegrini S. 2005. {\it Ap. J.} 624:155

\nhi 
Pellegrini S, Baldi A, Fabbiano G, Kim D-W. 2003a. {\it Ap. J.} 597:175

\nhi 
Pellegrini S, Cappi M, Bassani L, Della Ceca R, Palumbo GGC. 2000a. 
{\it Astron. Astrophys.} 360:878

\nhi 
Pellegrini S, Cappi M, Bassani L, Malaguti G, Palumbo GGC, Persic M. 2000b.
{\it Astron. Astrophys.} 353:447

\nhi 
Pellegrini S, Fabbiano G, Fiore F, Trinchieri G, Antonelli A. 2002. 
{\it Astron. Astrophys.} 383:1

\nhi 
Pellegrini S, Siemiginowska A, Fabbiano G, Elvis M, Greenhill L, et al. 2007. 
{\it Ap. J.} 667:749

\nhi 
Pellegrini S, Venturi T, Comastri A, Fabbiano G, Fiore F, et al.
2003b. {\it Ap. J.} 585:677

\nhi 
Peng CY, Ho LC, Impey CD, Rix H-W. 2002. {\it Astron. J.} 124:266

\nhi 
Penston MV, Fosbury RAE. 1978. {\it MNRAS} 183:479

\nhi
P\'equignot D. 1984. {\it Astron. Astrophys.} 131:159

\nhi 
Peterson BM, Bentz MC, Desroches L-B, Filippenko AV, Ho LC, et al.  2005. 
{\it Ap. J.} 632:799.  Erratum. 2005. {\it Ap. J.} 641:638

\nhi 
Petre R, Mushotzky RF, Serlemitsos PJ, Jahoda K, Marshall FE. 1993. 
{\it Ap. J.} 418:644

\nhi 
Phillips MM. 1979. {\it Ap. J.} 227:L121

\nhi 
Phillips MM, Charles PA, Baldwin JA. 1983. {\it Ap. J.} 266:485

\nhi 
Phillips MM, Jenkins CR, Dopita MA, Sadler EM, Binette L. 1986. 
{\it Astron. J.} 91:1062
 
\nhi 
Pogge RW. 1989. {\it Ap.~J.~Suppl.} 71:433

\nhi 
Pogge RW, Maoz D, Ho LC, Eracleous M. 2000. {\it Ap. J.} 532:323

\nhi 
Proga D, Begelman MC. 2003. {\it Ap. J.} 592:767

\nhi 
Ptak A, Serlemitsos PJ, Yaqoob T, Mushotzky R. 1999. {\it Ap.~J.~Suppl.}
120:179

\nhi 
Ptak A, Terashima Y, Ho LC, Quataert E. 2004. {\it Ap. J.} 606:173 

\nhi 
Ptak A, Yaqoob T, Mushotzky R, Serlemitsos P, Griffiths R. 1998. {\it Ap. J.} 
501:L37

\nhi 
Ptak A, Yaqoob T, Serlemitsos PJ, Kunieda H, Terashima Y. 1996. {\it Ap. J.} 
459:542

\nhi 
Quataert E. 2003. {\it Astron. Nachr. Suppl.} 324:435

\nhi 
Quataert E, Di Matteo T, Narayan R, Ho LC. 1999. {\it Ap. J.} 525:L89

\nhi 
Quillen AC, McDonald C, Alonso-Herrero A, Lee A, Shaked S, et al.
2001. {\it Ap. J.} 547:129

\nhi 
Ravindranath S, Ho LC, Peng CY, Filippenko AV, Sargent WLW. 2001. 
{\it Astron. J.} 122:653

\nhi 
Rees MJ, Phinney ES, Begelman MC, Blandford RD. 1982. {\it Nature} 295:17

\nhi 
Reeves JN, Turner MJL. 2000. {\it MNRAS} 316:234

\nhi 
Renzini A, Greggio L, di Serego Alighieri S, Cappellari M, Burstein
 D, Bertola F. 1995. {\it Nature} 378:39

\nhi 
Reynolds CS, Di Matteo T, Fabian AC, Hwang U, Canizares CR. 1996. {\it MNRAS} 
283:L111

\nhi 
Rice MS, Martini P, Greene JE, Pogge RW, Shields JC, et al. 2006. {\it Ap. J.} 
636:654

\nhi 
Richstone DO. 2004. In {\it Carnegie Observatories Astrophysics Series, Vol 1:
Coevolution of Black Holes and Galaxies}, ed. LC Ho, p. 280. Cambridge:
Cambridge Univ. Press

\nhi 
Rieke GH, Lebofsky MJ, Kemp JC. 1982. {\it Ap. J.} 252:L53

\nhi 
Rinn AS, Sambruna RM, Gliozzi M. 2005. {\it Ap. J.} 621:167

\nhi 
Roberts TP, Schurch NJ, Warwick RS. 2001. {\it MNRAS} 324:737

\nhi 
Roberts TP, Warwick RS. 2000. {\it MNRAS} 315:98

\nhi 
Rola C, Terlevich E, Terlevich R. 1997. {\it MNRAS} 289:419

\nhi 
Rose JA, Searle L. 1982. {\it Ap. J.} 253:556

\nhi 
Rose JA, Tripicco MJ. 1984. {\it Ap. J.} 285:55

\nhi 
Rubin VC, Ford WK Jr, Thonnard N. 1980. {\it Ap. J.} 238:471

\nhi 
Rupke D, Veilleux S, Kim D-C, Sturm E, Contursi A, et al. 2007. In {\it The 
Central Engine of Active Galactic Nuclei}, ed. LC Ho, J-M Wang, p. 525.
San Francisco: ASP

\nhi 
Sabra BM, Shields JC, Ho LC, Barth AJ, Filippenko AV 2003. {\it Ap. J.} 584:164

\nhi 
Sadler EM, Jenkins CR, Kotanyi CG. 1989. {\it MNRAS} 240:591

\nhi 
Sambruna RM, Gliozzi M, Eracleous M, Brandt WN, Mushotzky RF. 2003. 
{\it Ap. J.} 586:L37

\nhi 
Sandage A, Bedke J. 1994. {\it The Carnegie Atlas of Galaxies}.
Washington, DC: Carnegie Inst. of Washington

\nhi 
Sandage A, Tammann, GA. 1981. {\it A Revised Shapley-Ames Catalog of
 Bright Galaxies}.  Washington, DC: Carnegie Inst. of Washington

\nhi 
Sargent WLW, Filippenko AV. 1991. {\it Astron. J.} 102:107

\nhi 
Sarzi M, Shields JC, Pogge RW, Martini P. 2007.  In {\it The Central Engine of 
Active Galactic Nuclei}, ed. LC Ho, J-M Wang, p. 643.  San Francisco: ASP

\nhi 
Sarzi M, Rix H-W, Shields JC, Ho LC, Barth AJ, et al.
2005. {\it Ap. J.} 628:169

\nhi 
Satyapal S, Dudik RP, O'Halloran B, Gliozzi M. 2005. {\it Ap. J.} 633:86

\nhi 
Satyapal S, Sambruna RM, Dudik RP. 2004. {\it Astron. Astrophys.} 414:825

\nhi 
Satyapal S, Vega D, Dudik RP, Abel NP, Heckman T. 2008. {\it Ap. J.} in press

\nhi 
Satyapal S, Vega D, Heckman T, O'Halloran B, Dudik R. 2007. {\it Ap. J.} 663:L9

\nhi 
Schmidt M, Green RF. 1983. {\it Ap. J.} 269:352

\nhi 
Schmitt HR. 2001. {\it Astron. J.} 122:2243

\nhi 
Schmitt HR, Kinney AL, Ho LC, eds. 1999. {\it The AGN/Normal Galaxy 
Connection} Oxford: Elsevier Science Ltd.

\nhi 
Schulz H, Fritsch C. 1994. {\it Astron. Astrophys.} 291:713

\nhi
Serote-Roos M, Gon\c{c}alves AC. 2004. {\it Astron. Astrophys.} 413:91

\nhi 
Seth AC, Ag\"ueros M, Lee D, Basu-Zych A. 2008. {\it Ap. J.} in press

\nhi 
Shakura NI, Sunyaev RA. 1973. {\it Astron. Astrophys.} 24:337

\nhi 
Shang Z, Brotherton MS, Green RF, Kriss GA, Scott J, et al. 2005. 
{\it Ap. J.} 619:41

\nhi 
Shields GA. 1978. {\it Nature}. 272:706

\nhi 
Shields GA, Wheeler JC. 1978. {\it Ap. J.} 222:667

\nhi 
Shields JC. 1992. {\it Ap. J.} 399:L27

\nhi 
Shields JC, Rix H-W, McIntosh DH, Ho LC, Rudnick G, et al.
2000. {\it Ap. J.} 534:L27

\nhi 
Shields JC, Rix H-W, Sarzi M, Barth AJ, Filippenko AV, et al. 2007. 
{\it Ap. J.} 654:125

\nhi 
Shields JC, Sabra BM, Ho LC, Barth AJ, Filippenko AV. 2002. In {\it Mass 
Outflow in Active Galactic Nuclei: New Perspectives}, ed. DM Crenshaw, SB 
Kraemer, IM George, p. 105. San Francisco: ASP

\nhi 
Shields JC, Walcher CJ, B\"oker, T, Ho LC, Rix H-W, van~der~Marel RP. 2008. 
{\it Ap. J.} in press

\nhi 
Shih DC, Iwasawa K, Fabian AC. 2003. {\it MNRAS} 341:973

\nhi 
Siemiginowska A, Czerny B, Kostyunin V. 1996. {\it Ap. J.} 458:491

\nhi 
Sikora M, Stawarz L, Lasota J-P. 2007. {\it Ap. J.} 658:815

\nhi 
Slee OB, Sadler EM, Reynolds JE, Ekers RD. 1994. {\it MNRAS} 269:928

\nhi
Smith JDT, Draine BT, Dale DA, Moustakas J, Kennicutt RC, et al. 2007. 
{\it Ap. J.} 656:770

\nhi
So\l tan A. 1982. {\it MNRAS} 200:115

\nhi
Soria R, Fabbiano G, Graham A, Baldi A, Elvis M, et al.
2006. {\it Ap. J.} 640:126

\nhi
Spinoglio L, Malkan MA. 1992. {\it Ap. J.} 399:504

\nhi
Springel V, Di~Matteo T, Hernquist L. 2005. {\it MNRAS} 361:776

\nhi
Starling RLC, Page MJ, Branduardi-Raymont G, Breeveld AA, Soria R, Wu K. 2005. 
{\it MNRAS} 356:727

\nhi
Stasi\'nska G. 1984. {\it Astron. Astrophys.} 135:341

\nhi 
Stasi\'nska G, Cid Fernandes R, Mateus A, Sodr\'e L Jr, Asari NV. 2006. 
{\it MNRAS} 371:972

\nhi 
Stauffer JR. 1982a. {\it Ap.~J.~Suppl.} 50:517

\nhi 
Stauffer JR. 1982b. {\it Ap. J.} 262:66

\nhi 
Stauffer JR, Spinrad H. 1979. {\it Ap. J.} 231:L51

\nhi 
Stone JD, Pringle JE. 2001. {\it MNRAS} 322:461

\nhi 
Storchi-Bergmann T, Baldwin JA, Wilson AS. 1993. {\it Ap. J.} 410:L11

\nhi 
Storchi-Bergmann T, Ho LC, Schmitt HR, eds. 2004. {\it IAU Symp. 222, 
Interplay among Black Holes, Stars and ISM in Galactic Nuclei} Cambridge: 
Cambridge Univ. Press

\nhi
Storchi-Bergmann T, Pastoriza MG. 1990. {\it PASP} 102:1359

\nhi
Strateva IV, Brandt WN, Schneider DP, Vanden Berk DG, Vignali C. 2005. 
{\it Astron. J.} 130:387

\nhi
Sturm E, Rupke D, Contursi A, Kim D-C, Lutz D, et al. 2006. {\it Ap. J.} 
653:L13

\nhi
Sturm E, Schweitzer M, Lutz D, Contursi A, Genzel R, et al. 2005. 
{\it Ap. J.} 629:L21

\nhi
Sugai H, Malkan MA. 2000. {\it Ap. J.} 529:219

\nhi 
Sulentic JW, Marziani P, Dultzin-Hacyan D. 2000. 
{\it Annu. Rev. Astron. Astrophys.} 38:521

\nhi 
Swartz DA, Ghosh KK, Suleimanov V, Tennant AF, Wu K. 2002. {\it Ap. J.} 574:382

\nhi 
Szokoly GP, Bergeron J, Hasinger G, Lehmann I, Kewley L, et al. 2004.
{\it Ap.~J.~Suppl.} 155:271

\nhi 
Tan JC, Blackman EG. 2005. {\it MNRAS} 362:983

\nhi 
Taniguchi Y, Shioya Y, Murayama T. 2000. {\it Astron. J.} 120:1265

\nhi 
Terashima Y, Ho LC, Ptak AF. 2000. {\it Ap. J.} 539:161

\nhi 
Terashima Y, Ho LC, Ptak AF, Mushotzky RF, Serlemitsos PJ, et al.
2000a. {\it Ap. J.} 533:729

\nhi 
Terashima Y, Ho LC, Ptak AF, Yaqoob T, Kunieda H, et al.
2000b. {\it Ap. J.} 535:L79

\nhi 
Terashima Y, Iyomoto N, Ho LC, Ptak AF. 2002. {\it Ap.~J.~Suppl.} 139:1

\nhi 
Terashima Y, Kunieda H, Misaki K, Mushotzky RF, Ptak AF, Reichert GA. 1998a.
{\it Ap. J.} 503:212

\nhi 
Terashima Y, Ptak A, Fujimoto R, Itoh M, Kunieda H, et al.
1998b. {\it Ap. J.} 496:210

\nhi 
Terashima Y, Wilson AS. 2003a. {\it Ap. J.} 560:139

\nhi 
Terashima Y, Wilson AS. 2003b. {\it Ap. J.} 583:145

\nhi 
Terlevich R, Melnick J. 1985. {\it MNRAS} 213:841

\nhi 
Tran HD. 2001. {\it Ap. J.} 554:L19

\nhi 
Tremaine S, Gebhardt K, Bender R, Bower G, Dressler A, et al. 2002. 
{\it Ap. J.} 574:740

\nhi 
Tremonti CA, Heckman TM, Kauffmann G, Brinchmann J, Charlot S, et al. 
2004. {\it Ap. J.} 613:898

\nhi 
Turner TJ, Pounds KA. 1989. {\it MNRAS} 240:833

\nhi 
Tzanavaris P, Georgantopoulos I. 2007. {\it Astron. Astrophys.} 468:129

\nhi 
Ulvestad JS, Greene JE, Ho LC. 2007. {\it Ap. J.} 661:L159

\nhi 
Ulvestad JS, Ho LC. 2001a. {\it Ap. J.} 558:561

\nhi 
Ulvestad JS, Ho LC. 2001b. {\it Ap. J.} 562:L133

\nhi 
Ulvestad JS, Ho LC. 2002. {\it Ap. J.} 581:925

\nhi 
Ulvestad JS, Wilson AS. 1989. {\it Ap. J.} 343:659

\nhi 
Vanden Berk DE, Richards GT, Bauer A, Strauss MA, Schneider DP, et al. 2001. 
{\it Astron. J.} 122:549

\nhi 
Van Dyk SD, Ho LC. 1998. In {\it IAU Symp. 184, The Central Regions of the 
Galaxy and Galaxies}, ed. Y Sofue, p. 489. Dordrecht: Kluwer

\nhi 
Varano S, Chiaberge M, Macchetto FD, Capetti A. 2004. {\it Astron. Astrophys.} 
428:401

\nhi 
Veilleux S, Osterbrock DE. 1987. {\it Ap.~J.~Suppl.} 63:295

\nhi 
Verdoes Kleijn GA, Baum SA, de Zeeuw PT, O'Dea CP. 2002. 
{\it Astron. J.} 123:1334

\nhi 
Verdoes Kleijn GA, van~der~Marel RP, Noel-Storr J. 2006. 
{\it Astron. J.} 131:1961

\nhi 
V\'eron P, Gon\c{c}alves AC, V\'eron-Cetty M-P. 1997. 
{\it Astron. Astrophys.} 319:52

\nhi 
V\'eron P, V\'eron-Cetty M-P. 1986. {\it Astron. Astrophys.} 161:145

\nhi
V\'eron-Cetty M-P, V\'eron P. 1986. {\it Astron. Astrophys. Suppl.} 66:335

\nhi 
V\'eron-Cetty M-P, V\'eron P. 2006. {\it Astron. Astrophys.} 455:773

\nhi 
Viegas-Aldrovandi SM, Gruenwald RB. 1990. {\it Ap. J.} 360:474

\nhi 
Voit GM. 1992. {\it Ap. J.} 399:495

\nhi 
Voit GM, Donahue M. 1997. {\it Ap. J.} 486:242

\nhi 
Wake DA, Miller CJ, Di~Matteo T, Nichol RC, Pope A, et al. 2004. 
{\it Ap. J.} 610:L85

\nhi 
Walcher CJ, B\"oker T, Charlot S, Ho LC, Rix H-W, et al. 
2006. {\it Ap. J.} 649:692

\nhi 
Walsh JL, Barth AJ, Ho LC, Filippenko AV, Rix H-W, et al. 2008. {\it Ap. J.} 
submitted

\nhi 
Wang J-M, Luo B, Ho LC. 2004. {\it Ap. J.} 615:L5

\nhi 
Wang J-M, Staubert R, Ho LC. 2002. {\it Ap. J.} 579:554

\nhi 
Weaver KA, Wilson AS, Henkel C, Braatz JA. 1999. {\it Ap. J.} 520:130

\nhi 
Weedman DW, Feldman FR, Balzano VA, Ramsey LW, Sramek RA, Wu C-C. 1981. 
{\it Ap. J.} 248:105

\nhi 
Whysong D, Antonucci R. 2004. {\it Ap. J.} 602:116

\nhi 
Willner SP, Ashby MLN, Barmby P, Fazio GG, Pahre M, et al. 2004. 
{\it Ap.~J.~Suppl.} 154:222 

\nhi 
Willner SP, Elvis M, Fabbiano G, Lawrence A, Ward MJ. 1985. {\it Ap. J.} 299:443

\nhi 
Wilson AS 1979. {\it Proc. Roy. Soc. London} A366:461

\nhi 
Worrall DM, Birkinshaw M. 1994. {\it Ap. J.} 427:134

\nhi 
Wo\'zniak PR, Zdziarski AA, Smith D, Madejski GM, Johnson WN. 1998. {\it MNRAS}
299:449

\nhi 
Wrobel JM. 1991. {\it Astron. J.} 101:127

\nhi 
Wrobel JM, Fassnacht CD, Ho LC. 2001. {\it Ap. J.} 553:L23

\nhi 
Wrobel JM, Heeschen DS. 1991. {\it Astron. J.} 101:148

\nhi 
Wrobel JM, Ho LC. 2006. {\it Ap. J.} 646:L95

\nhi 
Wrobel JM, Terashima Y, Ho LC. 2008. {\it Ap. J.} in press

\nhi 
Wu Q, Cao X. 2005. {\it Ap. J.} 621:130

\nhi 
Wu Q, Yuan F, Cao X. 2007. {\it Ap. J.} 669:96

\nhi 
Xu Y, Cao X-W. 2007. {\it ChJAA} 7:63

\nhi 
Yan R, Newman JA, Faber SM, Konidaris N, Koo D, Davis M. 2006. {\it Ap. J.} 
648:281

\nhi 
Yaqoob T, Serlemitsos PJ, Ptak A, Mushotzky RF, Kunieda H, Terashima Y. 1995. 
{\it Ap. J.} 455:508

\nhi 
Young AJ, Nowak MA, Markoff S, Marshall HL, Canizares CR. 2007. {\it Ap. J.} 
669:830

\nhi 
Yu Q, Tremaine S. 2002. {\it MNRAS} 335:965

\nhi 
Yuan F. 2007. In {\it The Central Engine of Active Galactic Nuclei}, ed. LC 
Ho, J-M Wang, p. 95.  San Francisco: ASP

\nhi 
Yuan F, Cui W. 2005. {\it Ap. J.} 629:408

\nhi 
Yuan F, Markoff S, Falcke H. 2002. {\it Astron. Astrophys.} 383:854

\nhi 
Yuan F, Markoff S, Falcke H, Biermann PL. 2002. {\it Astron. Astrophys.} 
391:139

\nhi 
Yuan F, Narayan R. 2004. {\it Ap. J.} 612:724

\nhi 
Zakamska NL, Strauss MA, Krolik JH, Collinge MJ, Hall PB, et al. 2003. 
{\it Astron. J.} 126:2125

\nhi 
Zamorani G, Henry JP, Maccacaro T, Tananbaum H, So\l tan A, et al. 1981. 
{\it Ap. J.} 245:357

\nhi 
Zezas A, Birkinshaw M, Worrall DM, Peters A, Fabbiano G. 2005. {\it Ap. J.} 
627:711

\nhi 
Zhang X-G, Dultzin-Hacyan D, Wang T-G. 2007. {\it MNRAS} 374:691

\nhi 
Zhang Y, Gu Q-S, Ho LC. 2008. {\it Astron. Astrophys.} submitted

%\end{thebibliography}

\vfill\eject

%\listoffigures

\end{document}